\documentclass[onecolumn,preprintnumbers,amsmath,amssymb]{revtex4}

\usepackage{graphicx}
\usepackage{dcolumn}
\usepackage{bm}
\usepackage{subfigure}

\begin{document}

\title{Aspects of cosmological expansion in F(R) gravity models}

\author{Stephen A. Appleby and Richard A. Battye \\ {\it Jodrell Bank Centre for Astrophysics, School of Physics and Astronomy,} \\ {\it
University of Manchester, Oxford Road, Manchester, M13 9PL}}

\date{\today}

\begin{abstract}

We study cosmological expansion in F(R) gravity using the trace of the field
equations. High frequency oscillations in the Ricci scalar, whose amplitude
increases as one evolves backward in time, have been predicted in recent
works. We show that the approximations used to derive this result very
quickly breakdown in any realistic model due to the non-linear nature of the
underlying problem. Using a combination of numerical and semi-analytic
techniques, we study a range of models which are otherwise devoid of known
pathologies. We find that high frequency asymmetric oscillations and a
singularity at finite time appear to be present for a wide range of initial
conditions. We show that this singularity can be avoided with a certain range of initial conditions, which we find by evolving the models forwards in time. In addition we show that the oscillations in
the Ricci scalar are highly suppressed in the Hubble parameter and scale
factor.

\end{abstract}

\maketitle

\section{Introduction}

It is now generally accepted that the Universe is currently undergoing a period of accelerated expansion. The most popular approach to modeling the current epoch is to postulate a new energy component of the Universe, dark energy, which has negative pressure and hence drives the acceleration. Possible dark energy candidates include the cosmological constant \cite{cc1}, which has well known fine tuning issues, quintessence fields (see for example \cite{qn1} and references therein) and elastic dark energy \cite{ede1}.

An alternative approach is to modify the theory of gravity, and there have been a number of attempts to construct late time accelerating solutions to the gravitational field equations
by modifying gravity at large distances/late times. One such class of models are found by considering a gravitational action that contains an arbitrary function $F(R)$ of the Ricci scalar. The first modified gravity model to be considered introduced quadratic terms into the gravitational action \cite{st1,st2,st3,vk1}. Such models can give rise to early-time inflationary solutions to the gravitational field equations, where the Hubble parameter is initially $H \sim M_{\rm pl}$, and slowly rolls towards a stable Minkowski vacuum state.

Modified gravity models which yield late-time acceleration have been considered recently.
An important example is the CDDTT model \cite{ca1}, for which $F(R) = R - \mu^{4}/R$. This function has a de-Sitter vacuum solution to the field equations, which might be associated with the current epoch of the Universe with a suitable tuning of the mass scale $\mu$. However, it has been shown
that this model cannot satisfy local gravity constraints \cite{c1}, and also has an instability \cite{d1}. In addition to the CDDTT model, a number of modified gravity functions have been proposed which give rise to late time acceleration \cite{db1,df1,teg1,cr6,at10,fa1,cap1,no10,cr4,am1,aa2,di1,sx1}, and some progress has been made in understanding general properties of these models, for example the behaviour of metric perturbations \cite{as1,hu1} and the cosmological evolution of generic $F(R)$ functions \cite{ame1,be1,nn1} have been considered.

Any $F(R)$ model must satisfy certain conditions in order to exclude the possibility of ghost degrees of freedom and other instabilities. These conditions are $F'(R) >0$, which ensures that there are no ghost degrees of freedom in the model, and $F''(R) >0$,
which prevents instabilities from arising in the early Universe \cite{hu}. It seems also sensible to impose $F(0)=0$, which ensures that there is no cosmological constant, and $F(R) \to R$
for large $R$, since we want any modification to gravity to only become significant at late times. In addition to these conditions, it has been shown that if a standard matter era (for which $a \propto t^{2/3}$) is to be a fixed point then further conditions must be satisfied \cite{am1}.

Recently, a number of models have been studied in the literature which satisfy the above constraints, at least for $R>0$. Specifically the following functions have been considered \cite{hu,st,ap},

\begin{align}\label{eq:cr1}  &F_{\rm HSS}(R) = R - {R_{\rm vac} \over 2}{c \left({R \over R_{\rm vac}}\right)^{2n} \over 1+ c\left({R \over R_{\rm vac}}\right)^{2n}} ,\\ \label{eq:cr201}
&F_{\rm AB}(R) = {R \over 2} + {R_{\rm vac} \over 4b}\log\left[\cosh\left({2bR \over R_{\rm vac}}\right)- \tanh b \sinh\left({2bR \over R_{\rm vac}}\right)\right] ,\end{align}

\noindent where $R_{\rm vac}$ is the current vacuum curvature of the Universe, $b,c$ are dimensionless constants, and $n >0$. In ref.\cite{hu}, the $n$ is defined differently to ($\ref{eq:cr1}$), and in ref.\cite{st} the precise formula was slightly different to ($\ref{eq:cr1}$). However, both models considered in \cite{hu,st} have the same expansion in the limit $R \gg R_{\rm vac}$, which will be the important point in this paper. The model $F_{\rm HSS}$ represents a modification to General Relativity that is a power law in $R$, whilst $F_{\rm AB}$ contains exponential corrections. These models both have late-time accelerating epochs, and
can satisfy local tests of gravity. They can also satisfy the conditions discussed in ref.\cite{am1}.

Although the models ($\ref{eq:cr1}$,$\ref{eq:cr201}$) possess many desirable features, an additional issue has arisen recently with regards to their suitability as viable
gravitational models \cite{st,ts10}. Specifically, it has been suggested that if we look for solutions to the gravitational field equations that are perturbations
around known General Relativistic solutions, then
the Ricci scalar oscillates over very short timescales, and the amplitude of these oscillations will increase without bound to the past \cite{st}. If true, this presents us with a number
of issues, amongst which are that the small perturbation will become much larger than the General Relativistic solution at some time in the past, violating the perturbative assumption, and $R$ will become
negative at some finite time. Such phenomena would most likely lead to significant problems in reproducing the standard Cosmology.

The aim of this paper is to study the two models ($\ref{eq:cr1}$,$\ref{eq:cr201}$) in detail. We find that since $R$ oscillates over such small timescales,
the linearized analysis presented in refs.\cite{st,sh20} will not necessarily yield a solution that is indicative of the full solution for all times.
We then solve the gravitational field equations numerically and semi-analytically and show that the Ricci scalar undergoes asymmetric oscillations, and will drift away from the General Relativistic limit. We note that non-linear oscillations have been considered in $F(R)$ models previously, see ref.\cite{stf}.

\section{\label{sec:1} Modified gravity formalism}

In this section we consider the modified gravity action

\begin{equation} S = \int \sqrt{-g} d^{4}x \left( M_{\rm pl}^{2}F(R) + {\cal L}_{\rm m} \right) , \end{equation}

\noindent which yields the following field equations

\begin{equation} F'(R) R_{\mu\nu} - {1 \over 2}g_{\mu\nu}F(R) +\left[ g_{\mu\nu}\Box - \nabla_{\mu}\nabla_{\nu}\right] F'(R) = {T_{\mu\nu} \over M_{\rm pl}^{2}} ,\end{equation}

\begin{equation}\label{eq:1} RF'(R) - 2 F(R) + 3 \Box F'(R) = {T \over M_{\rm pl}^{2}} ,\end{equation}

\noindent where primes denote differentiation with respect to $R$, $T_{\mu\nu}$ is the energy momentum tensor and $M_{\rm pl}^{2}$ is the reduced Planck mass.

The HSS and AB models can be expanded as

\begin{equation}\label{eq:2} F(R) \approx R - {R_{\rm vac} \over 2} + \chi(R) , \end{equation}

\noindent for $R >R_{\rm vac}$, where $R_{\rm vac}$ acts as a small `cosmological constant' (although there is no true constant in these models since globally $F(0)=0$). $\chi(R)$, $\chi'(R)$ and $\chi''(R)$ are all small functions of $R$, in the sense that the dimensionless parameters $\chi(R)/R$, $\chi'(R)$ and $R\chi''(R)$ all satisfy $\chi(R)/R \ll 1$, $\chi'(R) \ll 1$ and $R\chi''(R) \ll 1$ for $R > R_{\rm vac}$. For the HSS and AB models, we have

\begin{align}\label{eq:cr50} \chi_{\rm HSS} = {\epsilon^{2n+1}_{\rm HSS} \over R^{2n}}, \qquad    \chi_{\rm AB} = \epsilon_{\rm AB}\exp\left(-R/ \epsilon_{\rm AB}\right), \end{align}

\noindent where $R_{\rm vac}$ is the current vacuum curvature of the Universe, and $\epsilon_{\rm AB} = R_{\rm vac}/4b$ and $\epsilon_{\rm HSS} = R_{\rm vac}/ (2c)^{1/(2n+1)}$ are parameters that are smaller than $R_{\rm vac}$. These expansions are valid for $R \gg \epsilon_{\rm HSS}$ and $R \gg \epsilon_{\rm AB}$ for the HSS and AB models respectively. For the rest of this paper, we will drop the subscripts $_{\rm AB}$ and $_{\rm HSS}$, and use $\epsilon = \epsilon_{\rm HSS}$ or $\epsilon_{\rm AB}$ and $\chi = \chi_{\rm HSS}$ or $\chi_{\rm AB}$, which should be obvious in context. We  will explicitly state which model is being studied in each section.  Finally, we note that we will frequently make use of the notation $R_{\rm GR}$, which is the General Relativistic
solution to the field equations $R_{\rm GR} = -T/M_{\rm pl}^{2}$. In particular during the matter era we have $R_{\rm GR} = 4/3t^{2}$, and during the radiation era $R_{\rm GR} \propto t^{-3/2}$.

\section{\label{sec:am1} Perturbative analysis}

In this section, we review the perturbative analysis that has been used in refs.\cite{st,ts10,sh20} to derive approximate solutions to the gravitational field equations for the AB and HSS models.  The full field equations are a set of non-linear, fourth order differential equations for the scale factor, and solving these equations directly is difficult (although see for example ref.\cite{nnn1} for an attempt to do so for the CDDTT model). The approach taken in this section is to look for solutions to the field equations
that are perturbations around known General Relativistic solutions, that is we look for a solution to ($\ref{eq:1}$)
of the form $R = R_{GR} + \delta R $, where $\delta R \ll R_{GR}$ is some small perturbation.

This approach involves linearizing the equation in $\delta R$. In doing so, the following expansions are used

\begin{align}\label{eq:6} F(R) &\approx R_{GR} + \delta R - {R_{\rm vac} \over 2} + \chi(R_{GR}) + \chi'(R_{GR}) \delta R + {\cal O}(\delta R)^{2} ,\\
F'(R) &\approx 1 + \chi'(R_{GR}) + \chi''(R_{GR}) \delta R + {\cal O}(\delta R)^{2} ,\\ \label{eq:201} F''(R) &\approx \chi''(R_{GR}) + \chi'''(R_{GR})\delta R + {\cal O}(\delta R)^{2} . \end{align}

\noindent In which case ($\ref{eq:1}$) becomes

\begin{equation} \label{eq:4} 3\chi'' \Box \delta R + 6\chi'''\nabla_{\alpha}R_{GR} \nabla^{\alpha} \delta R - \delta R = \alpha(R_{GR}) ,\end{equation}

\noindent to linear order in $\delta R$. Unless otherwise stated, in ($\ref{eq:4}$) and
for the remainder of this paper $\chi$ is taken to be a function of $R_{\rm GR}$ only, $\chi = \chi(R_{\rm GR})$. The function $\alpha(R_{GR})$ is

 \begin{equation}\label{eq:fb1} \alpha(R_{GR}) = 2\chi - R_{GR}\chi' -3 \chi''' (\nabla R_{GR})^{2} - 3\chi'' \Box R_{GR} ,\end{equation}

 \noindent  which is a function of $R_{\rm GR}$ only and is small. The field equation ($\ref{eq:4}$) is an inhomogeneous, second order differential equation for $\delta R$. The solution is the linear sum of a particular solution to ($\ref{eq:4}$) and the solution to the homogeneous equation

\begin{equation} \label{eq:5} 3\chi'' \Box \delta R + 6\chi'''\nabla_{\alpha}R_{GR} \nabla^{\alpha} \delta R - \delta R = 0 ,\end{equation}

\noindent which can be written approximately as

\begin{equation} \label{eq:gh1}{3 \over a^{3}} {d \over dt} \left( a^{3} {d \over dt}\left( \chi'' \delta R \right)\right) + \delta R = 0 ,\end{equation}

\noindent where $a(t)$ is the scale factor.

\begin{figure}[htp]
     \centering
     \subfigure[]{
          \label{fig:1}
          \includegraphics[scale=0.4]{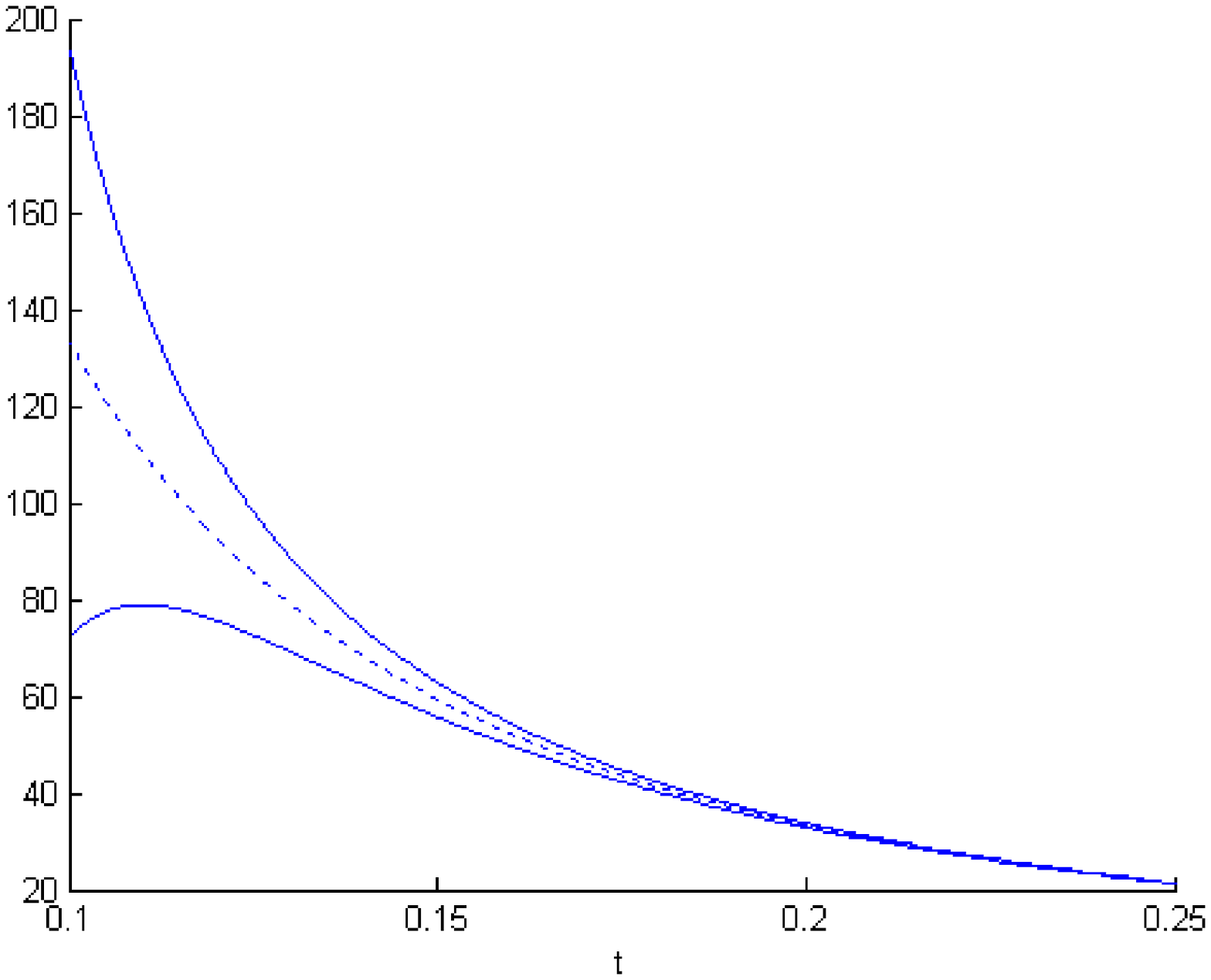}}
     \hspace{.2in}
     \subfigure[]{
          \label{fig:2}
                \includegraphics[scale=0.4]{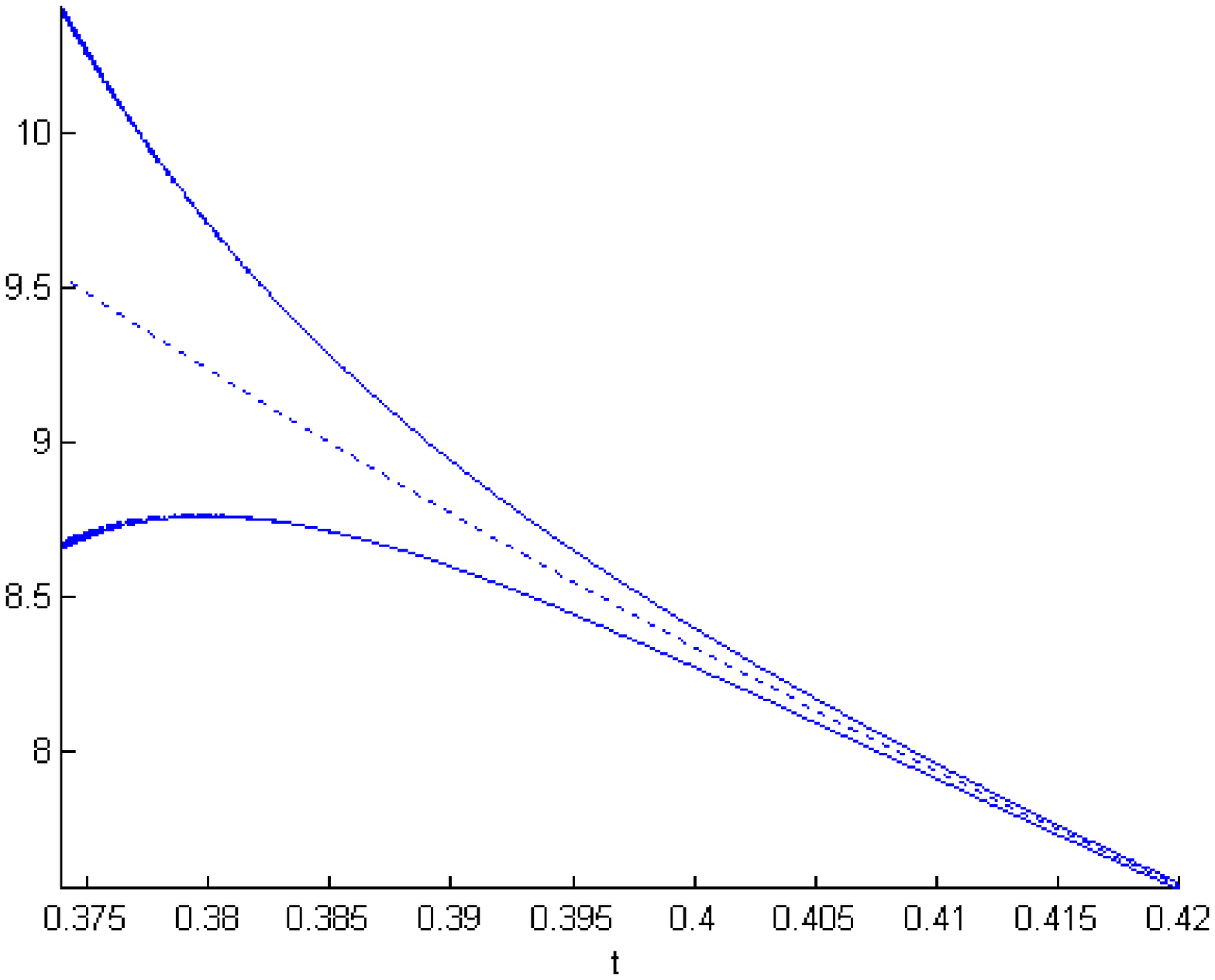}}
     \caption{(a) $R = R_{\rm GR} + \delta R$ as predicted in ref.\,\cite{st} for the HSS model. The solid lines are the upper and lower envelopes of the solution, and the dashed line is $R_{\rm GR}$. We note the turning point in the lower envelope, at which
     point the amplitude of $\delta R$ can become larger than $R_{\rm GR}$; (b) $R = R_{\rm GR} + \delta R$ for the AB model, where once again we observe a turning point in the lower envelope.}
     \label{fig:b1}
\end{figure}

\subsection{\label{sec:ss1} HSS model}

In ref.\cite{st}, a solution to ($\ref{eq:gh1}$) was derived for the HSS model. Using $\chi = \chi_{\rm HSS}$ and $\epsilon = \epsilon_{\rm HSS}$, the equation for $\delta R$ reads

\begin{equation}\label{eq:l2} \delta \ddot{R} + \left( 3H - {4(n+1)\dot{R}_{\rm GR} \over R_{\rm GR}} \right) \delta \dot{R}
+ {R_{\rm GR}^{2n+2} \over 6n(2n+1)\epsilon^{2n+1}} \delta R \approx 0 ,\end{equation}

\noindent A WKB solution can obtained for the matter era,

\begin{equation}\label{eq:a1} \delta R_{\rm mat} =  t^{-3n-4}\left[A_{1} \sin\left(A_{2} t^{-2n-1}\right) +A_{3} \cos\left(A_{2} t^{-2n-1}\right)\right] ,\end{equation}

\noindent and for the radiation era

\begin{equation} \label{eq:ff1} \delta R_{\rm rad} = t^{-(9n/4)-3}\left[A_{4} \sin\left(A_{5} t^{-(3n+1)/2}\right) + A_{6} \cos\left(A_{5} t^{-(3n+1)/2}\right) \right] ,\end{equation}

\noindent where $A_{1},A_{2},A_{3},A_{4},A_{5},A_{6}$ are constants.

We have solved equation ($\ref{eq:l2}$) numerically, and $R = R_{\rm GR}+\delta R$ is shown in fig.\ref{fig:1}, taking a pure matter era as an example, so $R_{\rm GR} = 4/3t^{2}$. We have used time coordinates such that $R_{\rm vac}=1$, and have chosen $\epsilon = ((c_{2}/2c_{1})^{2n} /2c_{2})^{1/(2n+1)}R_{\rm vac}=0.1$ and $n=1$. We have evolved backwards in the time coordinate, over the range $t = (0.25,0.1)$, using the initial conditions $\delta R(t_{\rm i}) = 0.1$, $\delta \dot{R}(t_{\rm i}) = 0$, where $t_{\rm i} = 0.25$. By solving equation ($\ref{eq:l2}$) numerically, we have found that $\delta R$ oscillates symmetrically around zero, confirming the behaviour expected from the WKB solutions. However, in fig.\ref{fig:1} we have not exhibited the actual oscillations of $R$ explicitly, since they are of too high frequency to be resolved, and so have only presented $R_{\rm GR}$ and the upper and lower envelopes of the oscillations.

From fig.\ref{fig:1}, we see that the lower envelope appears to have a turning point, indicating that in this approach after this time the oscillatory component $\delta R$ will come to dominate over $R_{\rm GR}$, and hence the Ricci scalar will become negative at some point in the past.

\subsection{AB model}

In refs. \cite{ts10,sh20}, a similar analysis has been considered for the model
$F(R) = R - R_{\rm vac}/2 + \chi_{\rm AB}$.
If we use $\chi_{\rm AB}$ in ($\ref{eq:gh1}$), we find the following equation

\begin{equation}\label{eq:lh1} \delta \ddot{R} + \left( 3H -{2 \dot{R}_{\rm GR} \over \epsilon}\right) \delta \dot{R} + {\epsilon e^{R_{\rm GR}/\epsilon}\over 3} \delta R \approx 0,\end{equation}

\noindent which has the WKB solution

\begin{equation}\label{eq:ab1} \delta R_{\rm mat} = {\exp(1/\epsilon t^{2}) \over t} \left[B_{1} \sin\left(\int \exp(2/3\epsilon t^{2})dt \right) +B_{2} \cos\left(\int \exp(2/3\epsilon t^{2})dt \right)\right] ,\end{equation}

\noindent during the matter era and

\begin{equation} \label{eq:ab2} \delta R_{\rm rad} =t^{-3/4}\exp(3\alpha/4\epsilon t^{3/2})\left[ B_{3}  \sin \left( \int \exp(\alpha/2\epsilon t^{3/2}) dt \right) +B_{4}  \cos \left( \int \exp(\alpha/2\epsilon t^{3/2}) dt \right) \right],\end{equation}

\noindent for the radiation era, where $\alpha, B_{1}, B_{2},B_{3}, B_{4}$ are constants. We have solved equation ($\ref{eq:lh1}$) numerically, taking $R_{\rm GR} = 4/3t^{2}$, using time coordinates with $R_{\rm vac} = 1$ and choosing $\epsilon = 0.32$. The resulting $R = R_{\rm GR} + \delta R$ is shown in fig.\ref{fig:2}, using the initial conditions $\delta R(0.42) = 0.01$, $\delta \dot{R}(0.42) = 0$ over the time range $t=(0.42,0.375)$. Once again, we have shown only the envelopes of the oscillations of $R$. These solutions exhibit similar oscillatory and growing behaviour as found for the HSS model.

\subsection{Issues with the perturbative analysis}

Using this approach for both the HSS and AB models, we find similar behaviour for $R$. Specifically, $\delta R$ undergoes rapid oscillations, and both the amplitude and frequency of these oscillations increases to the past. Since the amplitude of $\delta R$ grows without bound as $t \to 0$, it follows that $\delta R$ will violate the condition $\delta R \ll R_{\rm GR}$ at some point in the past, at which point the perturbative analysis will break down. Beyond this, we cannot assume that $R = R_{\rm GR} + \delta R$ is a solution to the gravitational field equations.

This oscillatory behaviour presents us with a number of problems. The first is that since $\delta R$ grows to the past, it will eventually satisfy  $|\delta R| \sim R_{\rm GR}$.  Beyond this point, $R$ will periodically be negative, and when $R$ is negative we can no longer use the expansion $F(R) \approx R - R_{\rm vac}/2 + \chi(R)$.  This is a problem particularly relevant for the HSS model, since for $R<0$ there will be a point at which $F''(R) = 0$, which is a singular point in the field equations. The second problem associated with the oscillatory behaviour of $\delta R$ is that the frequency grows to the past without bound. It has been pointed out in ref.\cite{st} that this issue can be ameliorated by introducing an additional term $\propto R^{2}/M^{2}$ into the action, where $M$ is a mass scale. This will provide a cut off in the frequency growth at $\omega = M$.

\begin{figure}[b]
     \centering
     \subfigure[]{
          \label{fig:3}
          \includegraphics[scale=0.4]{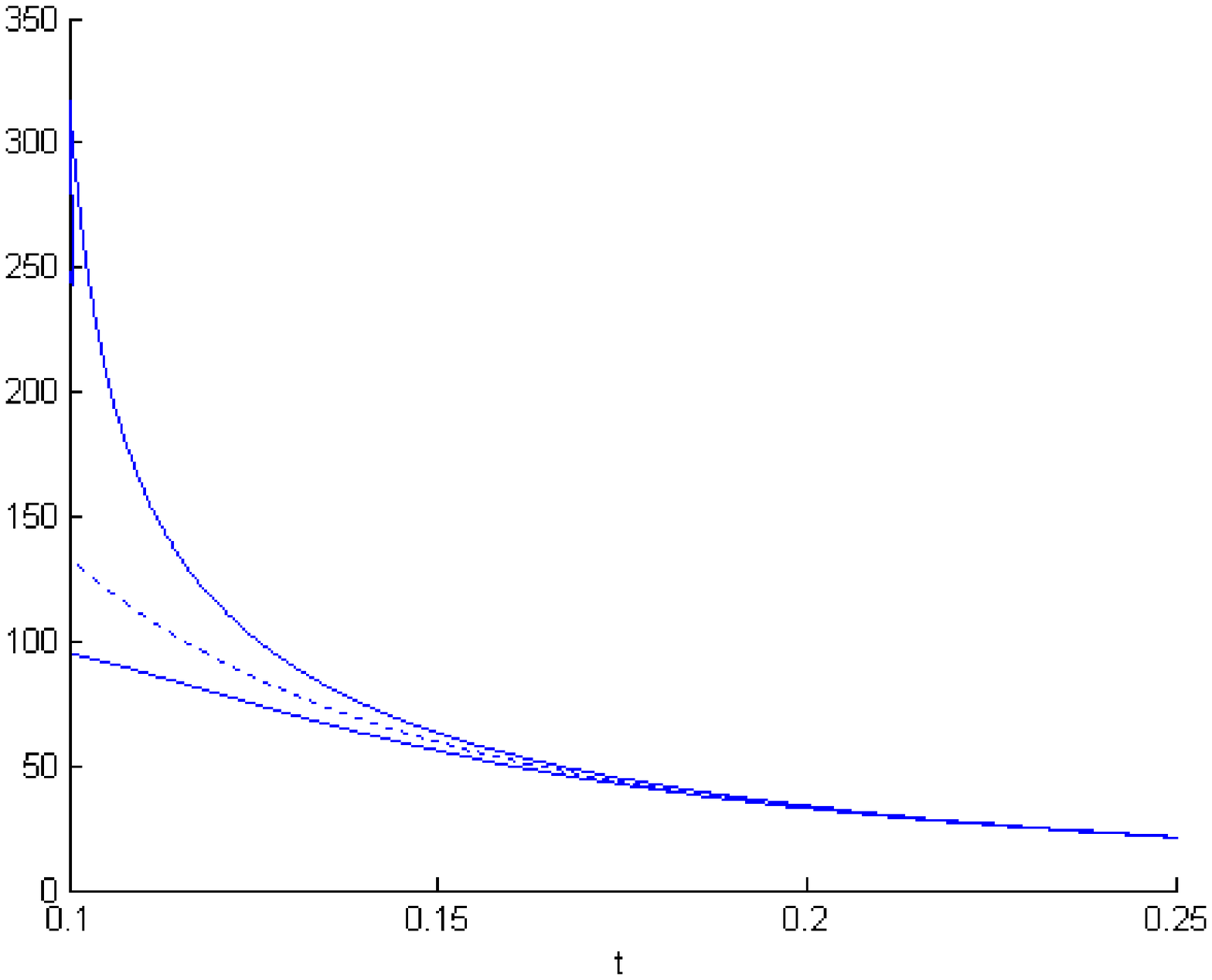}}
     \hspace{.2in}
     \subfigure[]{
          \label{fig:4}
                \includegraphics[scale=0.4]{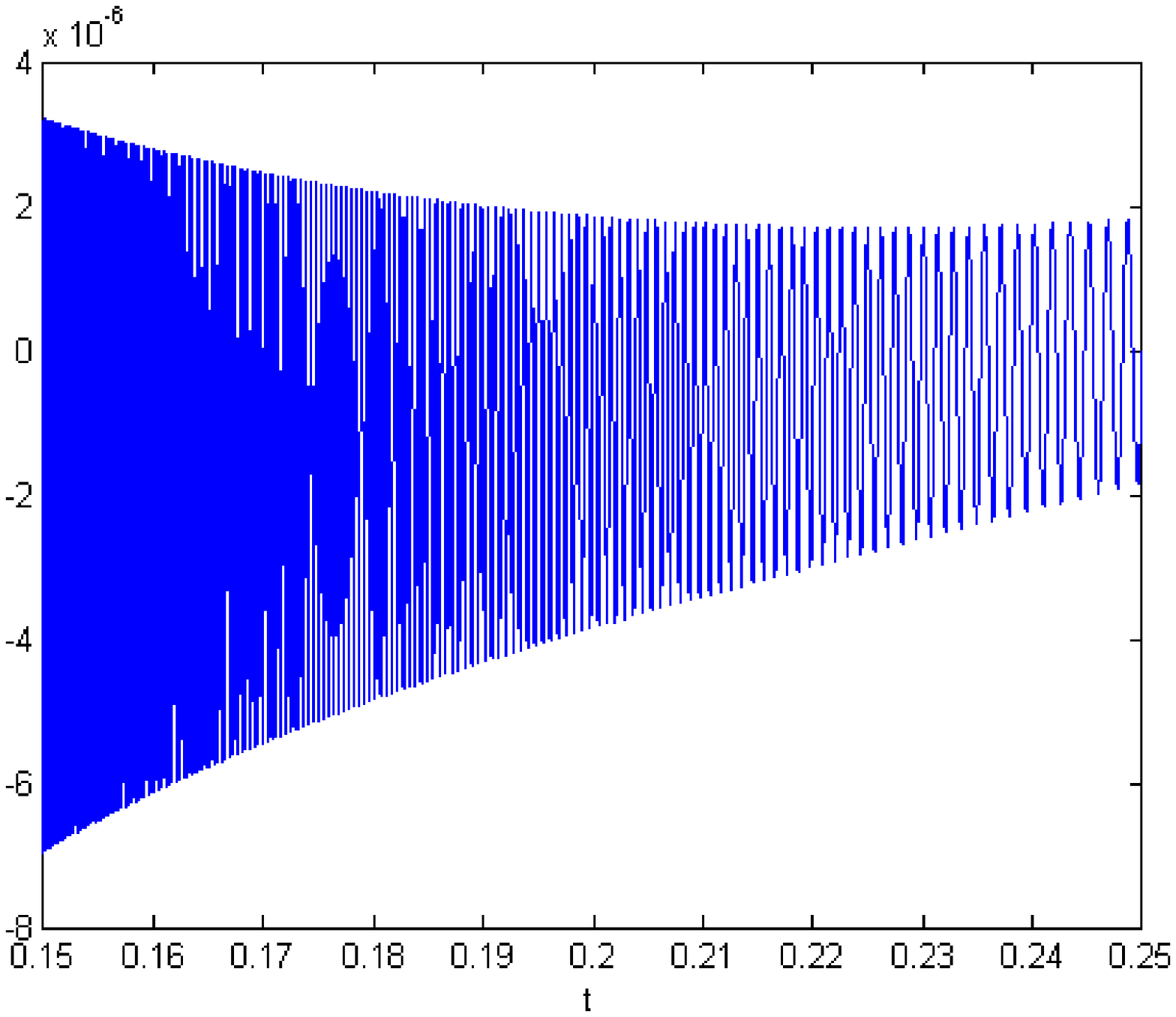}}\\
 \subfigure[]{
          \label{fig:5}
          \includegraphics[scale=0.4]{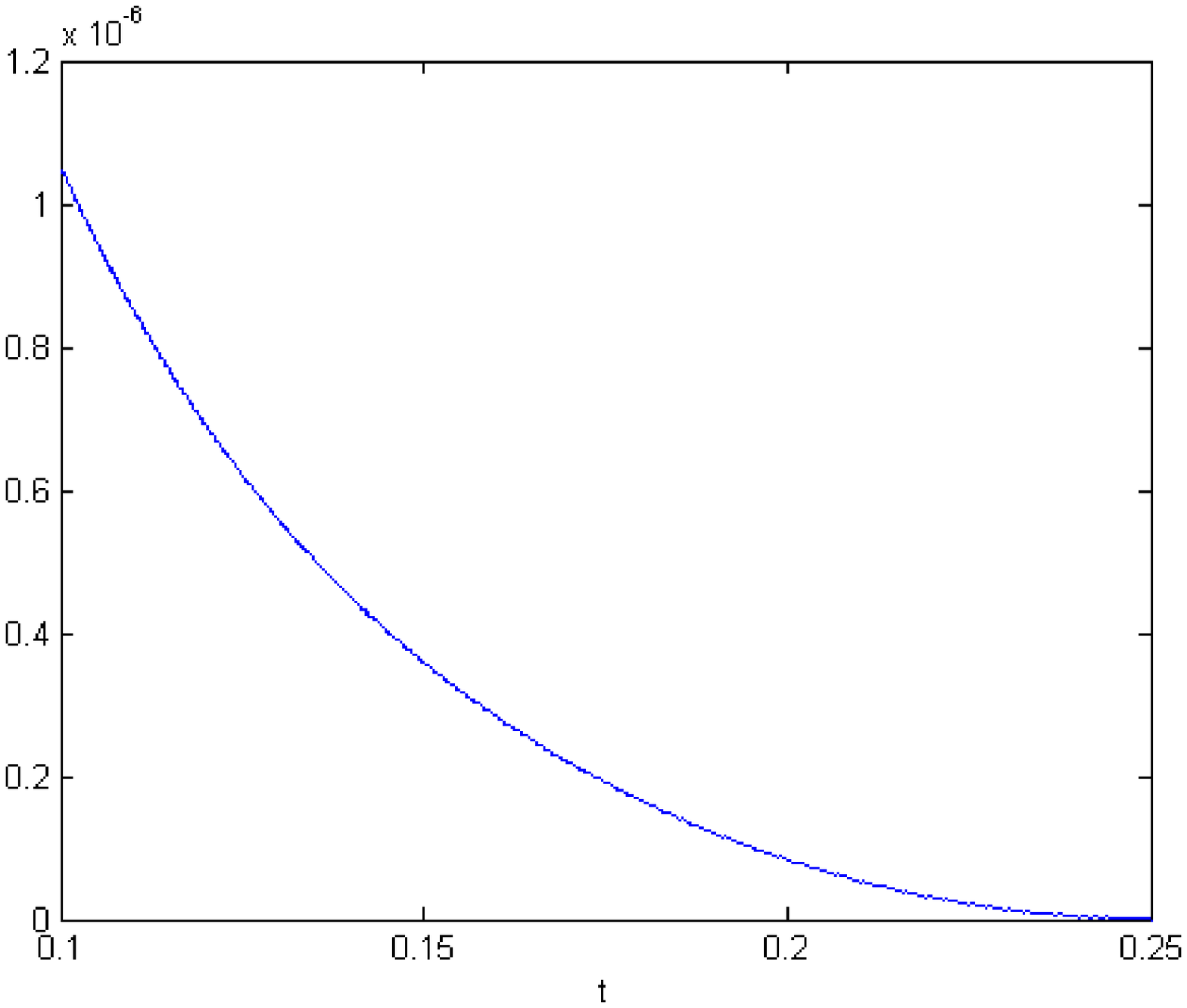}}
     \hspace{.2in}
 \subfigure[]{
          \label{fig:6}
          \includegraphics[scale=0.4]{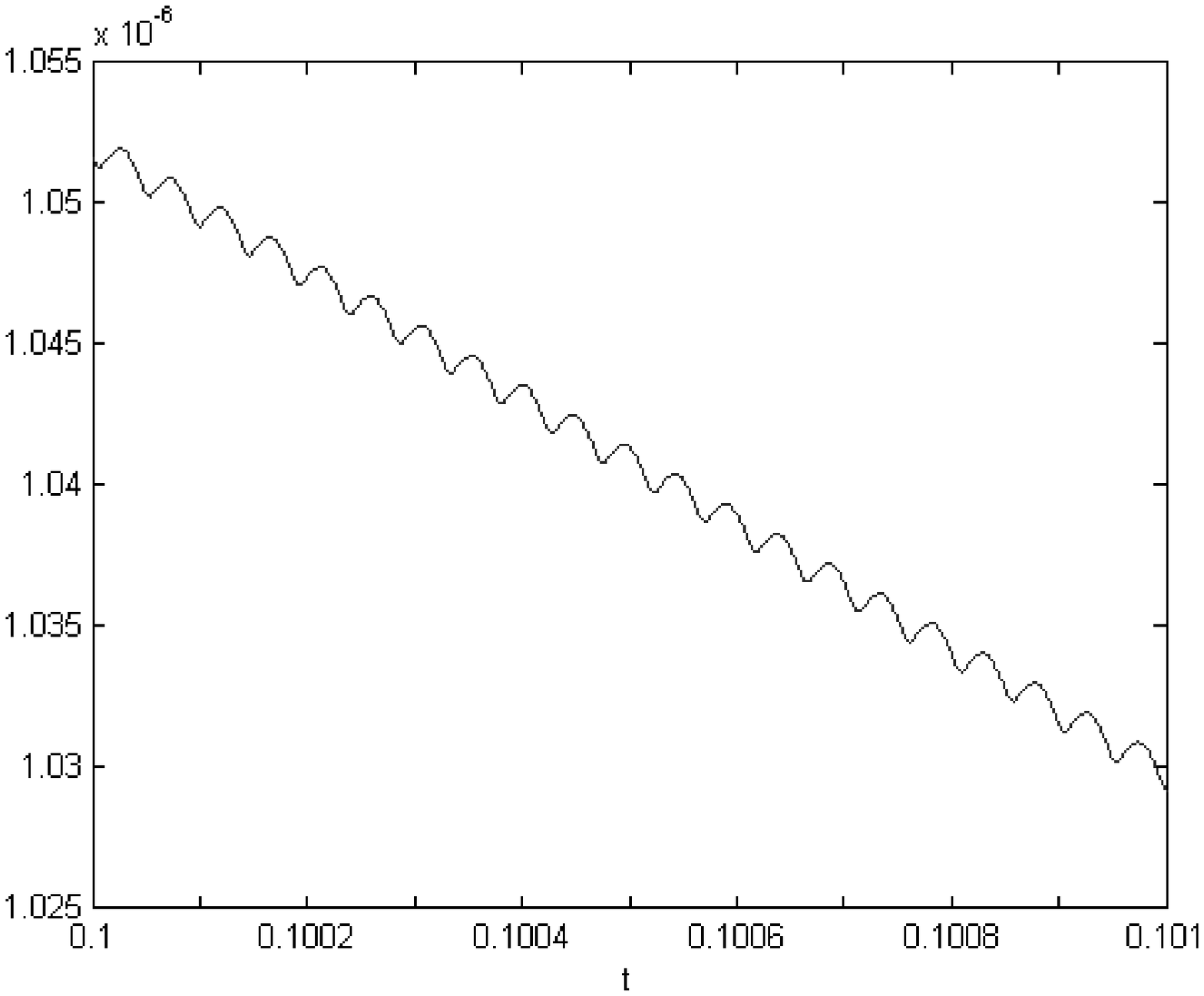}}
     \hspace{.2in}
     \caption{(a) the envelope functions for $R$ in the HSS model, with $\epsilon=0.1$, obtained by solving the full field equations numerically as described in the text, with perturbed initial conditions. We note that $R$ differs significantly from the linearized approximation $R = R_{\rm GR} + \delta R$, obtained in the previous section; (b) $\delta H = (H - H_{\rm GR})/H$. It is clear that $H$ oscillates, and the amplitude of these oscillations grows to the past; (c) the deviation of the scale factor from its General Relativistic limit, $\delta a = (a - a_{\rm GR})/a$. We see that $a$ will deviate from $a_{\rm GR}$ as we evolve backwards in time, however $\delta a$ is highly suppressed compared to the deviations in $R$; (d) $\delta a$ over a smaller range of $t$ close to the end point of the evolution. This shows the oscillatory behaviour of the scale factor which eventually develops.}
     \label{fig:f2}
\end{figure}

\begin{figure}[htp]
     \centering
     \subfigure[]{
          \label{fig:7}
          \includegraphics[scale=0.4]{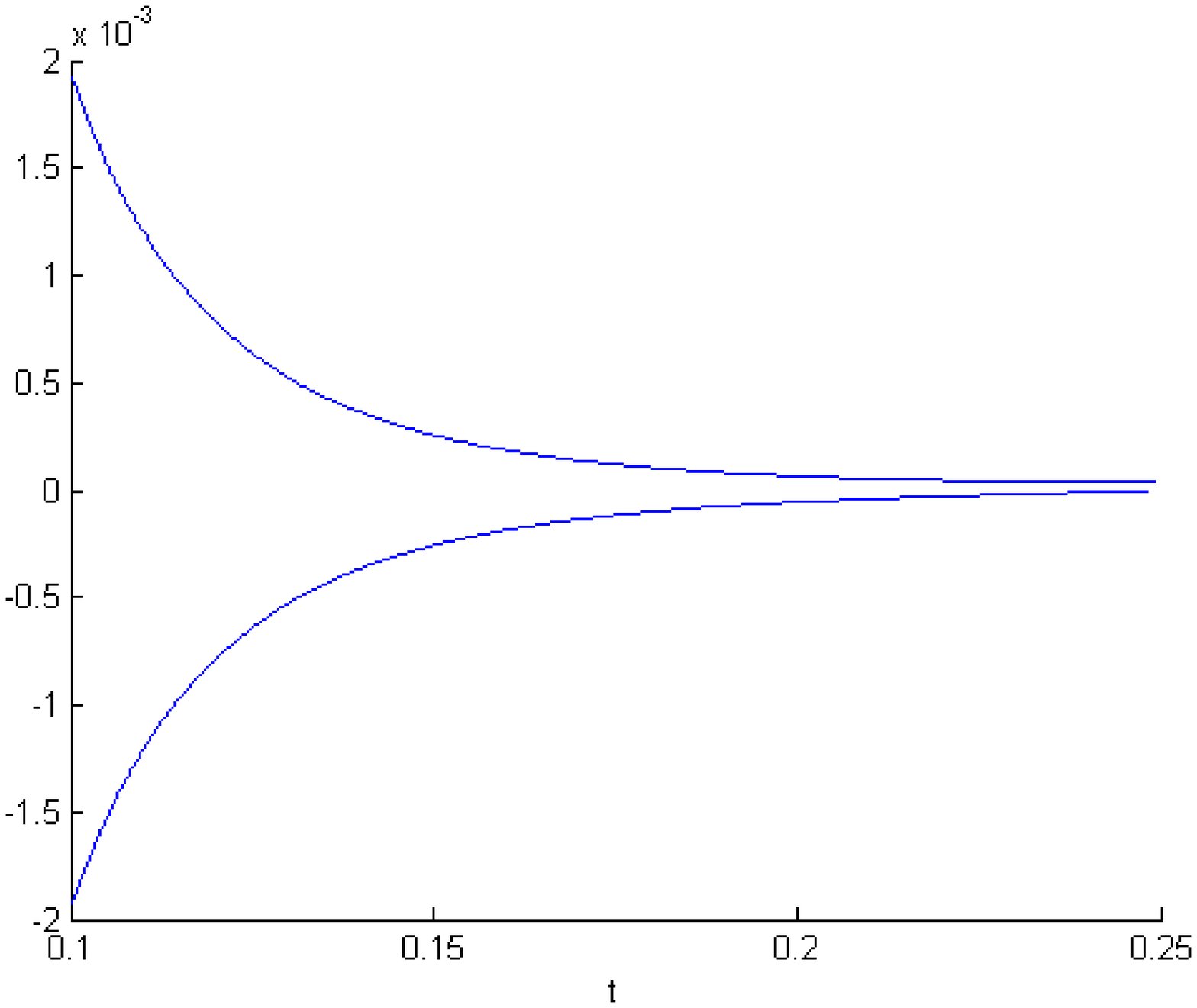}}
     \hspace{.2in}
     \subfigure[]{
          \label{fig:8}
                \includegraphics[scale=0.4]{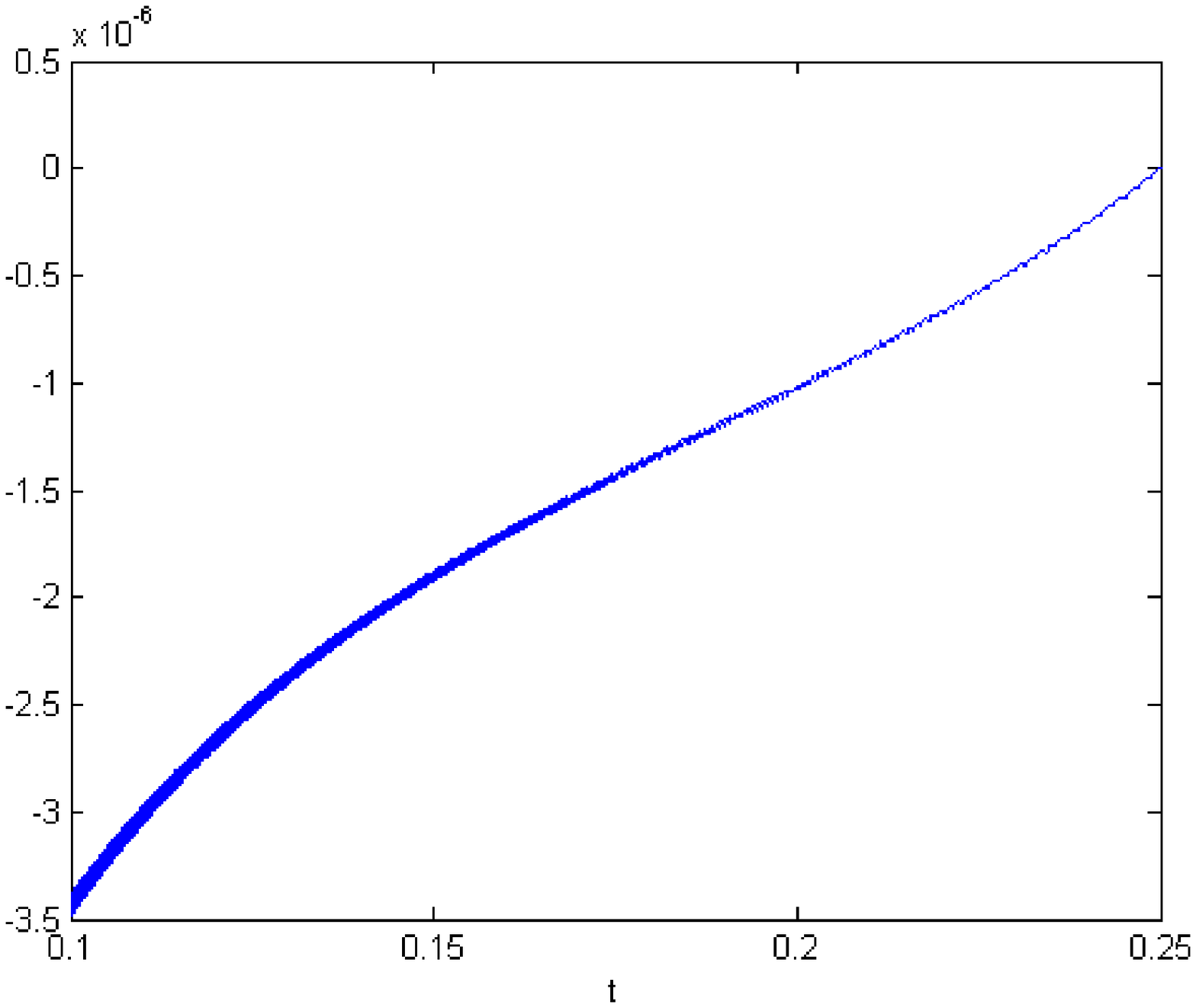}}\\
 \subfigure[]{
          \label{fig:9}
          \includegraphics[scale=0.4]{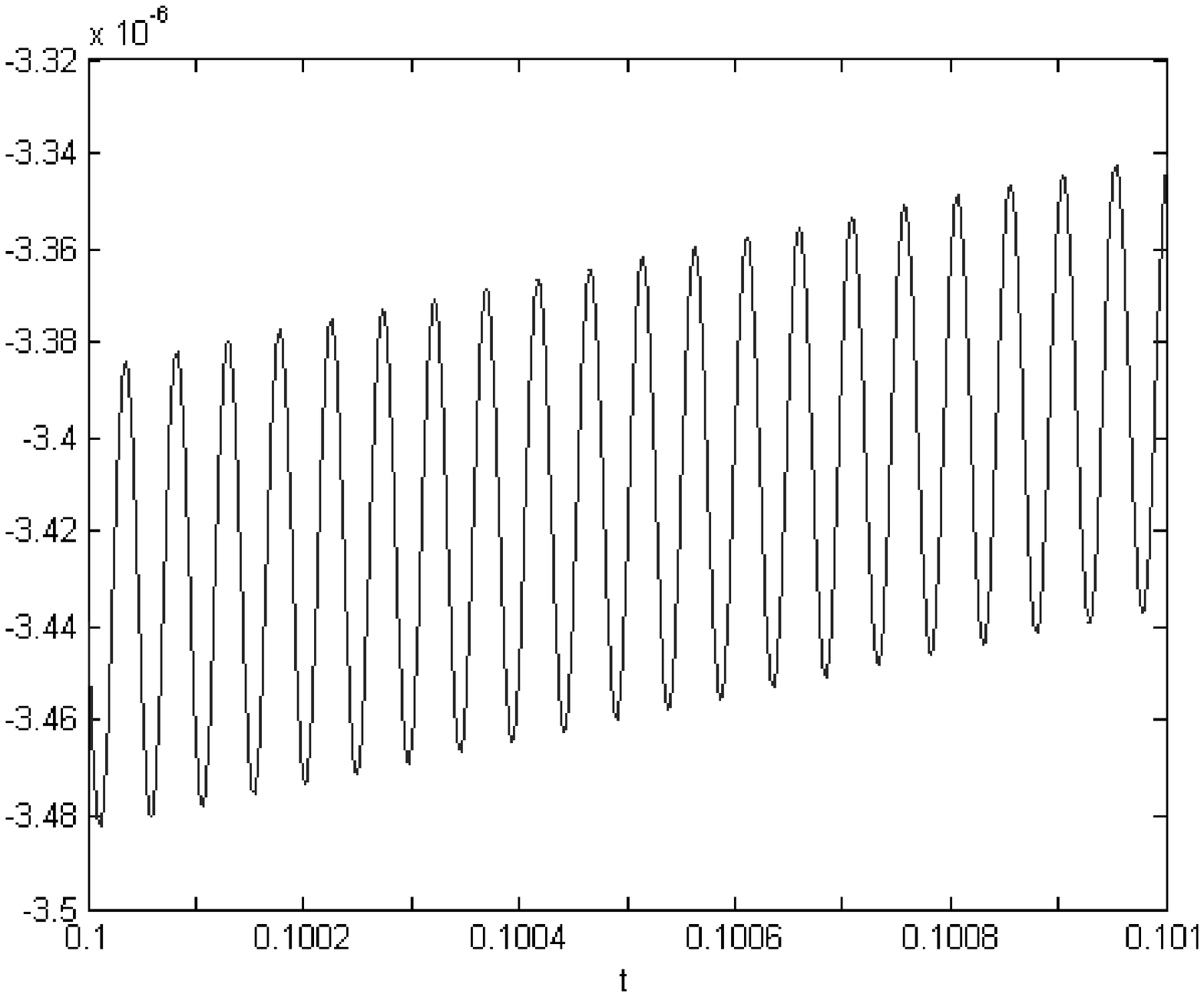}}
     \hspace{.2in}
 \subfigure[]{
          \label{fig:10}
          \includegraphics[scale=0.4]{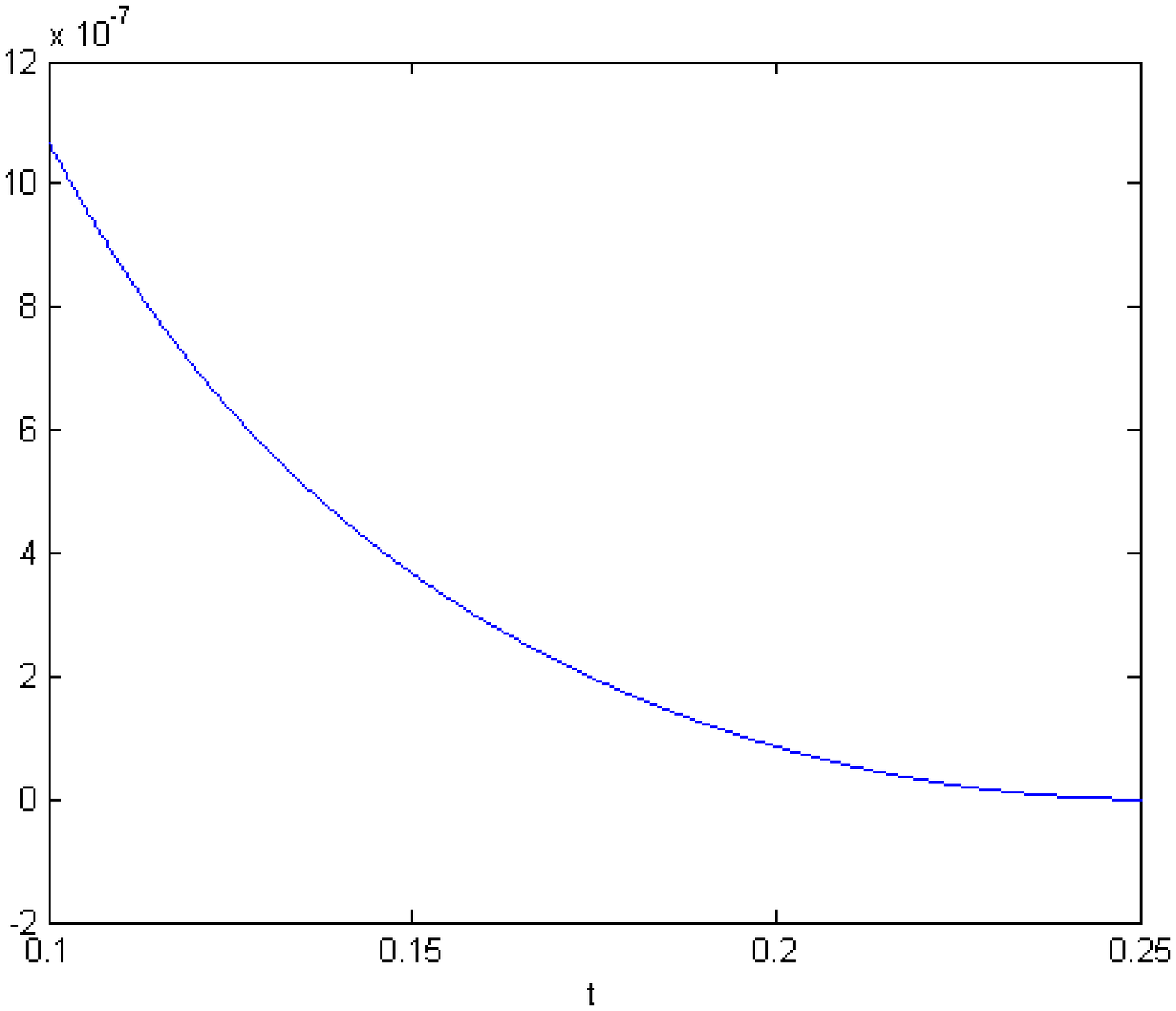}}
     \hspace{.2in}
     \caption{(a) the envelopes of $\delta R = (R - R_{\rm GR})/R$ for the HSS model, obtained by solving the full field equations numerically, with unperturbed initial conditions. Since $R = R_{\rm GR}$ is not a solution to the field equations, we find that $\delta R \neq 0$; (b) $\delta H = (H - H_{\rm GR})/H$. We see that $H$ oscillates, and the amplitude of these oscillations grows to the past. Further, $\delta H$ does not oscillate around zero, indicating that the Hubble parameter will deviate from $H_{\rm GR}$ as we evolve backwards in time; (c) $\delta H$ over a small time regime to explicitly show the oscillatory behaviour of $H$; (d) $\delta a = (a - a_{\rm GR})/a$. We see that $a$ will deviate from $a_{\rm GR}$ as we evolve backwards in time, however $\delta a$ is highly suppressed.}
     \label{fig:f3}
\end{figure}

\begin{figure}[htp]
     \centering
     \subfigure[]{
          \label{fig:11}
          \includegraphics[scale=0.4]{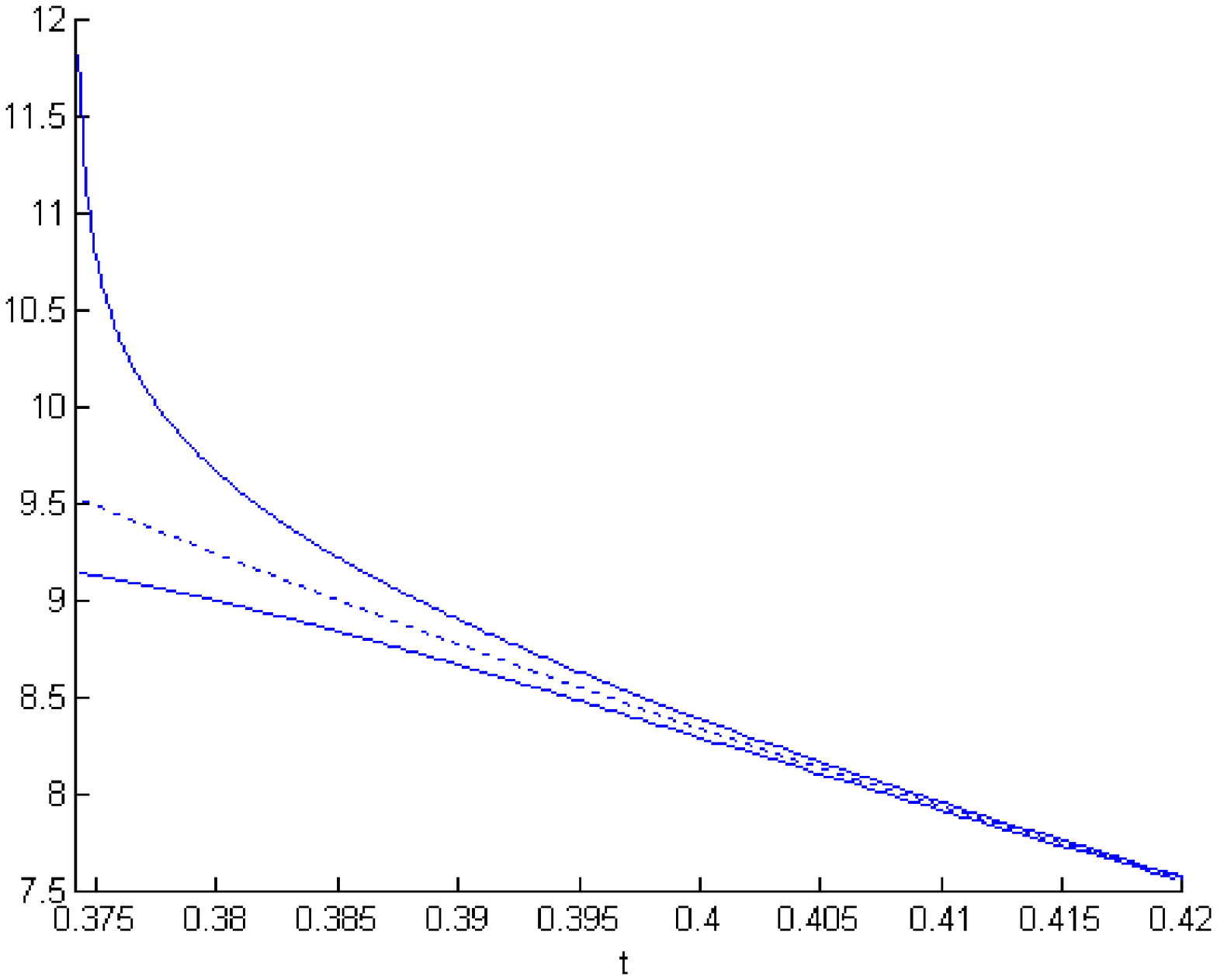}}
     \hspace{.2in}
     \subfigure[]{
          \label{fig:12}
                \includegraphics[scale=0.4]{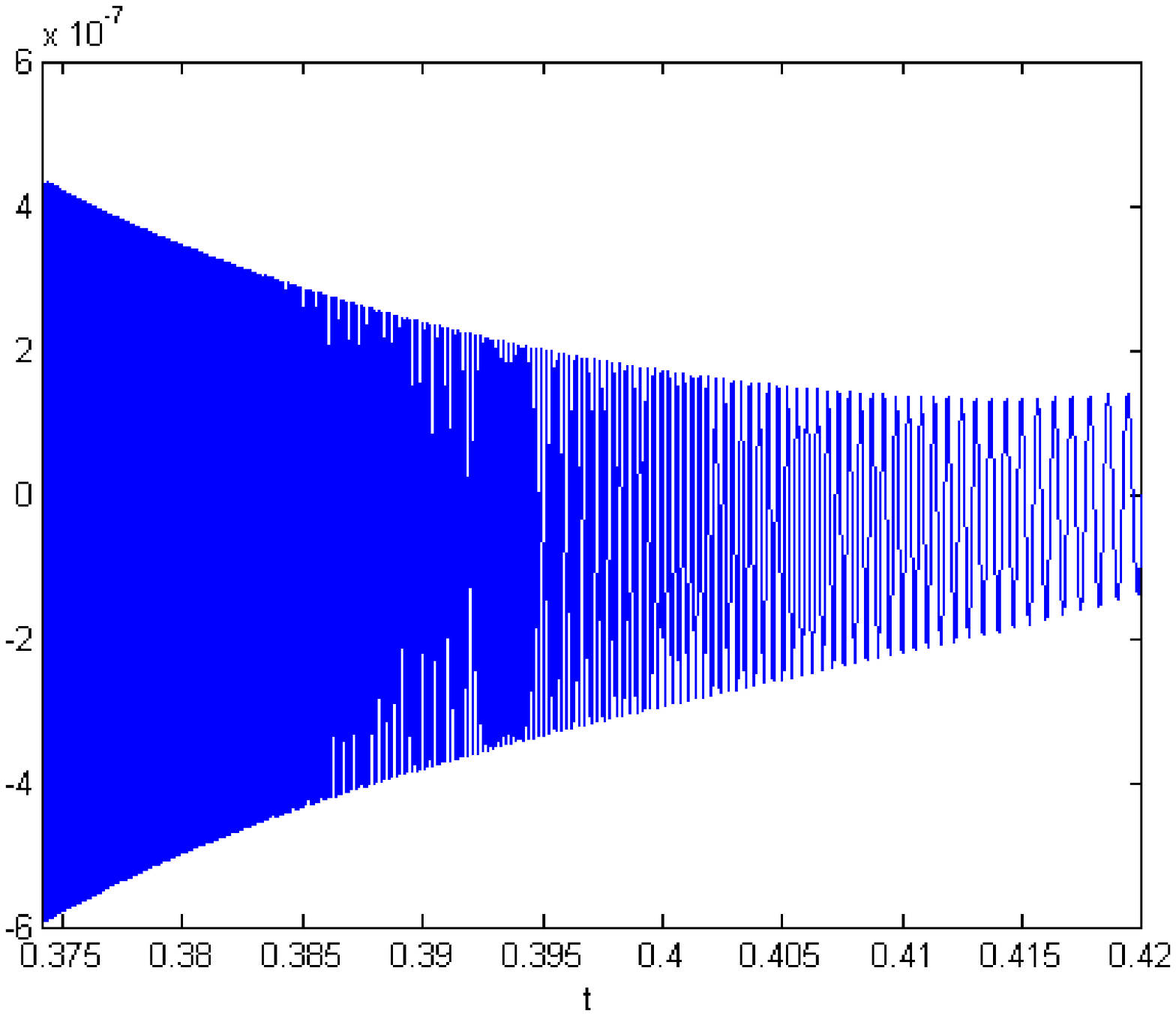}}
 \subfigure[]{
          \label{fig:13}
          \includegraphics[scale=0.4]{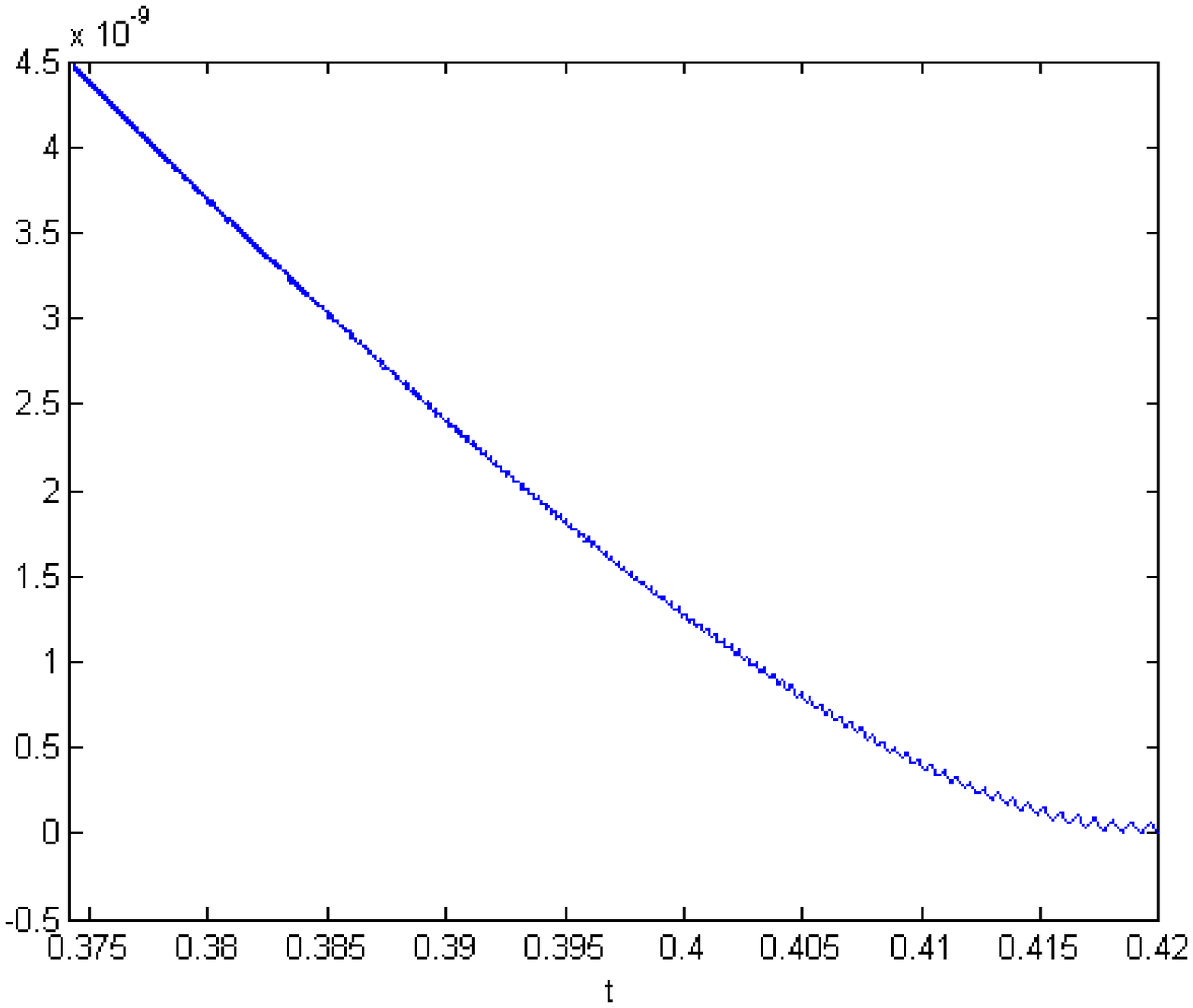}}
     \hspace{.2in}
 \subfigure[]{
          \label{fig:14}
          \includegraphics[scale=0.4]{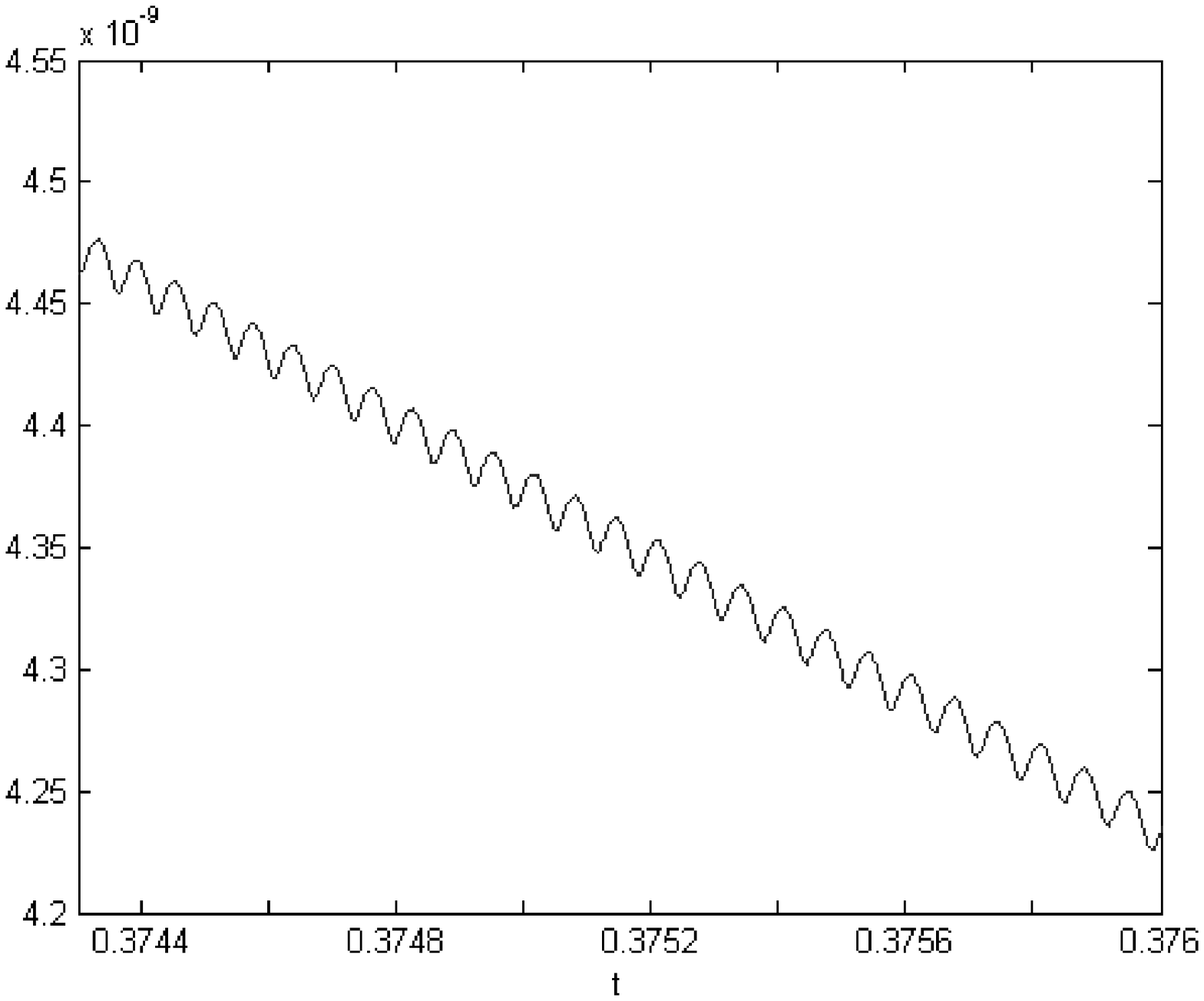}}
     \hspace{.2in}
     \caption{(a) the envelope functions for $R$ in the AB model, with $\epsilon = 0.32$, obtained by solving the full field equations numerically  as described in the text, with perturbed initial conditions. As in the HSS model, we see that $R$ differs significantly from the linearized approximation $R = R_{\rm GR} + \delta R$, obtained in the previous section; (b) $\delta H = (H - H_{\rm GR})/H$, which oscillates (not around $\delta H =0$); (c) $\delta a = (a - a_{\rm GR})/a$. We see that $a$ will deviate from $a_{\rm GR}$ as we evolve backwards in time, however $\delta a$ is highly suppressed; (d) $\delta a$ over a smaller range of $t$, which explicitly shows the oscillatory behaviour of the scale factor.}
     \label{fig:f4}
\end{figure}

\begin{figure}[htp]
     \centering
     \subfigure[]{
          \label{fig:15}
          \includegraphics[scale=0.4]{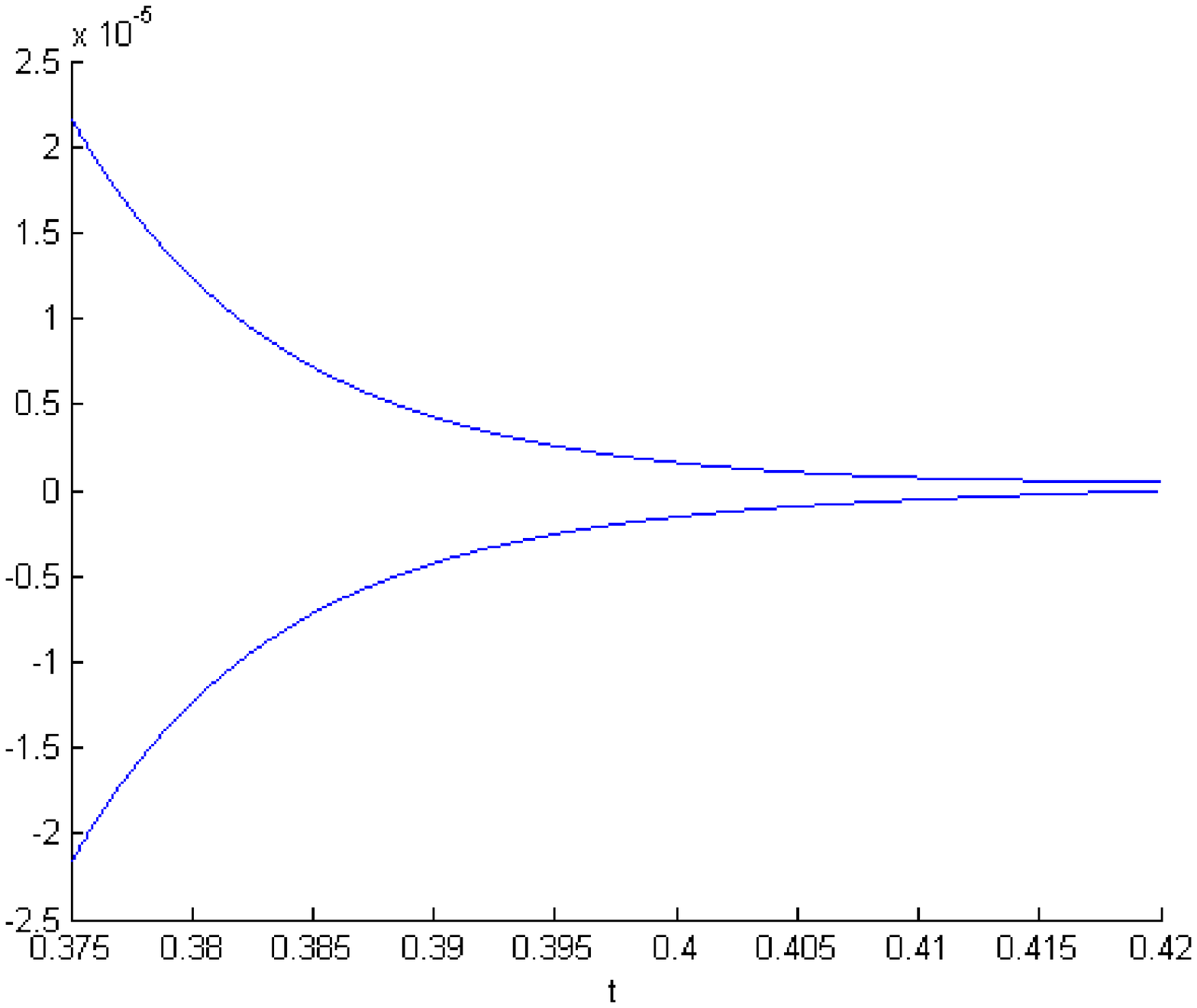}}
     \hspace{.2in}
     \subfigure[]{
          \label{fig:16}
                \includegraphics[scale=0.4]{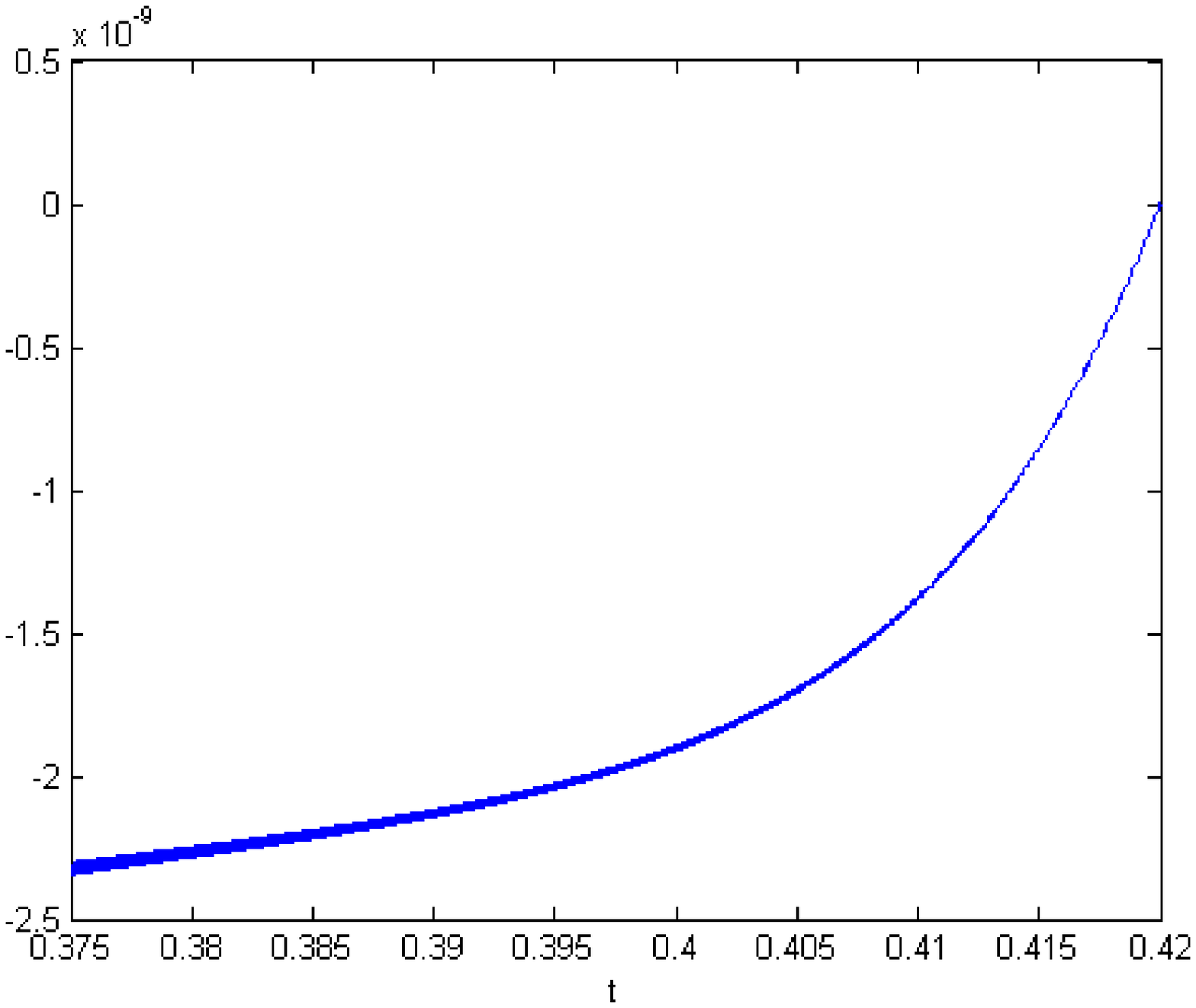}}\\
 \subfigure[]{
          \label{fig:17}
          \includegraphics[scale=0.4]{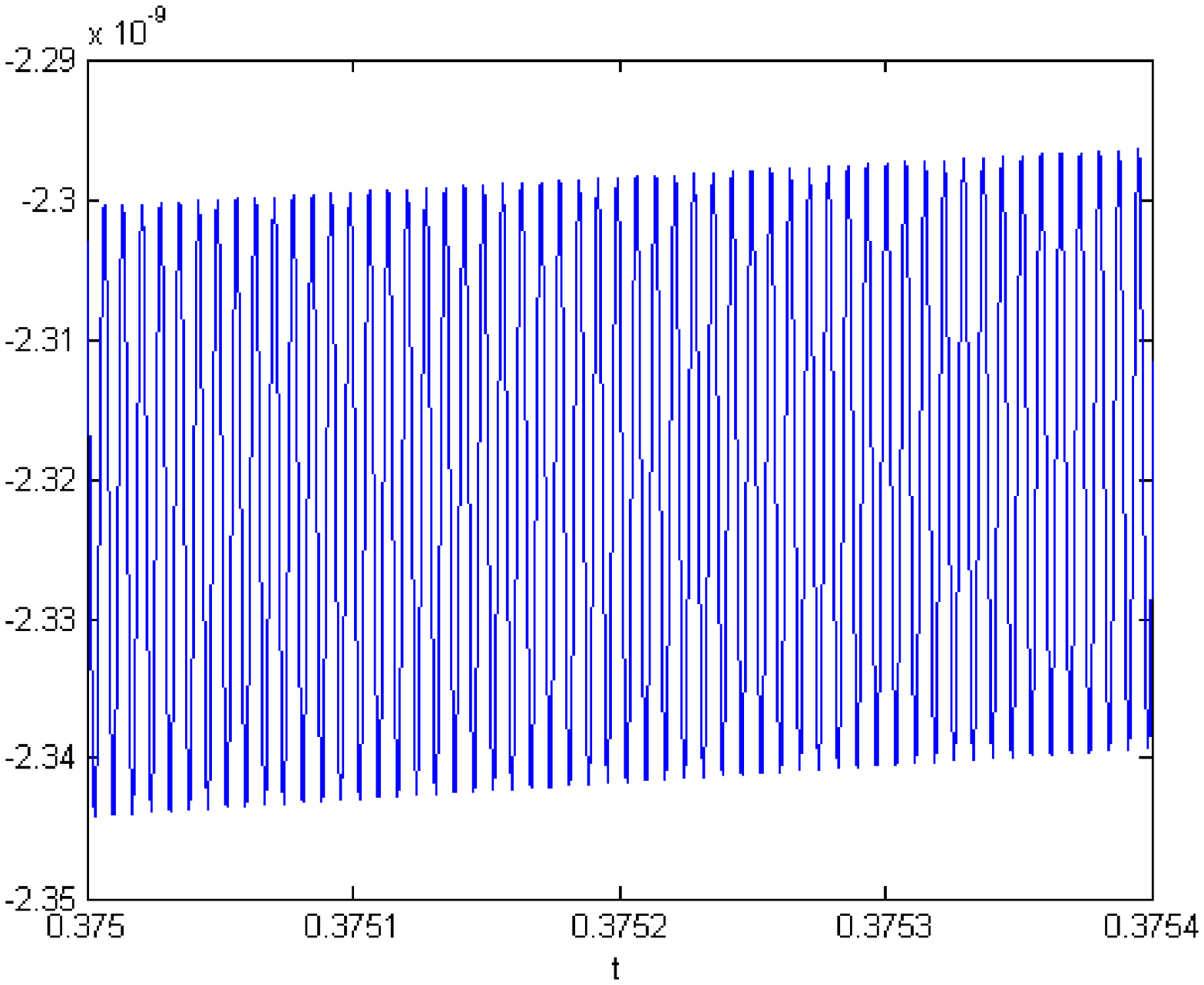}}
     \hspace{.2in}
 \subfigure[]{
          \label{fig:18}
          \includegraphics[scale=0.4]{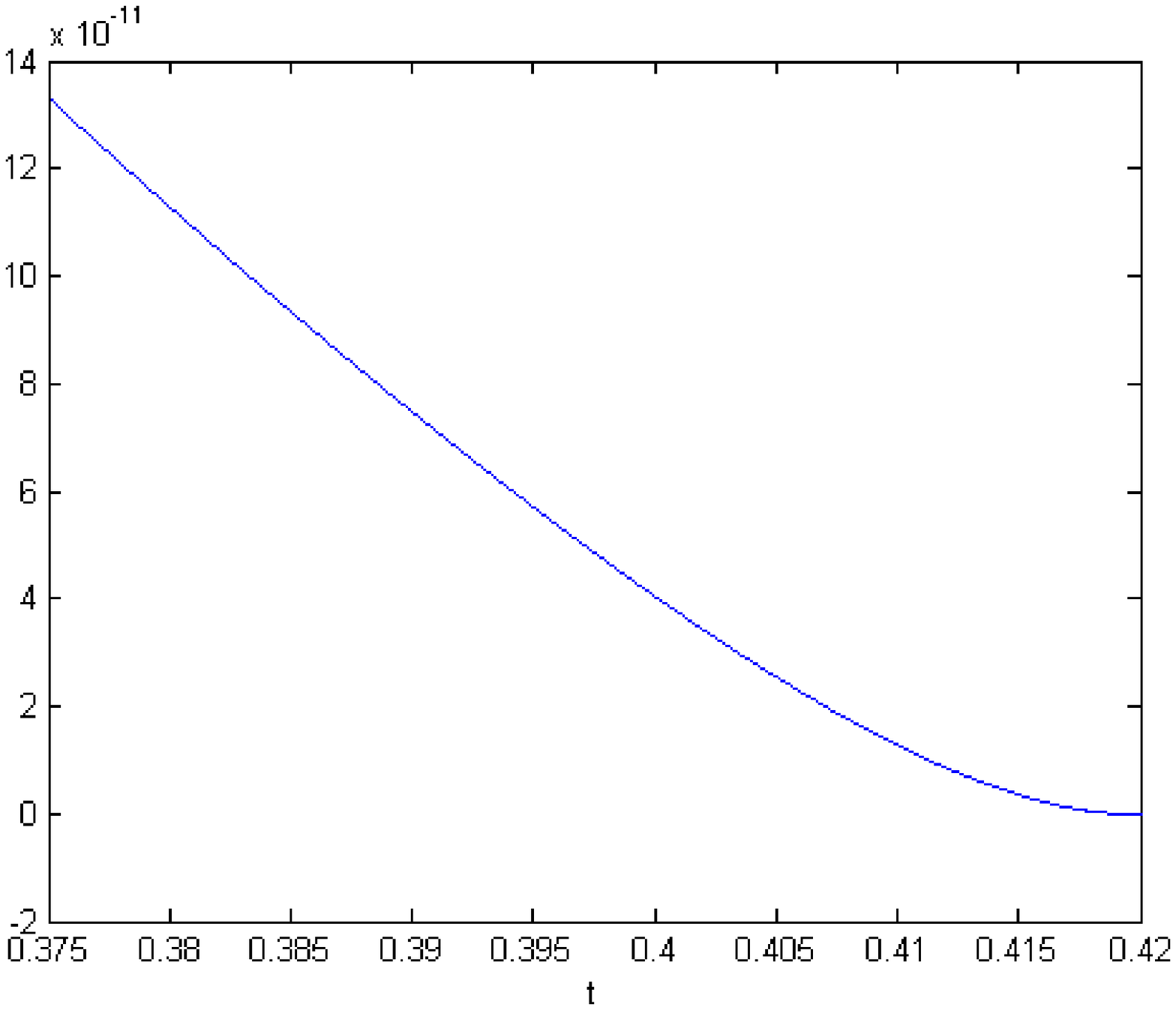}}
     \hspace{.2in}
     \caption{(a) envelope functions for $\delta R = (R - R_{\rm GR})/R$ for the AB model, with unperturbed initial conditions. Since $R = R_{\rm GR}$ is not a solution to the field equations, we find that $\delta R$ oscillates; (b) $\delta H = (H - H_{\rm GR})/H$. It is clear that $H$ oscillates, and the amplitude of these oscillations grows to the past. Further, $\delta H$ does not oscillate around zero, indicating that the Hubble parameter will deviate from $H_{\rm GR}$ as we evolve backwards in time; (c) $\delta H$ over a small time regime, explicitly showing the oscillatory behaviour of $H$; (d) $\delta a = (a - a_{\rm GR})/a$. We see that $a$ will deviate from $a_{\rm GR}$ as we evolve backwards in time, however $\delta a$ is also highly suppressed.}
     \label{fig:f5}
\end{figure}

\section{\label{sec:s10} Non-perturbative numerical Solution for the Ricci Scalar}

In this section, we first show that the above linearized analysis is only valid over a very limited range of $R$, and that the gravitational field equations are generically non-linear. To see this, we first write ($\ref{eq:1}$) as

\begin{equation}\label{eq:cf1} \ddot{R} + 3H\dot{R} + {\chi''' \over \chi''}\dot{R}^{2} + {R+T/M_{\rm pl}^{2} \over 3 \chi''} - {R \chi' - 2\chi \over 3\chi''} \approx 0 ,\end{equation}

\noindent where we have expanded the function $F(R) \approx R - R_{\rm vac}/2 + \chi(R)$, and neglected the term $R_{\rm vac}/2$. Both of these assumptions are acceptable if we consider a regime where $R > R_{\rm vac}$. In section \ref{sec:am1}, we linearized this equation in $\delta R$, and in doing so we neglected the term $(F''' / F'')(\delta \dot{R})^{2}$, which was assumed to be second order. However, since $\delta R$ oscillates with high frequency $\omega$, it follows that the term $ (\delta \dot{R})^{2}$ will not remain small, since it will grow like $\omega^{2}$ to the past. Hence the accuracy of the above linearized analysis will begin to reduce when $(F''' / F'')(\delta \dot{R})^{2}$ is of the same order as $3H\delta \dot{R}$. Beyond this point, we can no longer use the linearized equation for $\delta R$, and we must include these non-linear terms.

For the AB model, we can explicitly show the range of $\delta R$ over which the linearized analysis is valid. To see this, we consider the expansions ($\ref{eq:6}$-$\ref{eq:201}$). As an example, we take $F'(R)$, which was expanded as
$F'(R) \approx F'(R_{GR}) + F''(R_{GR})\delta R$. For the AB model this reads
$F'(R) \approx 1 - e^{-R_{\rm GR}/\epsilon} + {{1 \over \epsilon}e^{-R_{GR}/\epsilon} \delta R}$.
We compare this to our actual function $F'(R_{GR} + \delta R)$, given by
$F'(R_{GR} + \delta R) \approx 1 - e^{-R_{\rm GR}/\epsilon}e^{-\delta R/\epsilon}$,
and we see that the expansion of $F'(R_{\rm GR} + \delta R)$ is only valid in the very limited range $\delta R \ll \epsilon = R_{\rm vac}/4b$. Once this condition is violated, we can no longer use the linearized equation ($\ref{eq:4}$) for $\delta R$. Since in general we will have $R_{\rm GR} \gg \epsilon = R_{\rm vac}/4b$, then it follows that the linearized analysis will break down for the AB model long before $\delta R \sim R_{\rm GR}$.

The above reasoning suggests that although the linearized analysis successfully predicts the oscillatory behaviour of these models, we can only trust the solution for a limited range of $R$, and beyond this the field equation for $R$ is inherently non-linear. We conclude that we cannot find a solution to ($\ref{eq:1}$) for which $R \approx R_{\rm GR}$ for all $R$. Therefore in this section, we do not consider solutions to the field equations that are perturbations around the General Relativistic Ricci scalar, but rather look for full numerical solutions to ($\ref{eq:1}$). However, it is not just a second order differential equation for $R$, but rather a fourth order non-linear equation for the scale factor $a(t)$. Therefore we will treat ($\ref{eq:1}$), $\dot{H} + 2H^{2} = R/6$ and $H = \dot{a}/a$ as a set of coupled non-linear differential equations for $R$, $H$ and $a$. We stress that we are not making the assumption that $R \approx R_{\rm GR} + \delta R$, and the only assumption that we will make is that during the matter era $R_{\rm GR} \gg R_{\rm vac}$. We do so because we wish to compare the full numerical results obtained here with the results of the linearized analysis of the section \ref{sec:am1}, where the vacuum curvature was neglected. We have checked that introducing the vacuum curvature does not significantly effect our results, since we can absorb it into the Energy momentum tensor.

We look for solutions to these equations for two sets of initial conditions. First we perturb $R$ initially from its General Relativistic limit, and compare the full numerical solution to the linearized analysis of the previous section. We then set the initial conditions such that $R$, $a$ and $H$ are initially exactly their General Relativistic values, and observe how the HSS and AB models evolve backwards through the matter era.

\subsection{HSS model}

For the HSS model, using the example of a pure matter era for which $R_{\rm GR}= -T/M_{\rm pl}^{2} = 4/3t^{2}$, we find

\begin{equation}\label{eq:cf2} \ddot{R} + 3H\dot{R} -2(n+1) {\dot{R}^{2} \over R} +{1 \over 6n(2n+1)\epsilon^{2n+1}} \left(R + T/M_{\rm pl}^{2}\right) R^{2n+2} + {(n+1) \over 3n(2n+1)}R^{2} = 0  ,\end{equation}

\noindent together with the equations for $H$ and $a$. The trace of the energy-momentum tensor $T$ is related to $a$ and $H$ through the conservation equation $\nabla_{\alpha}T^{\alpha \nu} = 0$. Specifically, during the matter era we have $T/M_{\rm pl}^{2} \propto a^{-3}$. To derive ($\ref{eq:cf2}$) from ($\ref{eq:1}$), we have used $\chi_{\rm HSS}$ in ($\ref{eq:cf1}$). The only assumptions made are that $F(R) \approx R - R_{\rm vac}/2 + \chi_{\rm HSS}$ and that ignoring the $R_{\rm vac}$ term will not significantly effect our results, which is valid throughout our numerical calculation. It is clear that $R = -T/M_{\rm pl}^{2}$ is not a solution of these equations, and hence the HSS model will not exactly mimic General Relativity. As we will see, even if we set the initial conditions such that $R$, $H$ and $a$ are initially equal to their General Relativistic values, if we evolve this model backwards in time then they will deviate from $R_{\rm GR} = -T/M_{\rm pl}^{2}$.

We have solved the coupled differential equations ($\ref{eq:cf2}$), $\dot{H} + 2H^{2} = R/6$ and $H = \dot{a}/a$ numerically over the range $t = (0.25,0.1)$, taking $n=1$, $R_{\rm vac}=1$ and $\epsilon =0.1$ as before.
We have found that we can only evolve the Ricci scalar over a small range of $t$. As we will see in the next section, this is because the Ricci scalar will generically evolve to a singularity at some finite point in the past. For now, we will ignore this singular behaviour and solve ($\ref{eq:cf2}$) for $R$ over a small dynamical range. The Ricci scalar, obtained using a stiff differential equation solver, is shown in fig.\ref{fig:3}.
Here we have chosen the initial conditions $R(t_{\rm i}) = (4/3t^{2}_{\rm i}) +0.1$,  $\dot{R}(t_{\rm i}) = -8/3t^{3}_{\rm i}$, $H(t_{\rm i}) = 2/3t_{\rm i}$, and $a(t_{\rm i}) = (3/4)^{1/3}t_{\rm i}^{2/3}$, where $t_{\rm i}=0.25$, and evolved backwards in the time coordinate. These initial conditions are the same as those chosen in the previous section, so that we can compare the full solution found here to the solution found using the linearized approximation $R = R_{\rm GR} + \delta R$.
In fig.\ref{fig:3}, we have exhibited the envelope of the oscillations of the Ricci scalar, and the General Relativistic solution $R_{\rm GR}$. From the envelope functions, we see asymmetric oscillations of $R$ about $R_{\rm GR}$. We note that the solution obtained in this section is significantly different to the one obtained in the linearized analysis, shown in fig.\ref{fig:1}.
Specifically, in fig.\ref{fig:3} we find that there is no turning point in the lower envelope in the time range considered. Conversely, we find that the upper envelope function increases at a faster rate than predicted in the linearized analysis.

We have also plotted $\delta H = (H - H_{\rm GR})/H$ for this model in fig.\ref{fig:4}, which is the fractional difference between the Hubble parameter $H$ for the HSS model and the General Relativistic Hubble parameter during the matter era, $H_{\rm GR} = 2/3t$. We see that $ \delta H$ oscillates  asymmetrically and not exactly around its General Relativistic limit $\delta H=0$. The amplitude of these asymmetric oscillations is highly suppressed but increasing to the past, and unless $\epsilon$ is chosen to be sufficiently small these oscillations may come to dominate $H$. For the scale factor, we have again plotted the fractional difference $\delta a = (a - a_{\rm GR})/a$ in figs.\ref{fig:5} and \ref{fig:6}, and we see that $a$ deviates from the General Relativistic scale factor, but this deviation $\delta a$ is highly suppressed. However, as with the Hubble parameter, $\delta a$ grows to the past, and hence may become significant at some earlier time.

Having solved the system of equations for $R$, $H$ and $a$ by perturbing $R$ initially from the General Relativistic limit, we now solve with no initial perturbation, that is we take the initial conditions $R(t_{\rm i}) = 4/3t^{2}_{\rm i}$,  $\dot{R}(t_{\rm i}) = -8/3t^{3}_{\rm i}$, $H(t_{\rm i}) = 2/3t_{\rm i}$, and $a(t_{\rm i}) = (3/4)^{1/3}t_{\rm i}^{2/3}$, where $t_{i}=0.25$. The solution to equation ($\ref{eq:cf2}$) is shown in figs.3(a-d). Again we have taken $\epsilon = 0.1$, $R_{\rm vac}=1$ and evolved over the time regime $t=(0.25,0.1)$. We see that $R$ oscillates, and $H$ and $a$ deviate from their General Relativistic values. However, once again $\delta a$ and $\delta H$ are suppressed, and this model closely mimics General Relativity over this range of time.

\subsection{AB model}

We can perform similar calculations for the AB model. Taking once again $R_{\rm GR} = 4/3t^{2}$ as a specific example, we find

\begin{equation}\label{eq:ab100} \ddot{R} + 3H \dot{R} - {\dot{R}^{2} \over \epsilon} + {\epsilon \over 3} \left( R + T/M_{\rm pl}^{2} \right)e^{R/\epsilon} +{\epsilon^{2} \over 3} \left({R \over \epsilon}+2 \right) = 0  .\end{equation}

\noindent We obtain ($\ref{eq:ab100}$) by using $\chi_{\rm AB}$ in equation ($\ref{eq:cf1}$). We have solved this equation numerically, along with the equations for $H$ and $a$, using the same differential equation solver as for the HSS model, taking $\epsilon =  0.32$ and using
the initial conditions $R(t_{\rm i}) = 4/3t^{2}_{\rm i} + 0.01$, $\dot{R}(t_{\rm i}) = -8/3t_{\rm i}^{3}$, $H(t_{\rm i}) = 2/3t_{\rm i}$ and $a(t_{\rm i}) = (3/4)^{1/3}t_{\rm i}^{2/3}$. As in section \ref{sec:am1}, for the AB model we only solve equation ($\ref{eq:ab100}$) over the very small time range $t=(0.42,0.375)$, due to the presence of a singularity. By solving ($\ref{eq:ab100}$) over this limited dynamical range, we find asymmetric oscillations of $R$, $H$ and $a$. The envelope of the oscillations of $R$ are shown in fig.\ref{fig:11}. Once again, we see no turning point in the lower envelope, suggesting that $R$ does not become negative for these models. We also observe that the upper envelope grows faster than predicted by the linear analysis. The Hubble parameter and scale factor undergo asymmetric, suppressed oscillations which grow to the past, as seen in figs.4(b-d).

We also solve equation ($\ref{eq:ab100}$) without perturbing $R$, $H$ and $a$ from their General Relativistic limits initially, and evolving the system backwards over the same time regime $t=(0.42,0.375)$. The results are shown in figs.5(a-d). We find that $R$ oscillates, and $\delta H = (H - H_{\rm GR})/H$ and $\delta a = (a - a_{\rm GR})/a$ grow, indicating that $a$ and $H$ both diverge from their General Relativistic limits. However, this divergence is highly suppressed, as was found in the HSS model.

\section{\label{sec:2} Improved perturbative approach}

So far, we have considered two approaches to solving the modified gravitational field equations. It has been found that the linearized approach, in which the ansatz $R = R_{\rm GR} + \delta R$ with $\delta R \ll R_{\rm GR}$ is used in ($\ref{eq:1}$), will not give a solution that is indicative of the full solution, since non-linear terms cannot be neglected due to the rapid oscillations of the Ricci scalar. The second approach considered was to solve the full gravitational field equations numerically. In doing so, it was found that the Ricci scalar undergoes non-linear oscillations, very closely (but not exactly) about its General Relativistic limit. However, we found that solving the gravitational field equations over significant timescales is impossible due to some kind of singularity at a finite time.

In this section, we consider neither a linearized analysis nor a full numerical study of the field equations. Instead, under sensible assumptions, we show that the trace of the gravitational field equations for both the AB and HSS models reduces to a non-linear wave equation, and hence that $R$ undergoes asymmetric oscillations about its General Relativistic limit, in agreement with the results of section \ref{sec:s10}. Moreover we find that there is a singularity in $R$ at a finite time in both models. Since the approach taken is model specific, we consider the AB and HSS models separately.

\subsection{AB model}

\noindent We begin by substituting $\chi_{\rm AB}$ into ($\ref{eq:1}$), giving

\begin{equation}\label{eq:l1000} - 3\Box e^{-R/\epsilon} - R + R_{\rm vac} - (R+2\epsilon) e^{-R/\epsilon} = {T \over M_{\rm pl}^{2}},\end{equation}

\noindent Next, we define $z=e^{-R/\epsilon}$, so $R=-\epsilon \log z$ and ($\ref{eq:l1000}$) becomes

\begin{equation} \label{eq:l20000} -3 \Box z + \epsilon \log z + R_{\rm vac} + \epsilon z \left(\log z - 2\right) = {T \over M_{\rm pl}^{2}},\end{equation}

\noindent where we have assumed that $R_{\rm vac} \ll T/M_{\rm pl}^{2}$ and hence neglected the vacuum energy.
Next, by defining $z_{1} = z e^{-T/\epsilon M_{\rm pl}^{2}}$, we can write ($\ref{eq:l20000}$) as

\begin{equation} \label{eq:l5} -3\Box z_{1} e^{T/\epsilon M_{\rm pl}^{2}} + \epsilon \log z_{1} + \epsilon e^{T/\epsilon M_{\rm pl}^{2}}z_{1}\left( \log z_{1} + {T \over \epsilon M_{\rm pl}^{2}} - 2 \right) = 0 .\end{equation}

To solve this equation, we write it as

\begin{equation} \label{eq:l7} 3 e^{T/2\epsilon M_{\rm pl}^{2}} {d \over dt} \left( e^{T/2\epsilon M_{\rm pl}^{2}}{d z_{1} \over dt} + {z_{1} \over \epsilon M_{\rm pl}^{2}}e^{T/2\epsilon M_{\rm pl}^{2}}
{d T \over dt} \right) + 3\left( e^{T/2\epsilon M_{\rm pl}^{2}} {d z_{1} \over dt} + {z_{1} \over \epsilon M_{\rm pl}^{2}} e^{T/2\epsilon M_{\rm pl}^{2}}{d T \over dt} \right) {e^{T/2\epsilon M_{\rm pl}^{2}} \over 2\epsilon M_{\rm pl}^{2}}
{d T \over dt} +\end{equation} \begin{equation*} 9e^{T/2\epsilon M_{\rm pl}^{2}} a^{-1}{da \over dt}\left( e^{T/2\epsilon M_{\rm pl}^{2}} {dz_{1} \over dt} + {z_{1}\over \epsilon M_{\rm pl}^{2}} e^{T/2\epsilon M_{\rm pl}^{2}}{d T \over dt} \right) + \epsilon \log z_{1} + \epsilon e^{T/\epsilon M_{\rm pl}^{2}}z_{1}\left( \log z_{1} + {T \over \epsilon M_{\rm pl}^{2}} - 2 \right) = 0 .\end{equation*}

\noindent If we now introduce the fast time coordinate

\begin{equation}\label{eq:lo1} \lambda = \int e^{-T/2\epsilon M_{\rm pl}^{2}} dt ,\end{equation}

\noindent which is the timescale associated with the oscillations of the Ricci scalar, then we can write ($\ref{eq:l7}$) as

\begin{equation}\label{eq:l8} 3 z''_{1} + \left( {9 \over 2\epsilon M_{\rm pl}^{2}} T'  + 9 \bar{H} \right) z'_{1} + \left( {3 \over 2 \epsilon^{2}M_{\rm pl}^{4}}T'^{2} +
{3 \over \epsilon M_{\rm pl}^{2}} T'' + {9 \over \epsilon M_{\rm pl}^{2}} \bar{H} T' \right) z_{1} + \epsilon \log z_{1} + \epsilon e^{T/\epsilon M_{\rm pl}^{2}}z_{1}\left( \log z_{1} + {T \over \epsilon M_{\rm pl}^{2}} - 2 \right) = 0 ,\end{equation}

\noindent where primes denote derivatives with respect to $\lambda$, and we have defined $\bar{H} = a'/a$. If we solve equation ($\ref{eq:l8}$) for $z_{1}$, then we can deduce $R$ from the relation

\begin{equation}\label{eq:cr10} R = -{T \over M_{\rm pl}^{2}} - \epsilon \log z_{1} .\end{equation}

\noindent Note that ($\ref{eq:cr10}$) is of the form $R = R_{\rm GR} + R_{\rm osc}$, however we have not assumed that $R_{\rm osc} \ll R_{\rm GR}$ to deduce this expression.

If we now define $x = z_{1}-1$, then we can write ($\ref{eq:l8}$) as

\begin{equation}\label{eq:l90} 3 x'' + \left( {9 \over 2\epsilon M_{\rm pl}^{2}} T'  + 9 \bar{H} \right) x' + \epsilon \log (1+x) + x \left[ {3 \over 2 \epsilon^{2}M_{\rm pl}^{4}}T'^{2} +
{3 \over \epsilon M_{\rm pl}^{2}} T'' + {9 \over \epsilon M_{\rm pl}^{2}} \bar{H} T'  + \epsilon e^{T/\epsilon M_{\rm pl}^{2}}\left({T \over \epsilon M_{\rm pl}^{2}} - 2 \right)\right] + \end{equation} \begin{equation*} \epsilon e^{T/\epsilon M_{\rm pl}^{2}}(1+x) \log (1+x)  = -\left( {3 \over 2 \epsilon^{2}M_{\rm pl}^{4}}T'^{2} +
{3 \over \epsilon M_{\rm pl}^{2}} T'' + {9 \over \epsilon M_{\rm pl}^{2}} \bar{H} T' \right) - \epsilon e^{T/\epsilon M_{\rm pl}^{2}}\left(  {T \over \epsilon M_{\rm pl}^{2}} - 2 \right) .\end{equation*}

\noindent Equation ($\ref{eq:l90}$) is a non-linear, second order inhomogeneous differential equation for $x$, and the full solution is the sum of the particular solution to ($\ref{eq:l90}$) and the solution to the corresponding homogeneous equation given by setting the right hand side to zero. $x=0$ is now the General Relativistic limit, but it is clear that this is not an exact solution. We will find that $x$ undergoes asymmetric oscillations governed by the homogeneous equation, around the particular solution of ($\ref{eq:l90}$). We now calculate the oscillatory and drift components of $x$.

\subsubsection{Drift}

We begin by calculating the drift of $x$ away from $x=0$. To do so, we make the assumption that the derivative terms $x''$ and $x'$ in ($\ref{eq:l90}$) are subdominant and hence can be neglected. With this assumption, ($\ref{eq:l90}$) reduces to an algebraic equation for $x$,

\begin{equation}\label{eq:la90}  \epsilon \log (1+x) + x \left[ {3 \over 2 \epsilon^{2}M_{\rm pl}^{4}}T'^{2} +
{3 \over \epsilon M_{\rm pl}^{2}} T'' + {9 \over \epsilon M_{\rm pl}^{2}} \bar{H} T'  + \epsilon e^{T/\epsilon M_{\rm pl}^{2}}\left({T \over \epsilon M_{\rm pl}^{2}} - 2 \right)\right] +  \epsilon e^{T/\epsilon M_{\rm pl}^{2}}(1+x) \log (1+x) \end{equation} \begin{equation*} = -\left( {3 \over 2 \epsilon^{2}M_{\rm pl}^{4}}T'^{2} +
{3 \over \epsilon M_{\rm pl}^{2}} T'' + {9 \over \epsilon M_{\rm pl}^{2}} \bar{H} T' \right) - \epsilon e^{T/\epsilon M_{\rm pl}^{2}}\left(  {T \over \epsilon M_{\rm pl}^{2}} - 2 \right) .\end{equation*}

\noindent If we assume that $x \ll 1$, then by expanding $\log(1+x)$ in ($\ref{eq:la90}$) we find the following expression for $x$,

\begin{equation}\label{eq:l125} x = -\left(3e^{T/\epsilon M_{\rm pl}^{2}}\left[ {64 \over 9 \epsilon^{3} t^{6}} - {8 \over 3\epsilon^{2} t^{4}}   + {1 \over 3} \left(   {T \over \epsilon M_{\rm pl}^{2}} - 2 \right)\right] \right) .\end{equation}

\noindent This solution is valid if we only consider terms linear in $x$, that is we consider terms of order $e^{T/\epsilon M_{\rm pl}^{2}} \ll 1$ only. Taking the derivative of $x$, we find that $x'$ and $x''$ are given by

  \begin{align} x' &= e^{T/2\epsilon M_{\rm pl}^{2}} \dot{x} \\ x'' &= e^{T/\epsilon M_{\rm pl}^{2}} \left( {\dot{T} \over 2 \epsilon M_{\rm pl}^{2}}x + \dot{x} \right) .\end{align}

\noindent Since $x'$ and $x''$ are much smaller than $x$ (they are suppressed by a factor of $\exp(T/2\epsilon M_{\rm pl}^{2}) \ll 1$ and $\exp(T/\epsilon M_{\rm pl}^{2}) \ll 1$ relative to $x$ respectively), then it follows that our original assumption that we can neglect derivative terms of $x$ in ($\ref{eq:l90}$) is valid, and ($\ref{eq:l125}$) is an approximate solution to ($\ref{eq:l90}$).

\subsubsection{Oscillations}

The solution to the homogeneous equation, obtained by setting the left hand side of ($\ref{eq:l90}$) to zero, will give the oscillatory component of $x$ and hence $R$. We will first show that the last two terms on the left hand side of ($\ref{eq:l125}$), are highly suppressed, and hence can be neglected for $T/M_{\rm pl}^{2} \gg \epsilon$. To do so, we write $T''$, $(T')^{2}$ and $\bar{H} T'$ in terms of $t$,

\begin{equation} {(T')^{2} \over M_{\rm pl}^{4}}  =
\left({8 \over 3t^{3} e^{-T/2\epsilon M_{\rm pl}^{2}}}\right)^{2}, \qquad \qquad {T'' \over M_{\rm pl}^{2}}  =  -{8 \over 9t^{6}\epsilon}e^{T/\epsilon M_{\rm pl}^{2}}(9\epsilon t^{2} +4), \qquad \qquad {\bar{H}T' \over M_{\rm pl}^{2}} = {16 \over 9t^{4}}e^{T/\epsilon M_{\rm pl}^{2}}, \end{equation}

\noindent and the corresponding homogeneous equation to ($\ref{eq:l125}$) can be approximately written as

\begin{equation}\label{eq:l9} 3 x'' + \left( {9 \over 2\epsilon M_{\rm pl}^{2}} T'  + 9 \bar{H} \right) x' + \epsilon \log (1+x) + 3e^{T/\epsilon M_{\rm pl}^{2}}x\left[ {64 \over 9 \epsilon^{2} t^{6}} - {8 \over 3\epsilon t^{4}}   + {1 \over 3}\epsilon \left( \log (1+x) + {T \over \epsilon M_{\rm pl}^{2}} - 2 \right)\right]+ \epsilon e^{T/\epsilon M_{\rm pl}^{2}}\log (1+x) = 0 .\end{equation}

\noindent We see that the last two terms on the left hand side of ($\ref{eq:l9}$) have a common factor of $\exp(T/\epsilon M_{\rm pl}^{2}) \ll 1$ and hence can be neglected. We arrive at the following equation for $x$,

\begin{equation}\label{eq:l10}  x'' + \beta x' + {1 \over 3}\epsilon \log (1+x) \approx 0 ,\end{equation}

\noindent where $\beta = {3 \over 2\epsilon M_{\rm pl}^{2}} T' + 3 \bar{H}$ is a small function of $\lambda$. In ($\ref{eq:l10}$), we will assume that $H = H_{\rm GR}$ and $T = T_{\rm GR}$, since the oscillations of $H$ and $a$ are suppressed by a factor of $\exp(T/\epsilon M_{\rm pl}^{2})$ and  can therefore be neglected. With this assumption, ($\ref{eq:l10}$) reduces to an equation purely in terms of $x$ and its derivatives.

Equation ($\ref{eq:l10}$) describes an anharmonic, non-linear oscillator with a small negative damping term, and hence we have shown analytically that the Ricci scalar will undergo asymmetric oscillations, in agreement with the results of section \ref{sec:s10}. We henceforth describe the oscillator considered here as the logarithmic oscillator. The potential $V(x)$ for the logarithmic and simple harmonic oscillators are shown in figs.6(a,c). The logarithmic potential will give rise to wave solutions that are not symmetric.

Before continuing, we observe that if we expand $R$ as $R = R_{\rm GR} + \delta R$ for $\delta R \ll \epsilon$, then $x$ can be expanded as $x = \exp(-(R-R_{\rm GR})/\epsilon)-1 \approx  - \delta R/\epsilon$, and ($\ref{eq:l10}$) becomes

\begin{equation}\label{eq1:2} \delta R'' + \beta \delta R' + { 1 \over 3}\epsilon \delta R\approx 0 .\end{equation}

\noindent If we write ($\ref{eq1:2}$) in terms of $t$, then it reduces to the linearized equation ($\ref{eq:lh1}$).

We have solved the non-linear equation ($\ref{eq:l10}$) numerically, using the initial conditions $x_{\rm i}=\exp(-0.01/0.32)$, $x'_{\rm i}=0$, and evolved over the range $\lambda=(-1350,-13339)$. This range of $\lambda$ and choice of initial conditions are the same as were chosen in section \ref{sec:s10}, where the full gravitational field equations were solved numerically. In fig.$\ref{fig:fg8}$(a), we compare the Ricci scalar $R = R_{\rm GR} - \epsilon \log (1+x)$ obtained in this section to that found by solving the full gravitational field equations numerically in section \ref{sec:s10}. It is clear that the results obtained in this section closely mimic the full numerical solution. In fig.$\ref{fig:fg8}$(b) we compare $\delta R_{\rm a} = R - R_{\rm GR}= - \epsilon \log(1+x_{\rm osc} + x_{\rm dr})$ to $\delta R = R - R_{\rm GR}$, where $R$ is the Ricci scalar obtained by solving the full gravitational field equations numerically. We see that the fractional difference between $\delta R$ and $\delta R_{\rm a}$ remains small throughout the dynamical range considered.

\subsubsection{Existence of Singularity}

Using this approach, we have found that we can only evolve $x$ and hence $R$ backwards in the time coordinate $t$ over the same limited dynamical range as in section \ref{sec:s10}. However, by solving equation ($\ref{eq:l10}$) we can see why this is the case. In fig.\ref{fig:l26}, we have perturbed $x$ from $x=0$ and evolved backwards in the $\lambda$ coordinate. We see that the amplitude of the oscillations of $x$ increase due to the presence of the damping term in ($\ref{eq:l10}$), and after a finite time $x \to -1$. Since $R = -T_{\rm pl}^{2} - \epsilon \log (1+x)$ and $\dot{R} = \dot{T}/M_{\rm pl}^{2} - \epsilon \dot{x}/(1+x)$, then it follows that both $R$ and $\dot{R}$ will be singular at this point. We conclude that wide ranges of initial conditions $x_{\rm i}$ and $\dot{x}_{\rm i}$ will give rise to a singularity in the field equations as we evolve backwards in time. That is not to say that all initial conditions will lead to this problem, and there exists an extremely restricted range of $x_{\rm i}$ and $\dot{x}_{\rm i}$ at the beginning of the matter era that will not give rise to singular behaviour at some finite time in the past.

Instead of evolving backwards over the matter era, if we choose $x_{\rm i}$ appropriately at the beginning of the matter era and evolve forwards in time, $x$ (and hence $R$) will be well behaved and regular. To find the allowed initial conditions, we can first use the fact that $x_{\rm i} > -1$ is the lower bound. To obtain an upper bound, we assume that the damping term in ($\ref{eq:l10}$) is negligible and consider

 \begin{equation}\label{eq2:2} x''  + {1 \over 3}\epsilon\log (1+x) \approx 0 .\end{equation}

 \noindent This expression can be multiplied by $x'$ and integrated to obtain the following expression,

 \begin{equation} \label{eq100:1}{1 \over 2}(x')^{2} + {1 \over 3}\epsilon (1+x) \left[ \log (1+x) - 1 \right] = E ,\end{equation}

 \noindent where $E$ is a constant of integration. The values of $x$ at the peaks and troughs of its wavetrain can be obtained from ($\ref{eq100:1}$) by setting $x'=0$ and solving the algebraic equation

 \begin{equation}\label{eq2:3}E = {1 \over 3}\epsilon (1+x)\left[ \log (1+x) - 1 \right] .\end{equation}

 \noindent For a given $E$, there will be two solutions for $x$, corresponding to the maxima and minima of the wave. We have deduced that $x=-1$ is the minimum value that $x$ can take in order to avoid singularities, and substituting this into ($\ref{eq2:3}$) gives $E=0$. As we have stated, ($\ref{eq2:3}$) will yield two solutions for any particular $E$, and so finding the second solution for $E=0$ will give us the maximum value of $x$. This solution is $x = \exp(1)-1$, and hence we conclude that in order to obtain a matter era free from singularities, we must impose the condition that $-1<x < \exp(1)-1$ at the beginning of the matter era. In terms of $\delta R = R - R_{\rm GR}$, this corresponds to the range $-\epsilon < \delta R < \infty$.

Since we will generically encounter a singularity if we evolve $x$ backwards in the time coordinate $t$, we now solve equation ($\ref{eq:l10}$) by taking $x$ initially at some early time and evolving forwards in the $\lambda$ coordinate, which corresponds to evolving forwards in the standard time coordinate $t$. In doing so, $x$ and hence $R$ will remain regular. We take as initial conditions $x_{\rm i}=\exp(-2.6/0.32)-1$, $x'_{\rm i} = 0$ and evolve over the range $\lambda=(-13339,220)$.  By solving ($\ref{eq:l10}$), we obtain $x_{\rm osc}$, which is the oscillatory component of $x$. The full solution to ($\ref{eq:l90}$) is approximately given by $x \approx x_{\rm dr} + x_{\rm osc}$, where $x_{\rm dr}$ is the drift term given in ($\ref{eq:l125}$). In fig.\ref{fig:fg9}(a-d) we compare $\delta R_{\rm a}$ to $\delta R$. We see that there are two competing effects; the non-linear oscillations, which are initially large but quickly decay, and the drift away from $\delta R = 0$, which grows as we evolve forwards in the time coordinate. The approximate solution obtained in this section closely mimics the full numerical solution. However, the predicted drift begins to deviate from the full solution at late times. However, this only occurs when the approximation $T/M_{\rm pl}^{2} \gg \epsilon$ is no longer valid.

\subsubsection{Constant Energy Momentum Tensor}

As an aside, we conclude this section by considering a constant energy momentum tensor, that is we take $R_{\rm GR} = R_{0} = {\rm const} \gg \epsilon$. For this Ricci scalar, we have $\lambda = \exp(R_{0}/2\epsilon)t$, $T'=0$ and $\bar{H} = \sqrt{R_{0}/12}\exp(-R_{0}/2\epsilon)$. Using these in ($\ref{eq:l8}$), we obtain the following equation for $z_{1}$,

\begin{equation}\label{eq:jh2} z''_{1}+ 3\sqrt{R_{0}/12}e^{-R_{0}/2\epsilon}z'_{1} + {1 \over 3}\epsilon \log z_{1} \approx 0 .\end{equation}

\noindent The case of constant $R_{\rm GR}$ was discussed in ref.\cite{sh20} using the perturbative analysis of section \ref{sec:am1}, and it was found that the Ricci scalar contained an exponentially growing component, suggesting an instability in this model. Here, we solve ($\ref{eq:jh2}$) for $z_{1}$ and obtain the Ricci scalar from $R = R_{0} - \epsilon \log z_{1}$. The solution is presented in figs.\ref{fig:f7}(a-b), where we have taken $\epsilon = 0.4$, $R_{\rm vac}=1$, $R_{0} = 14 R_{\rm vac}$ and evolved over the time regime $\lambda = (0, 500)$. We see that $R$ undergoes damped, rapid oscillations about its General Relativistic limit. We find no exponential growth, contrary to the claim made in ref.\cite{sh20}.

\begin{figure}[tp]
     \centering
     \subfigure[]{
          \label{fig:19}
          \includegraphics[scale=0.4]{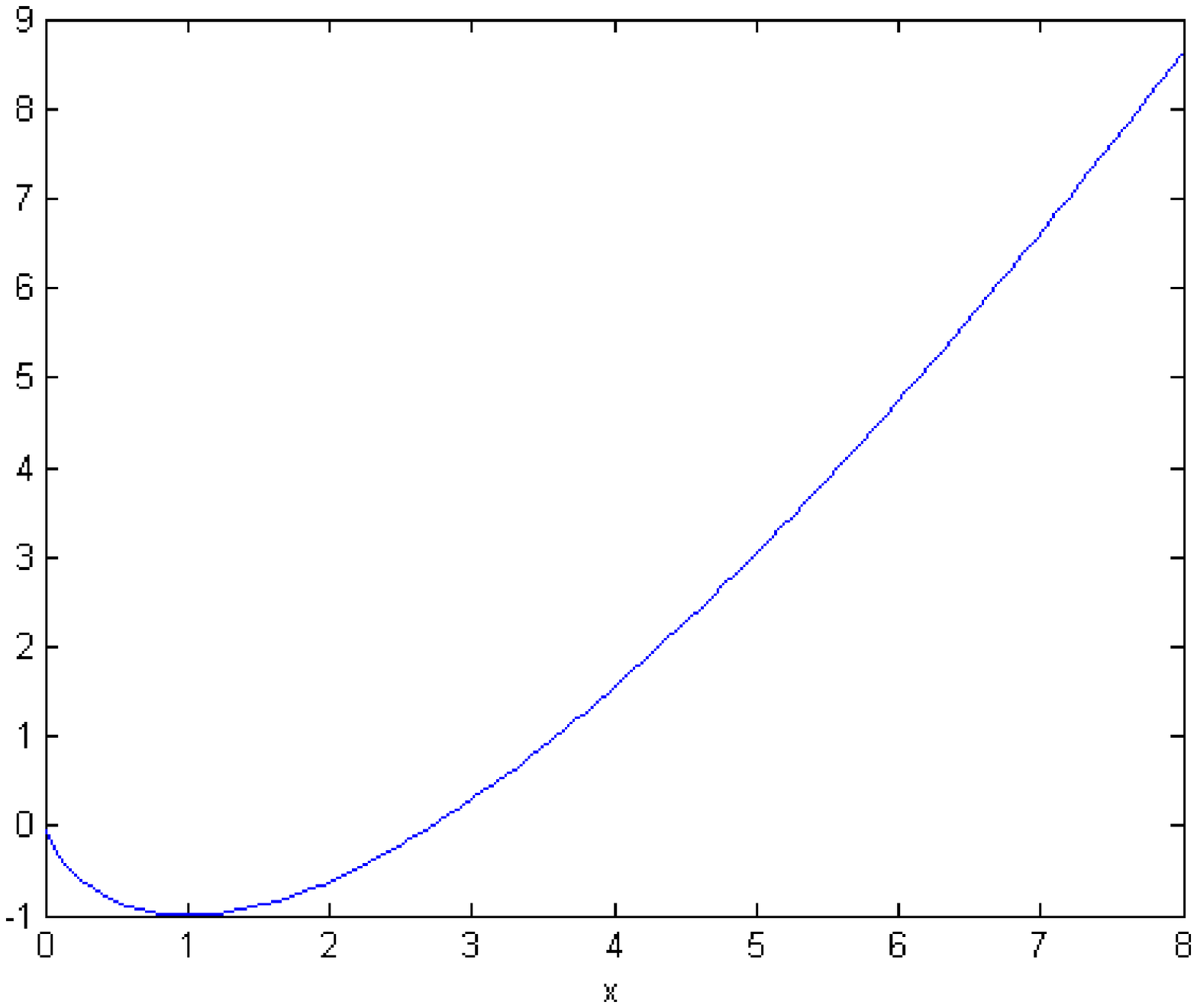}}
\hspace{.2in}
     \subfigure[]{
          \label{fig:xx21}
                \includegraphics[scale=0.4]{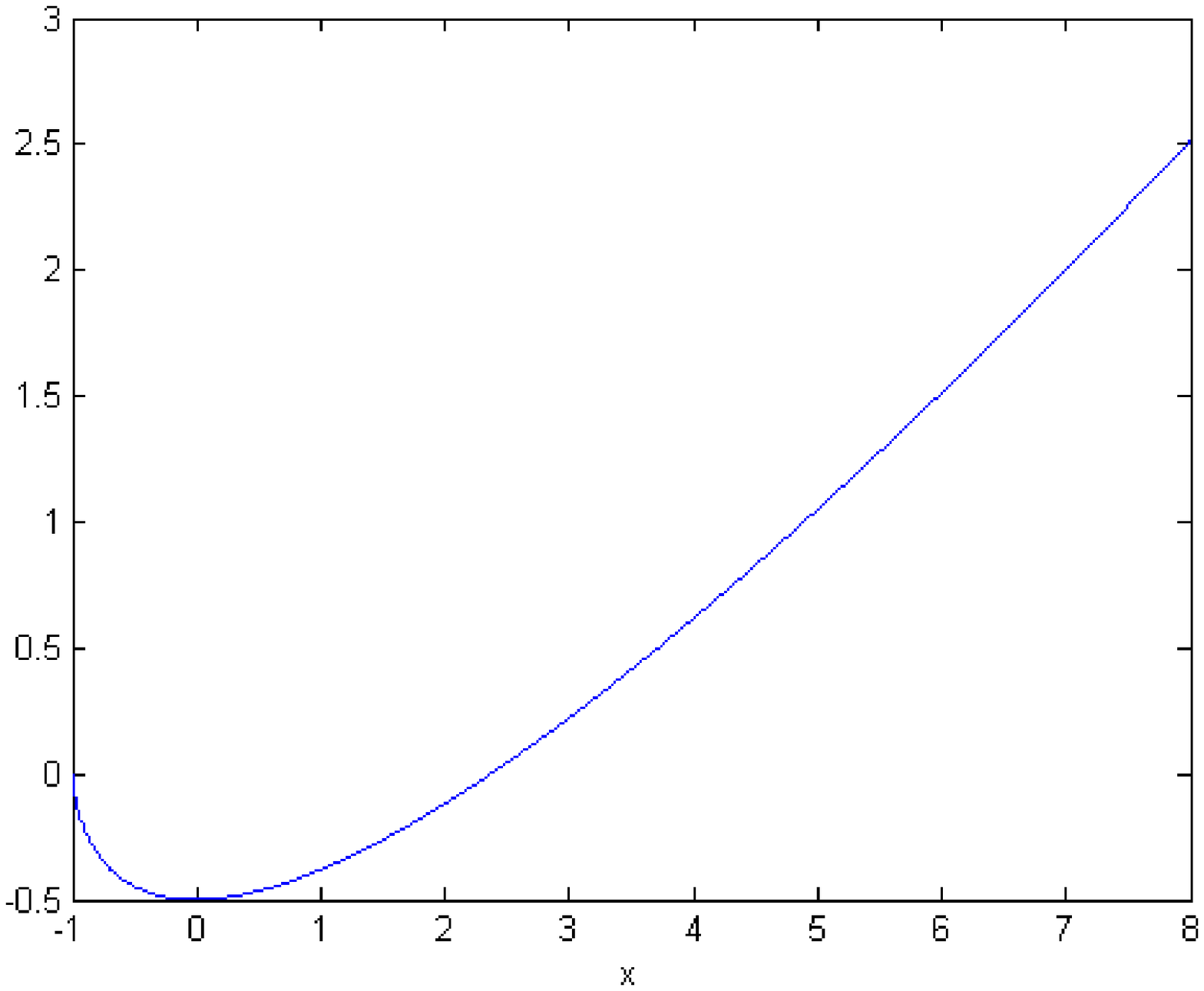}}
     \hspace{.2in}
     \subfigure[]{
          \label{fig:20}
                \includegraphics[scale=0.4]{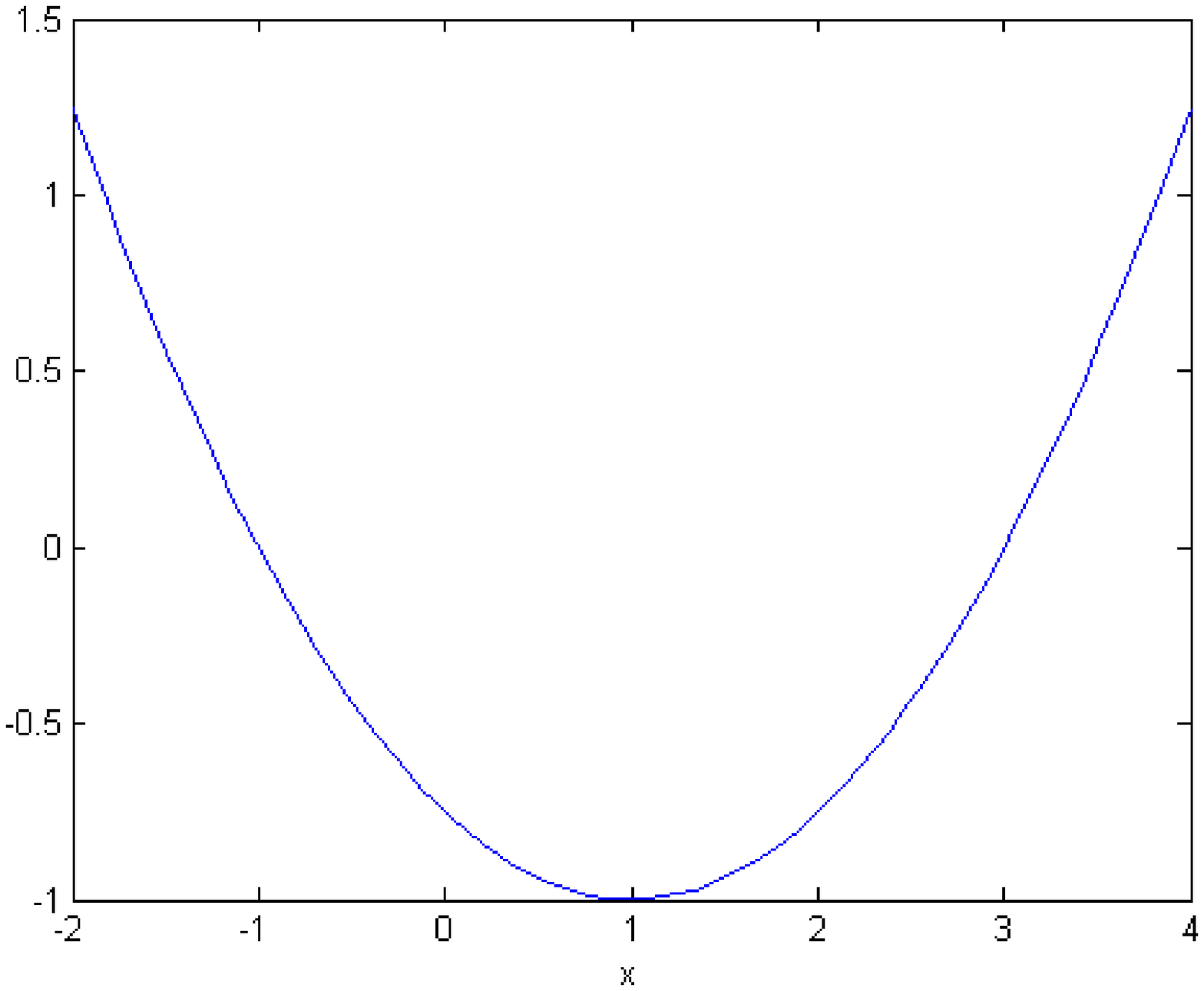}}
     \caption{(a) the logarithmic oscillator potential $V_{1} = -x + x\log x$; (b) the oscillator potential in the HSS model, $V_{2} = -(1+x)^{2/3}\left( 3/2 - (1+x)^{1/3}\right)$; (c) the simple harmonic oscillator potential. The asymmetry of the potential in (a) and (b) will lead to asymmetric solutions.}
     \label{fig:f6}
\end{figure}

\begin{figure}[htp]
     \centering
     \subfigure[]{
\label{fig:l23}
          \includegraphics[scale=0.4]{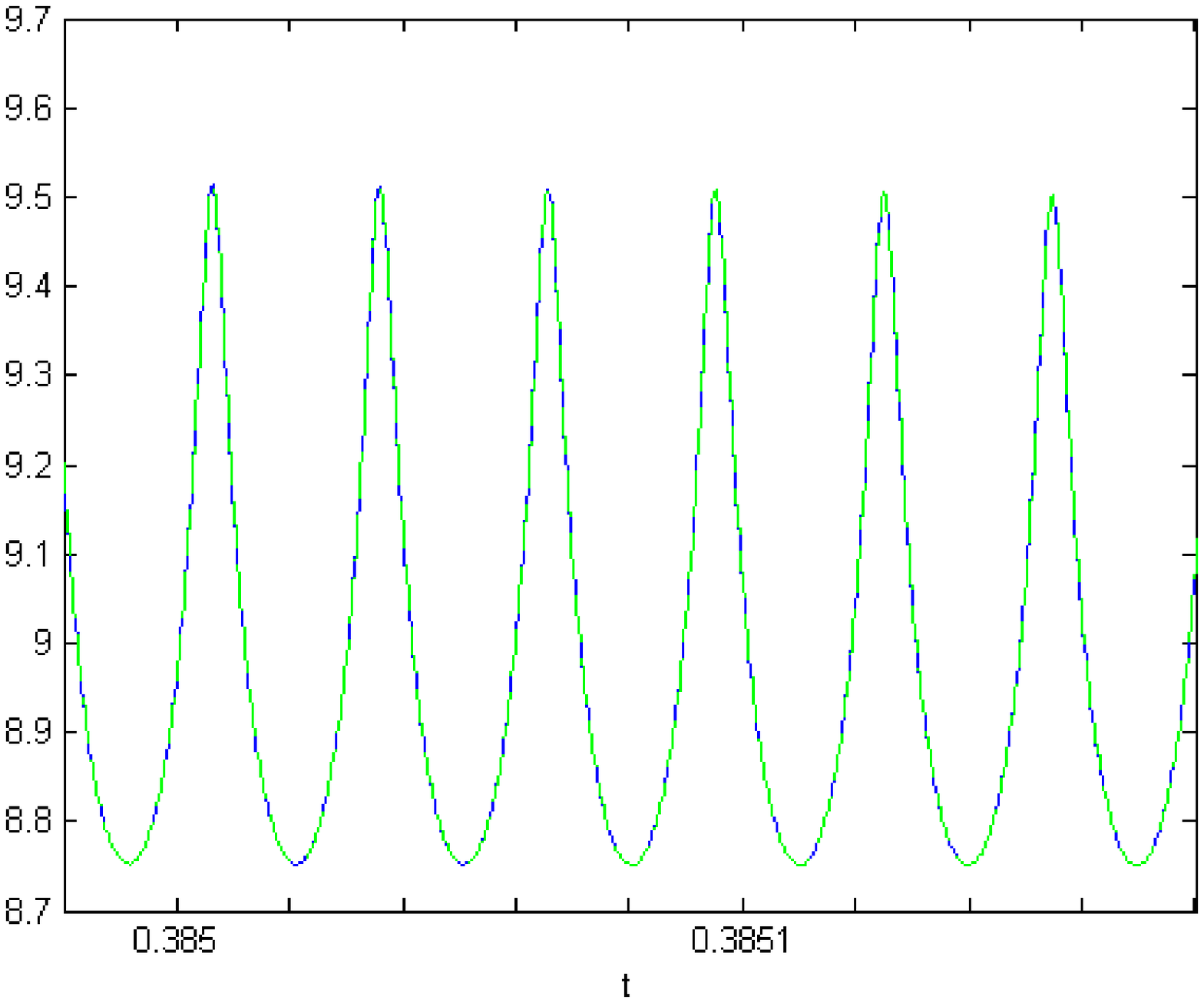}}
     \hspace{.2in}
     \subfigure[]{
          \label{fig:l24}
                \includegraphics[scale=0.4]{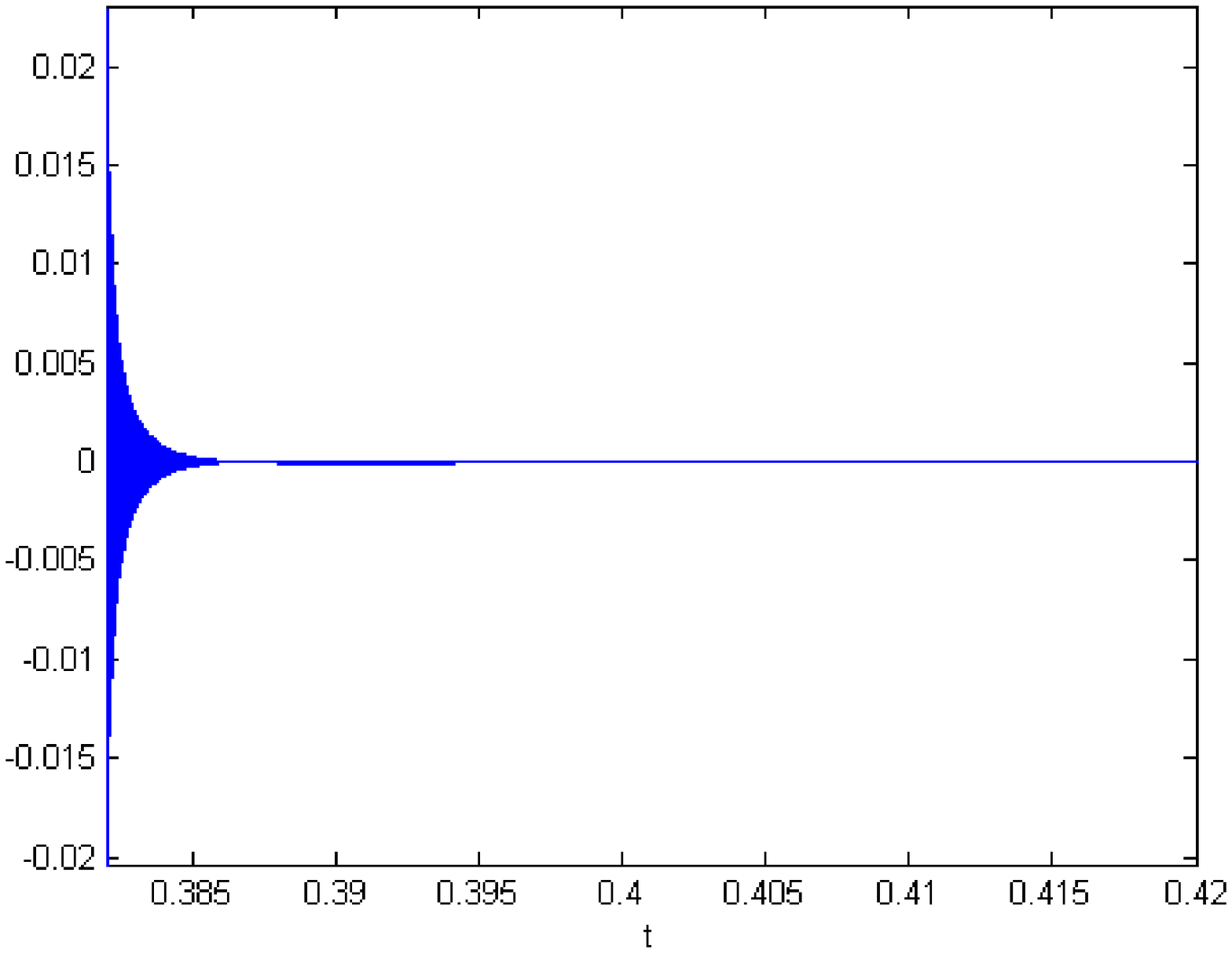}}
\hspace{.2in}
     \subfigure[]{
          \label{fig:l26}
                \includegraphics[scale=0.4]{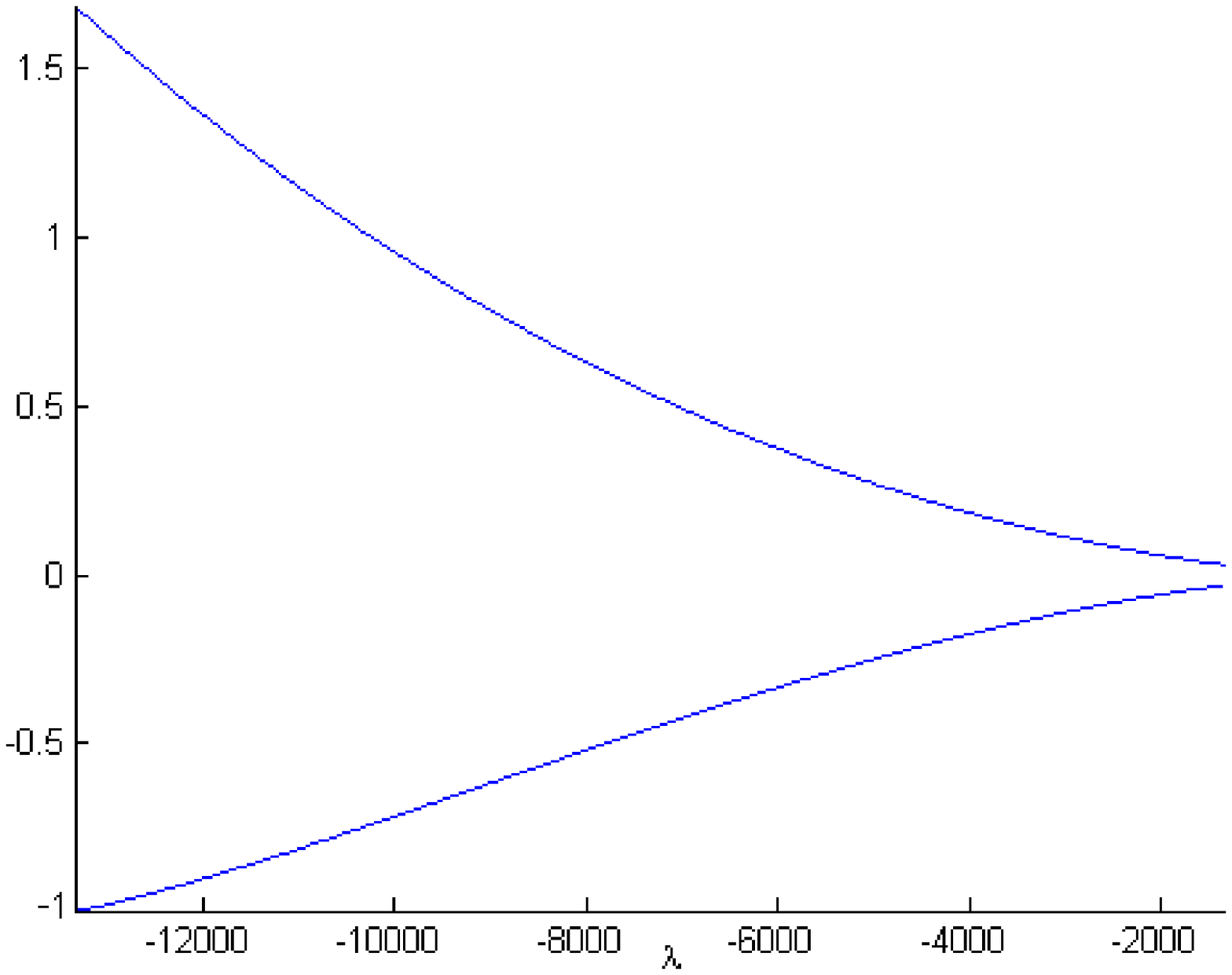}}
     \caption{(a) The Ricci scalar obtained by solving the full gravitational field equations for the AB model numerically (solid line) is exhibited along with the approximate solution $R = R_{\rm GR} - \epsilon \log (1+x)$ (dashed line), where $x$ is a solution to ($\ref{eq:l10}$). We see that the approximate solution closely mimics the full numerical solution; (b) The difference $(\delta R_{a} - \delta R)/\delta R$, as defined in the text. We see close agreement between the full and approximate solutions; (c) $x$ for the AB model (that is, the solution to ($\ref{eq:l10}$)). We see that after a finite time $x \to -1$, at which time $R \to \infty$. This singularity generically occurs when we evolve the AB model backwards through the matter era.}
     \label{fig:fg8}
\end{figure}

\begin{figure}[htp]
     \centering
     \subfigure[]{
\label{fig:ll23}
          \includegraphics[scale=0.4]{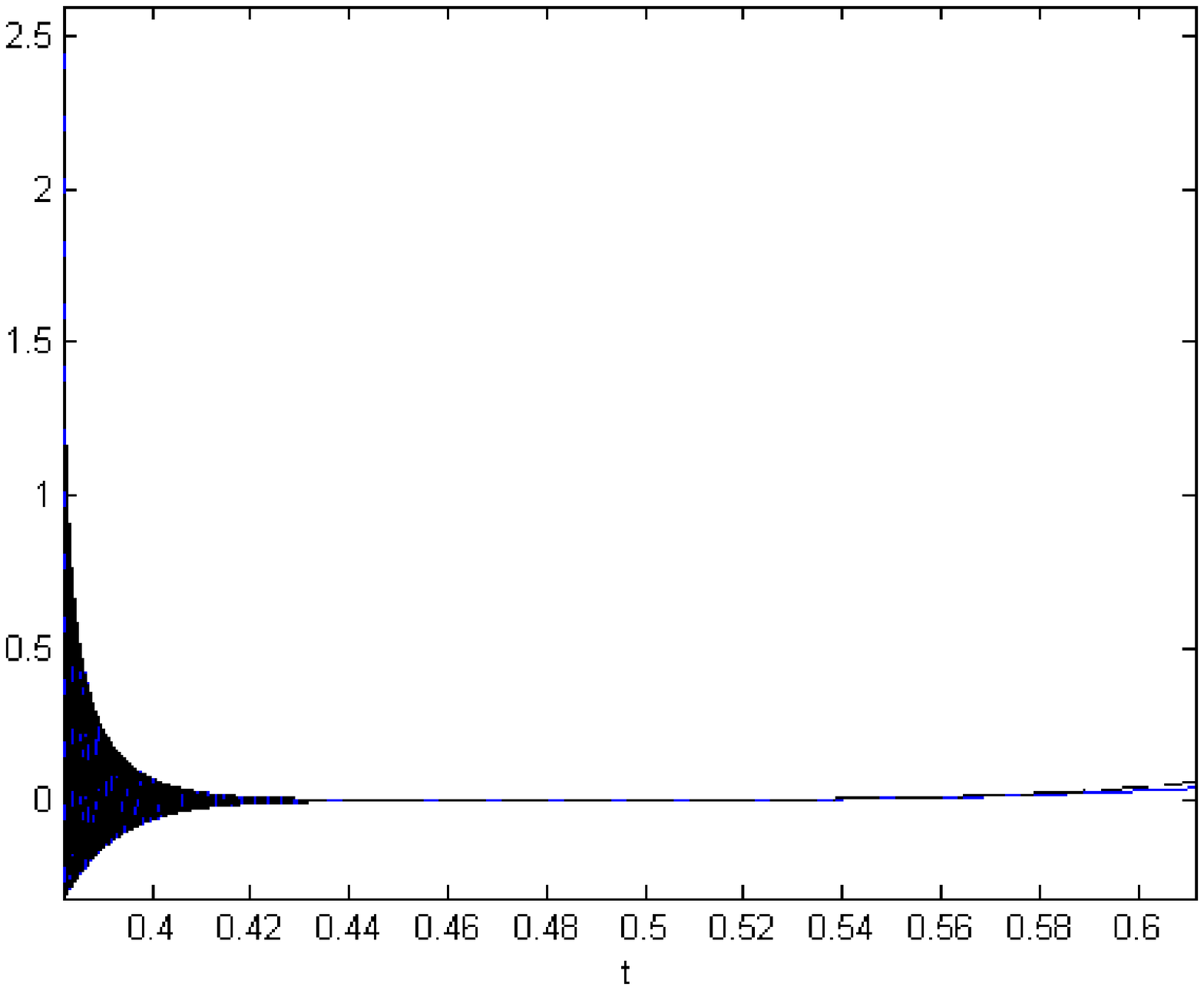}}
     \hspace{.2in}
     \subfigure[]{
          \label{fig:ll24}
                \includegraphics[scale=0.4]{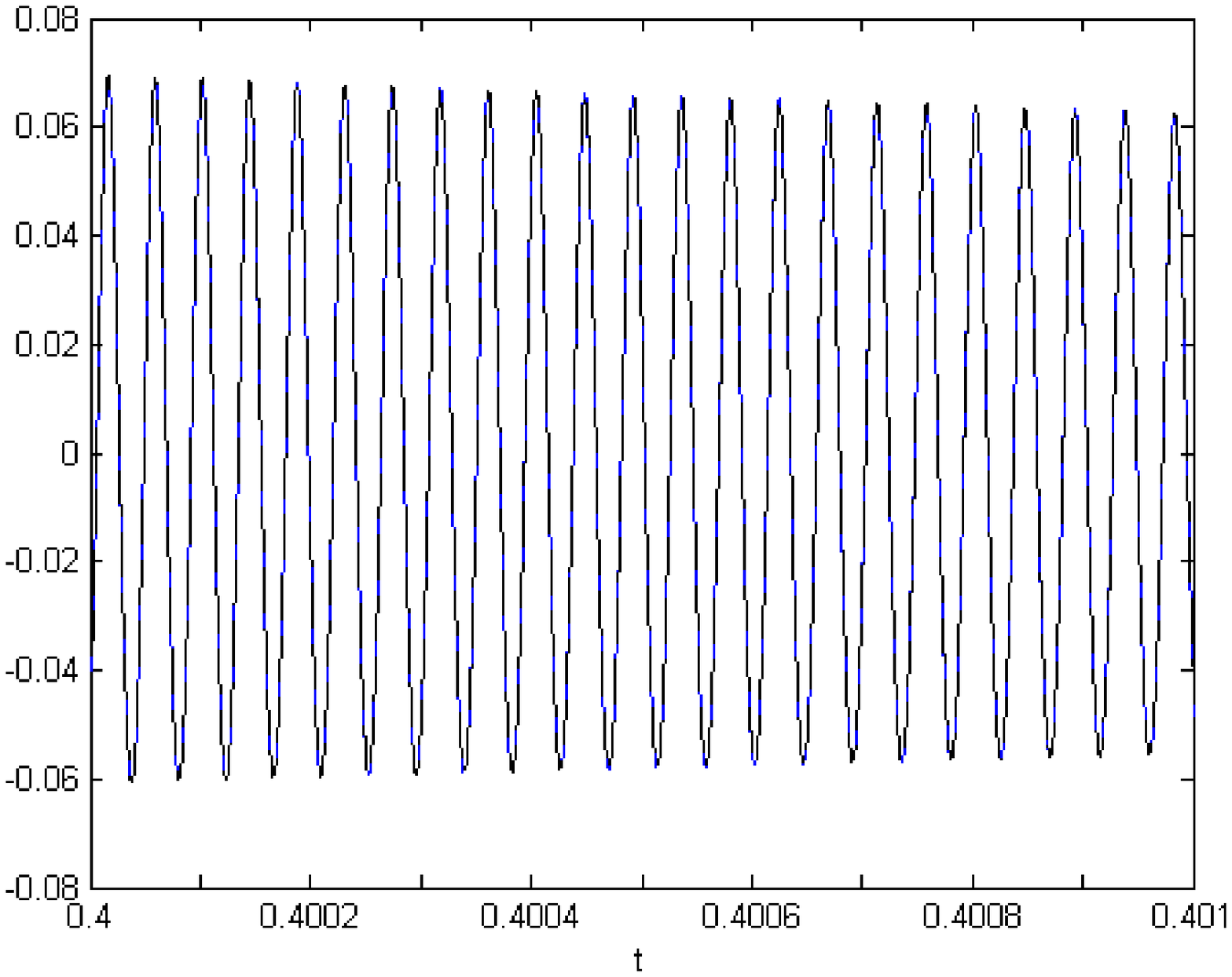}}
\hspace{.2in}
     \subfigure[]{
          \label{fig:ll25}
                \includegraphics[scale=0.4]{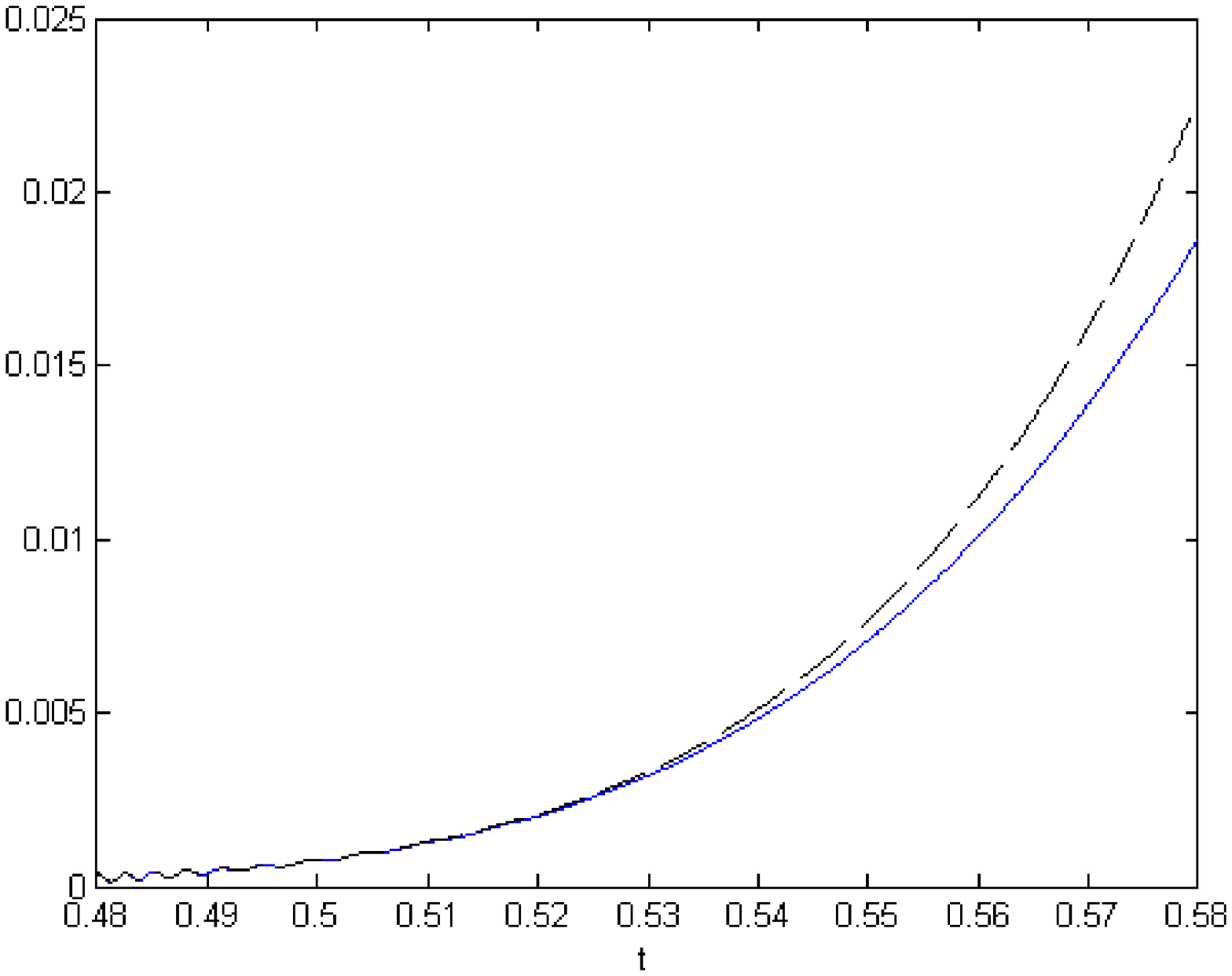}}
\hspace{.2in}
     \subfigure[]{
          \label{fig:ll26}
                \includegraphics[scale=0.4]{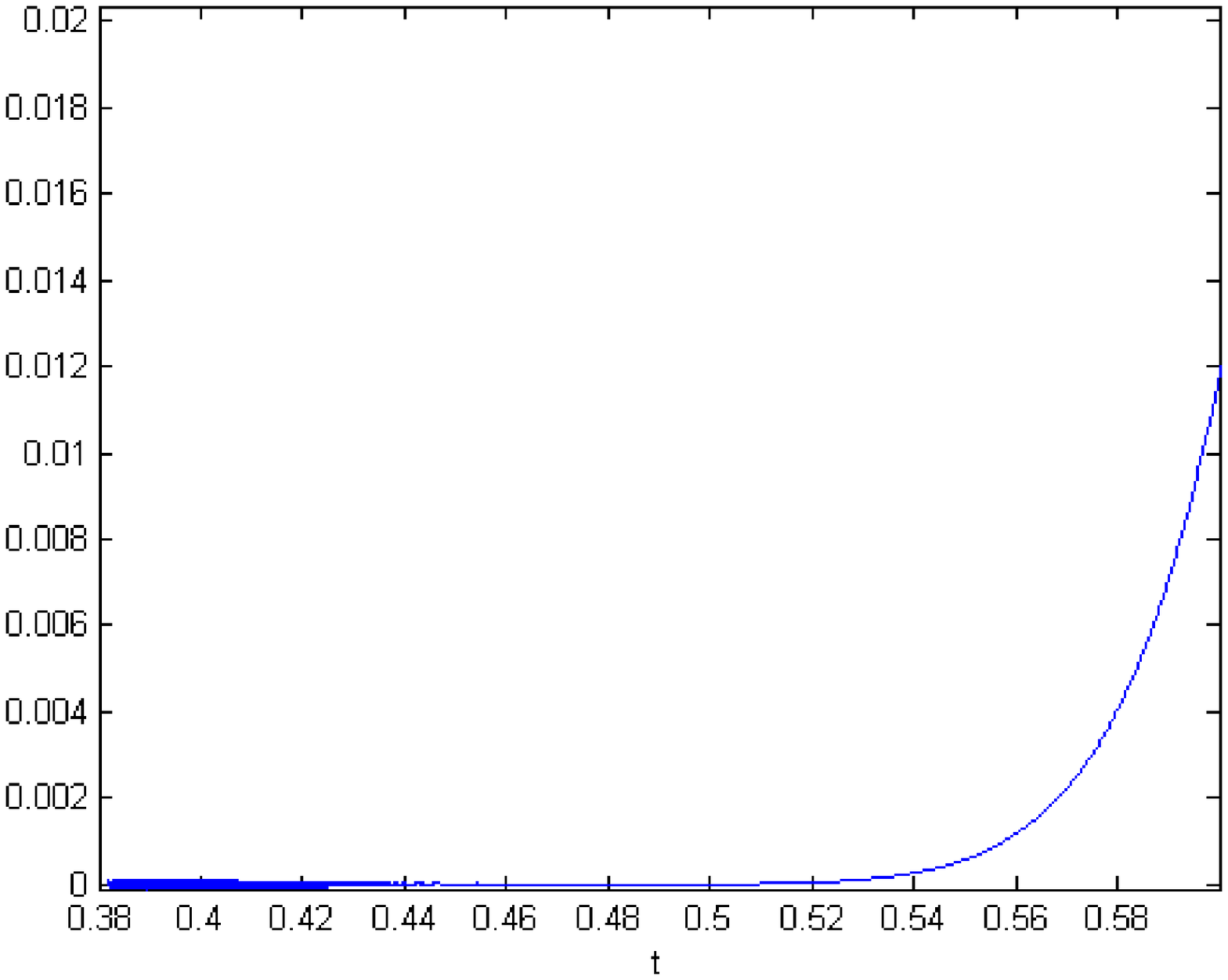}}
     \caption{(a) We take $x$ at an early time initially and evolve forwards in the time coordinate $\lambda$ (and hence forwards in $t$). The solid dark line is $\delta R = R - R_{\rm GR}$, obtained by solving the full field equations numerically, and the light dashed line is the approximate solution $\delta R_{\rm a} = - \epsilon \log(x_{\rm osc} + x_{\rm dr})$, where $x_{\rm osc}$ is the oscillatory component of $x$, given by the solution to ($\ref{eq:l10}$), and $x_{\rm dr}$ is given in ($\ref{eq:l125}$); (b) The full and approximate solutions $\delta R$ and $\delta R_{a}$ over a small time regime. We see close agreement between the two; (c) $\delta R$ and $\delta R_{a}$ over a different time regime, showing the drift away from $\delta R = 0$. Again we see a close agreement, except at late times where $R \sim R_{\rm vac}$; (d) The difference $\delta R_{a} - \delta R$. This difference is small for $R_{\rm GR} \gg \epsilon$, but increases when $R_{\rm GR} \sim R_{\rm vac}$.}
     \label{fig:fg9}
\end{figure}

\begin{figure}[tp]
     \centering
     \subfigure[]{
          \label{fig:21}
          \includegraphics[scale=0.4]{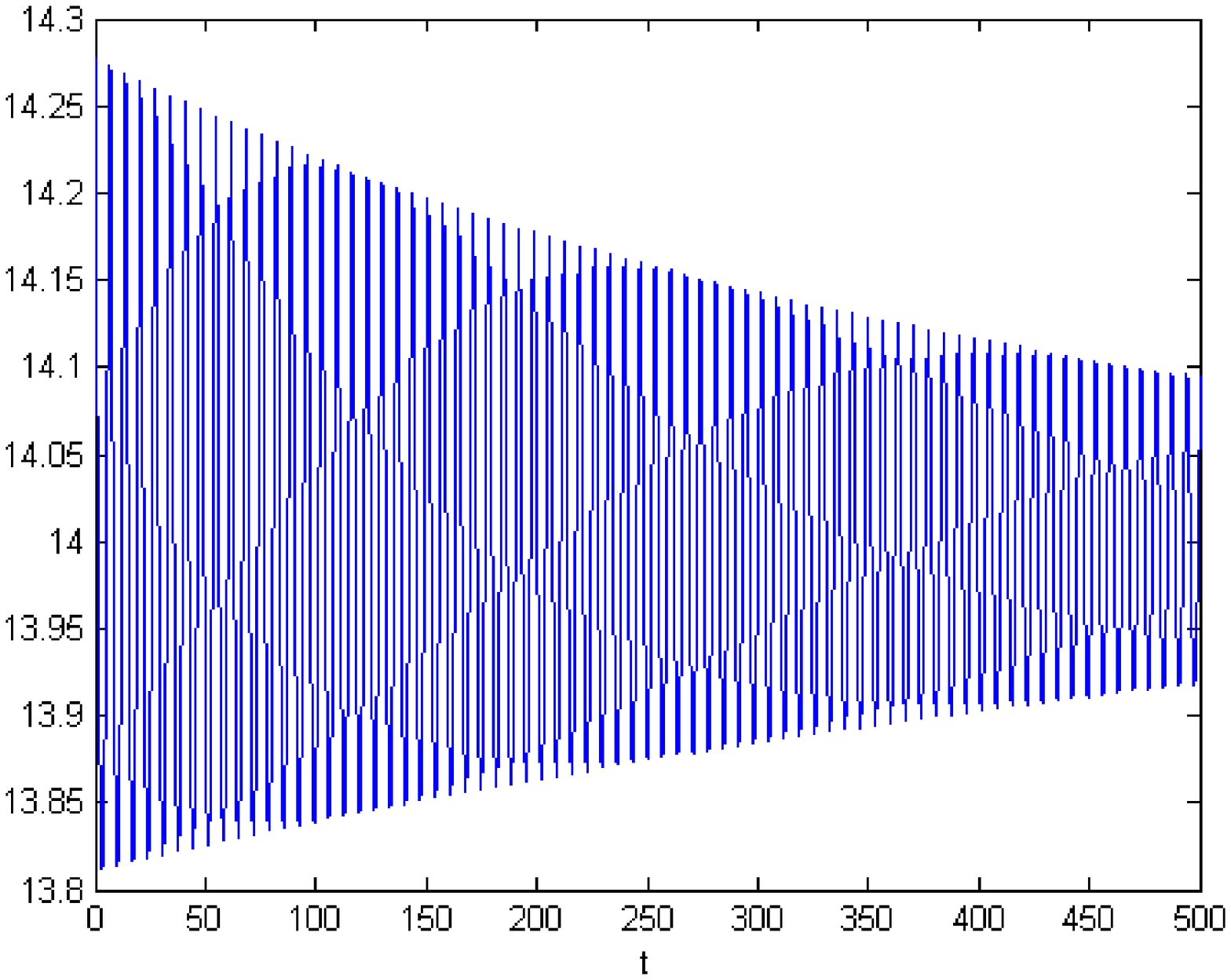}}
     \hspace{.2in}
     \subfigure[]{
          \label{fig:22}
                \includegraphics[scale=0.4]{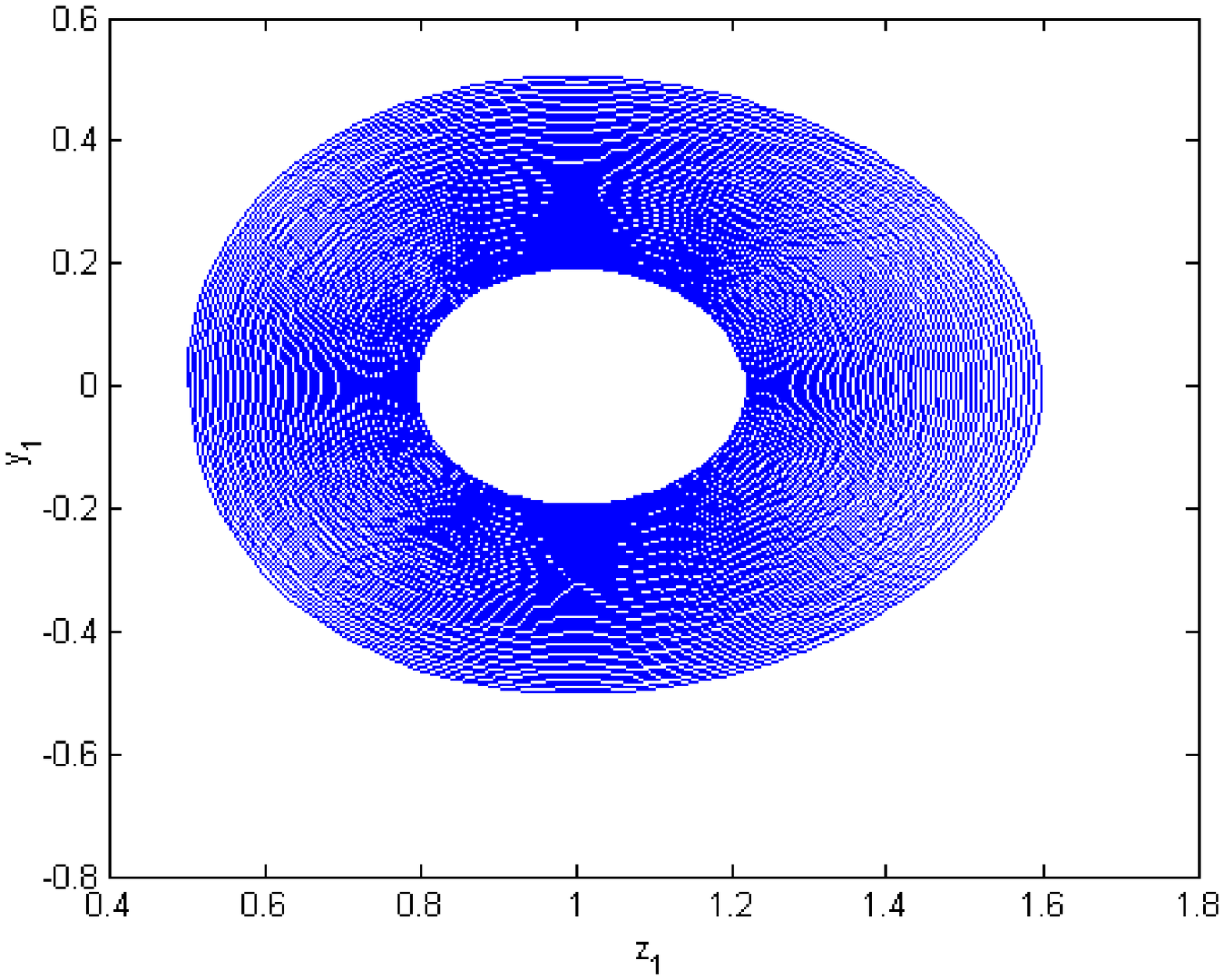}}
     \caption{(a) $R = R_{0} - \epsilon \log z_{1}$, for constant $R_{0} = 14$. We see that the Ricci scalar undergoes damped oscillations around $R_{0}$; (b) is a phase diagram of $y_{1} = \dot{z}_{1}$ against $z_{1}$. We see that $z_{1}$ undergoes damped oscillations around $z_{1}=1$, which is the General Relativistic limit.}
     \label{fig:f7}
\end{figure}

\begin{figure}[htp]
     \centering
     \subfigure[]{
\label{fig:lll23}
          \includegraphics[scale=0.4]{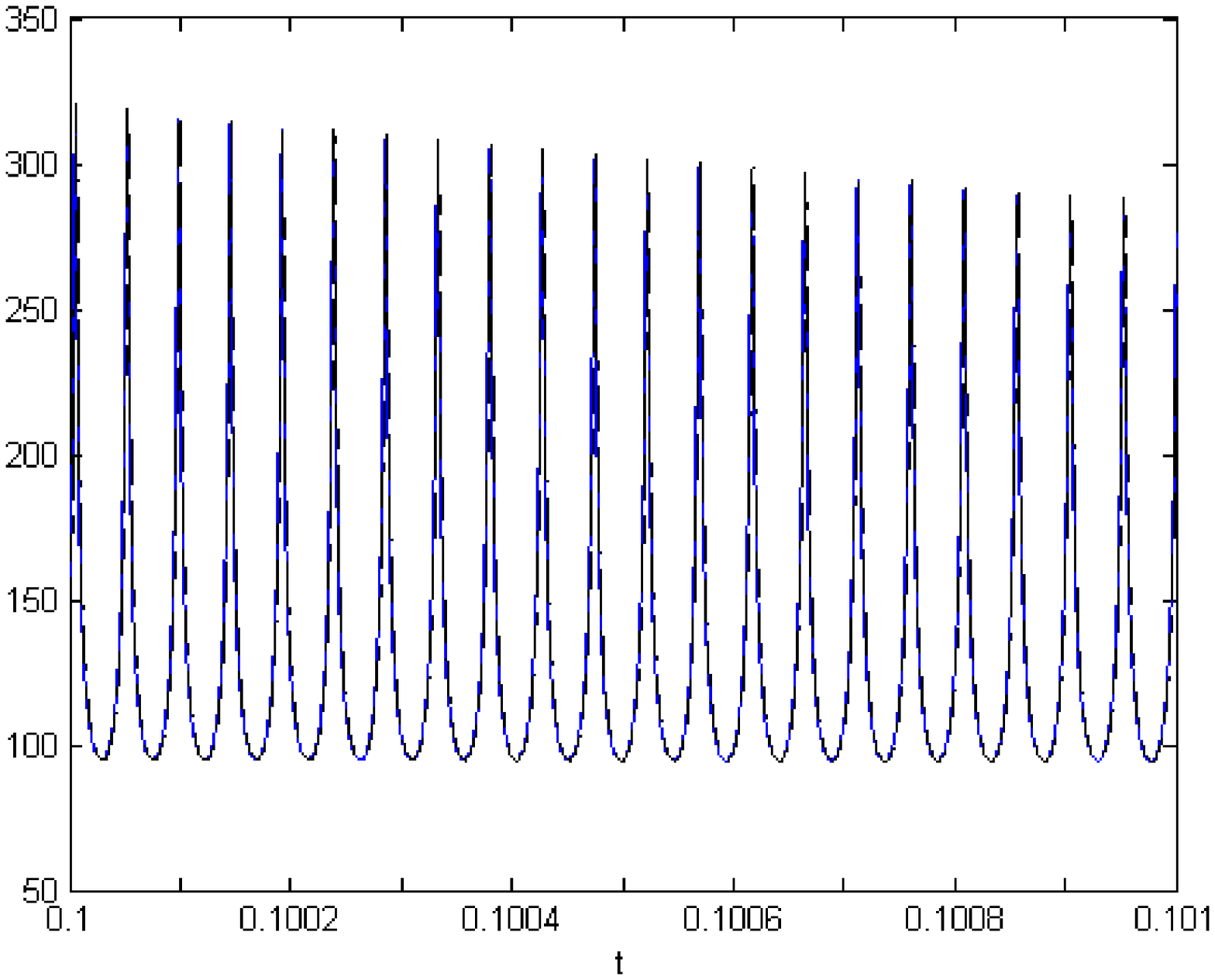}}
     \hspace{.2in}
     \subfigure[]{
          \label{fig:lll24}
                \includegraphics[scale=0.4]{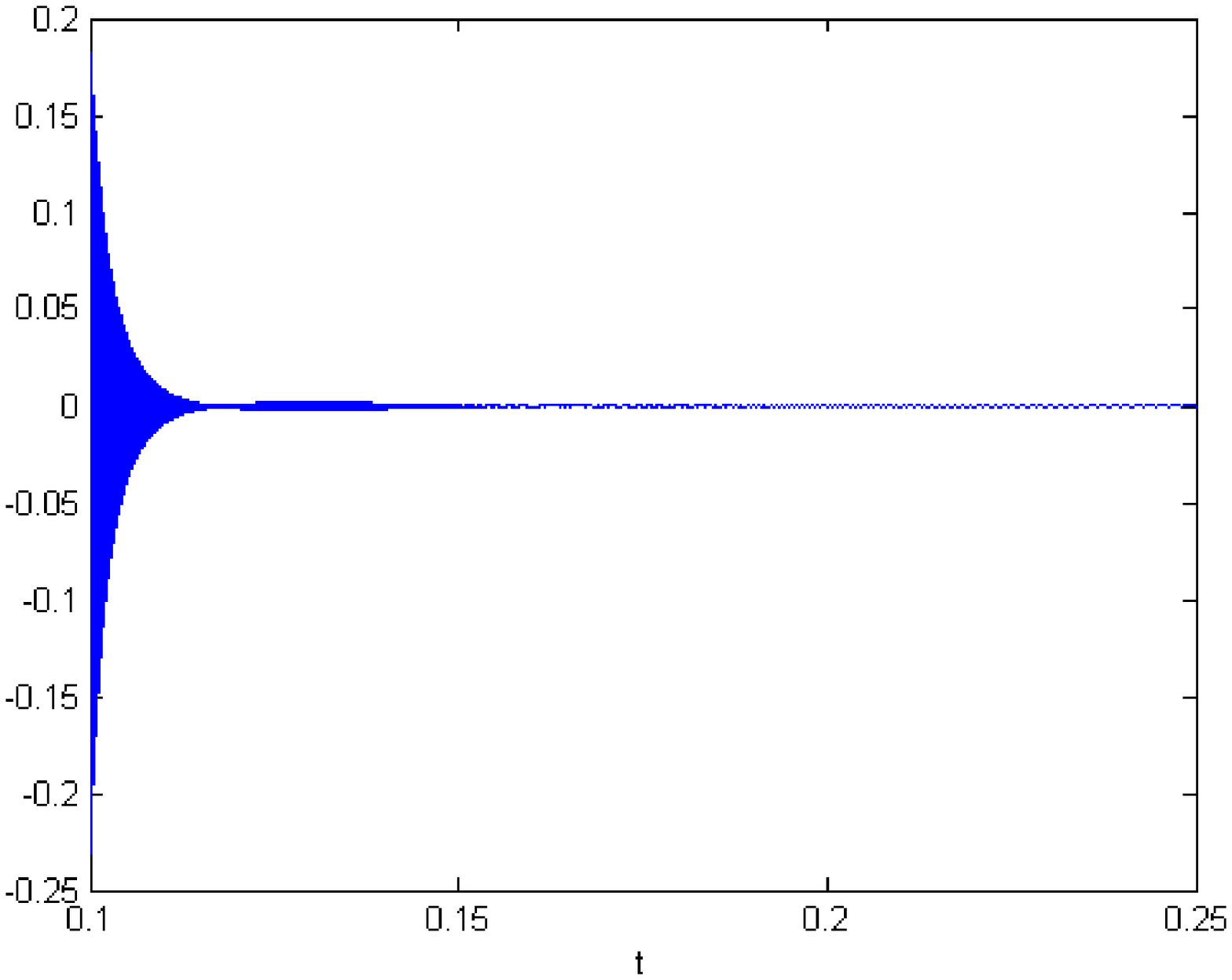}}
\hspace{.2in}
     \subfigure[]{
          \label{fig:lll26}
                \includegraphics[scale=0.4]{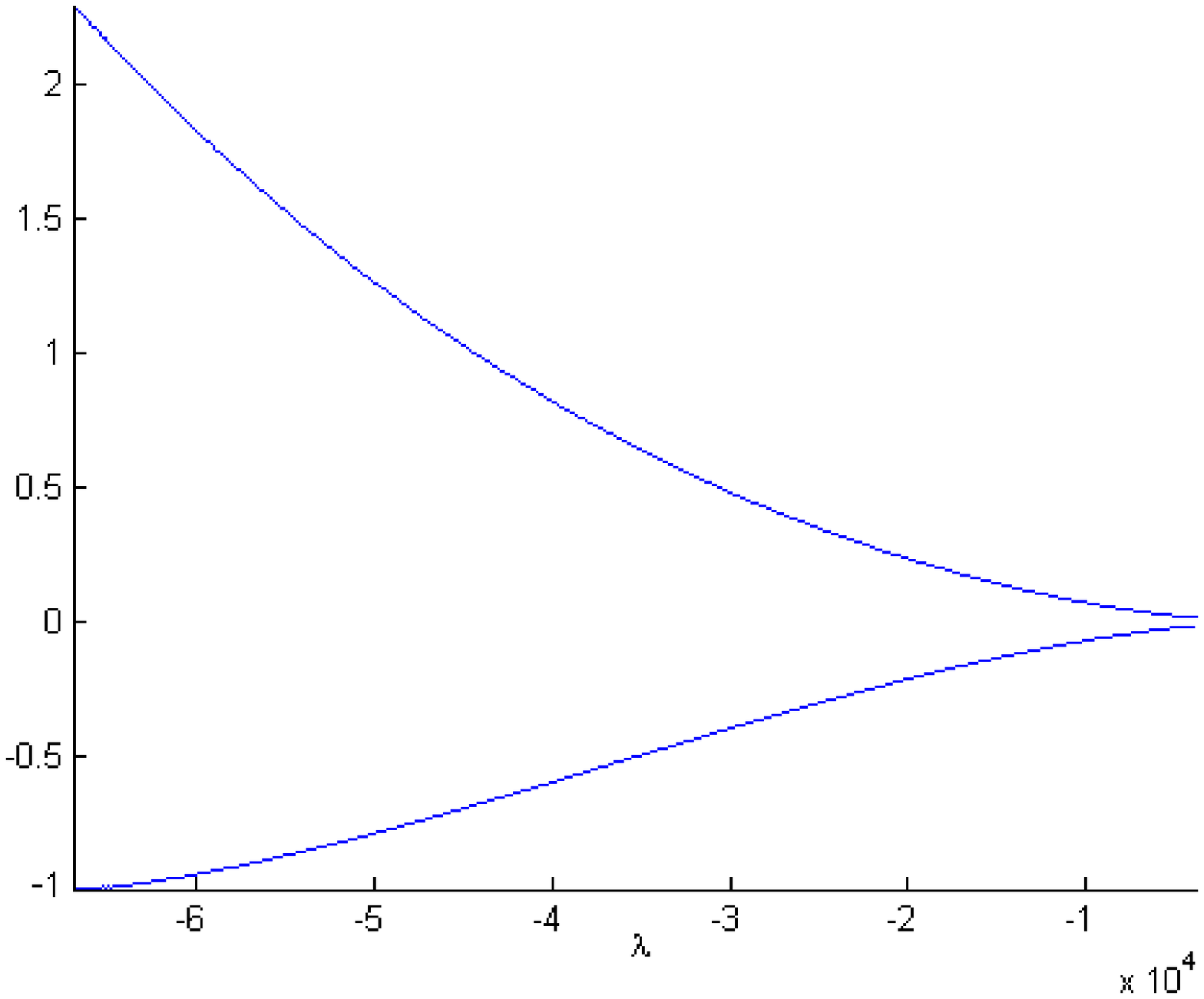}}
     \caption{(a) The Ricci scalar obtained by solving the full gravitational field equations for the HSS model numerically (solid line) is exhibited along with the approximate solution $R = R_{\rm GR}/ (1+x)^{1/3}$ (dashed line), where $x$ is a solution to ($\ref{eq:hu127}$). We see that the approximate solution closely mimics the full numerical solution; (b) The difference $(x_{a} - x)/x$, as defined in the text. We see close agreement between the full and approximate solutions; (c) $x$ for the AB model (that is, the solution to ($\ref{eq:l10}$)). We see that after a finite time $x \to -1$, at which time $R \to \infty$. This singularity generically occurs when we evolve the HSS model backwards through the matter era.}
     \label{fig:fg10}
\end{figure}

\begin{figure}[htp]
     \centering
     \subfigure[]{
\label{fig:llll23}
          \includegraphics[scale=0.4]{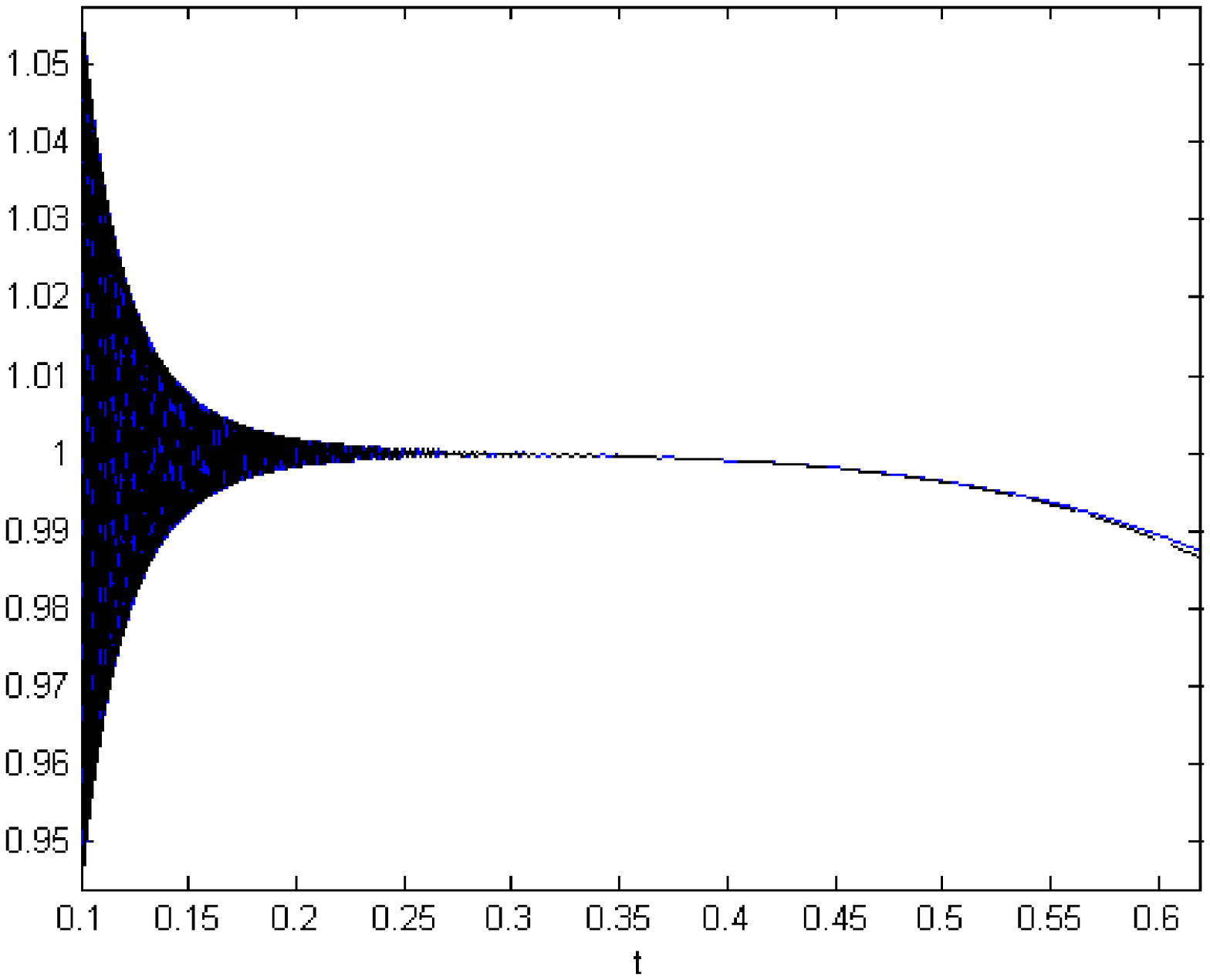}}
     \hspace{.2in}
     \subfigure[]{
          \label{fig:llll24}
                \includegraphics[scale=0.4]{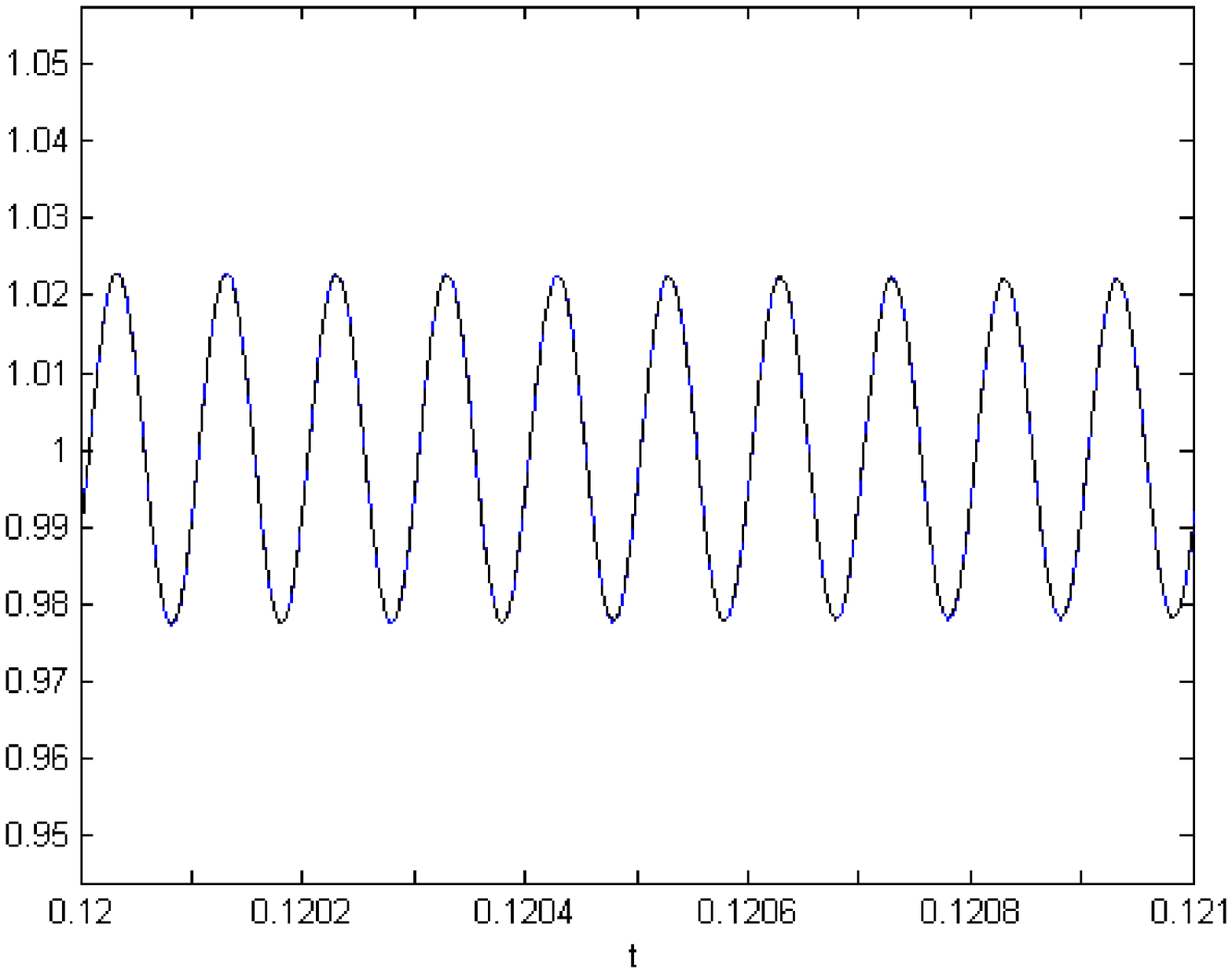}}
\hspace{.2in}
     \subfigure[]{
          \label{fig:llll25}
                \includegraphics[scale=0.4]{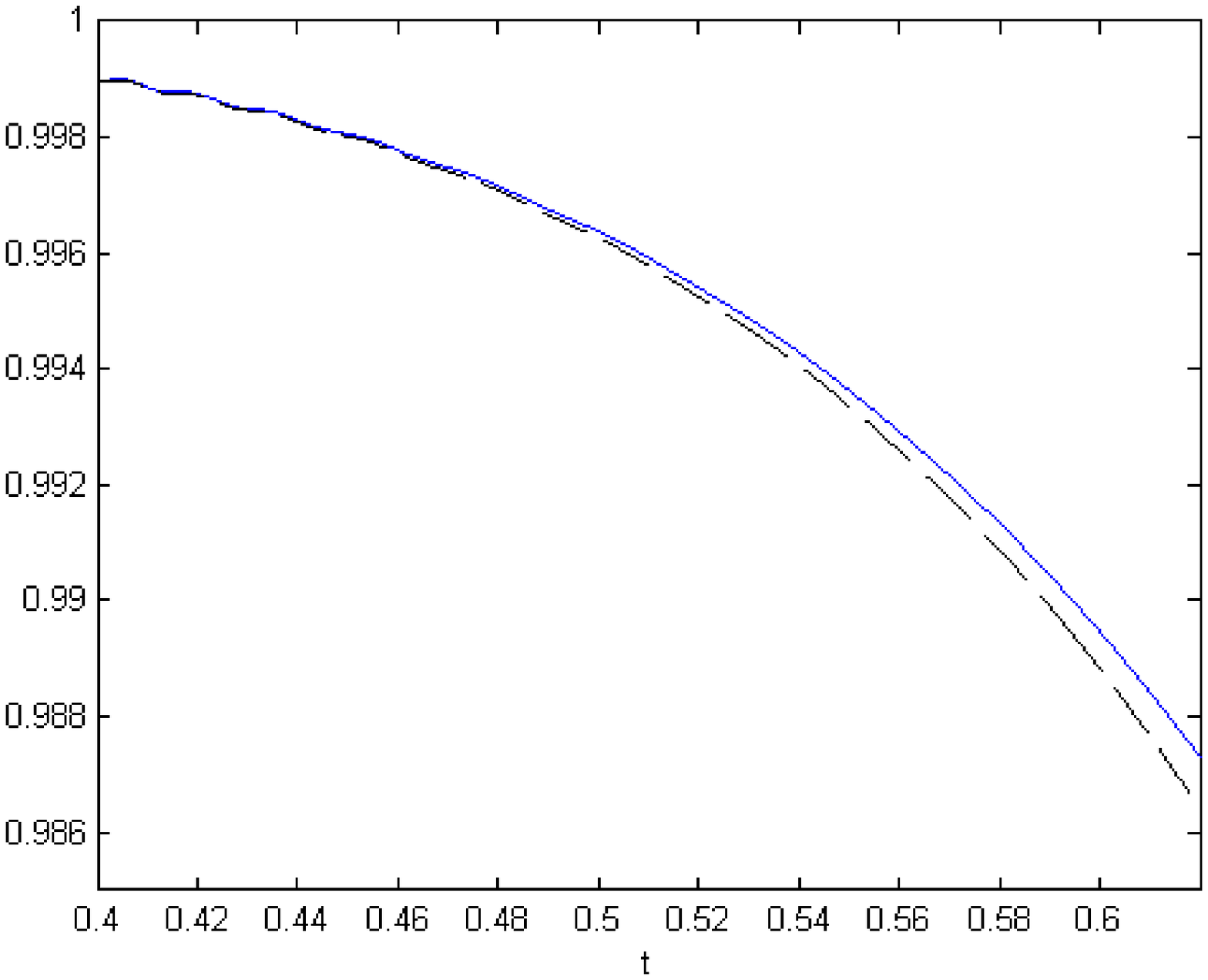}}
\hspace{.2in}
     \subfigure[]{
          \label{fig:llll26}
                \includegraphics[scale=0.4]{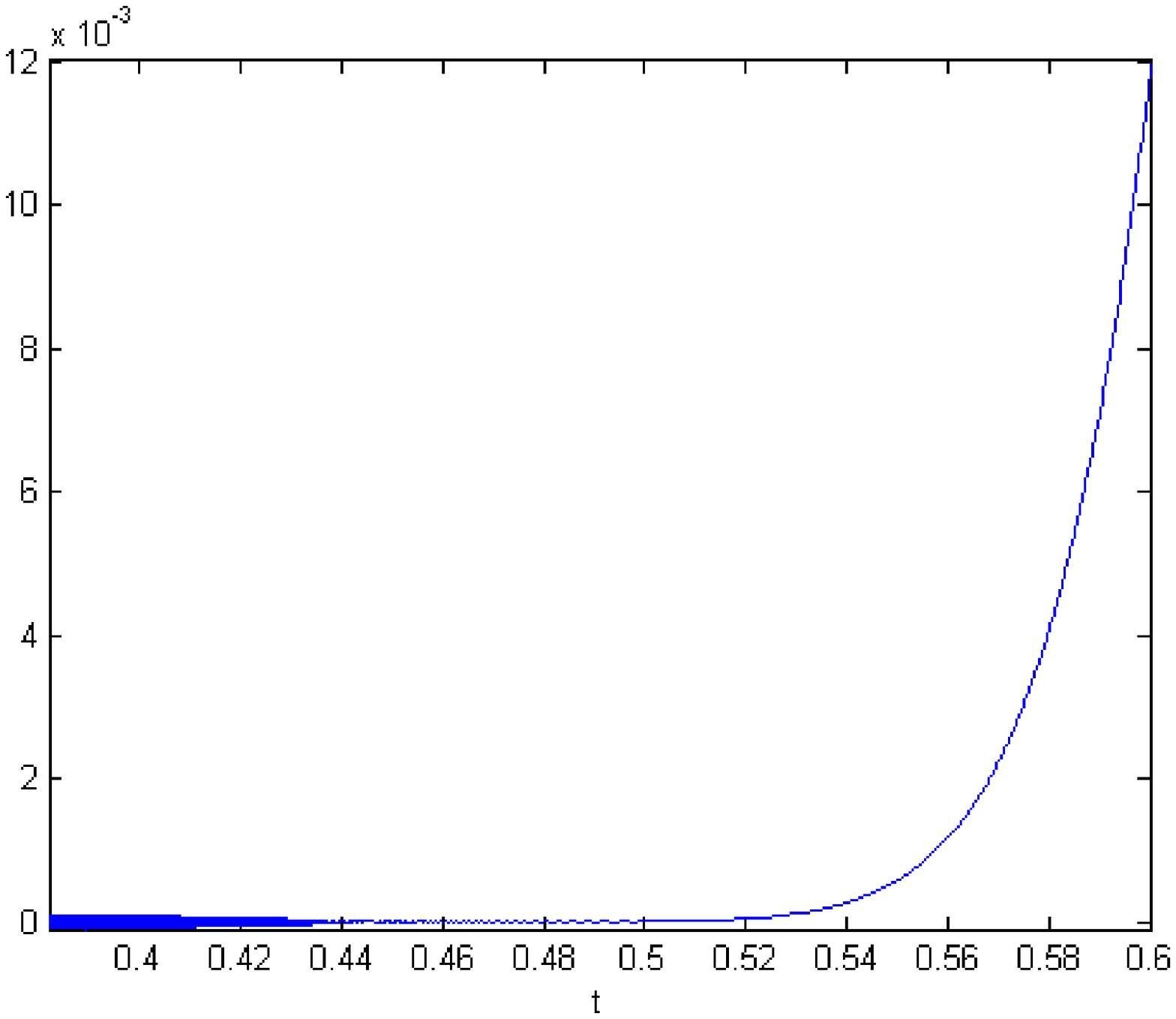}}
     \caption{(a) We take $x$ at an early time initially and evolve forwards in the time coordinate $\lambda$ (and hence forwards in $t$). The solid dark line is $x = (R_{\rm GR}/R)^{3}$, obtained by solving the full field equations numerically, and the light dashed line is the approximate solution $x_{\rm a} = x_{\rm osc} + x_{\rm dr}$, where $x_{\rm osc}$ is the oscillatory component of $x$, given by the solution to ($\ref{eq:hu127}$), and $x_{\rm dr}$ is given in ($\ref{eq:hu125}$); (b) The full and approximate solutions $x$ and $x_{a}$ over a small time regime. The oscillatory components of $x$ and $x_{a}$ show close agreement oscillations; (c) $x$ and $x_{a}$ over a different dynamical range. We see the drift away from $x = 1$; (d) The difference $x - x_{\rm a}$. This difference is small for $R_{\rm GR} \gg \epsilon$, but diverges when $R_{\rm GR} \sim R_{\rm vac}$.}
     \label{fig:fg11}
\end{figure}

\subsection{HSS Model}

In this section, we consider a similar calculation to that presented above, but now for the HSS model. We first re-write the trace of the gravitational field equations in terms of a dimensionless function and a fast time coordinate. We then derive approximate expressions to describe the oscillations and drift of the Ricci scalar away from its General Relativistic limit. We begin by substituting $\chi_{\rm HSS}$ into ($\ref{eq:1}$), which gives

\begin{equation}\label{eq:hu1}  \Box R - {2(n+1) \over R} (\nabla R)^{2} -\left(R+ {T \over M_{\rm pl}^{2}}   \right){R^{2n+2} \over 6n(2n+1)\epsilon^{2n+1}} - {n+1 \over 3n(2n+1)} R^{2} \approx 0 . \end{equation}

\noindent where we have neglected the term $R_{\rm vac}$. To transform equation ($\ref{eq:hu1}$) into an oscillator equation, we must make a series of field redefinitions. By first defining

\begin{equation} u={(-T/M_{\rm pl}^{2})^{2n+1} \over R^{2n+1}} >0, \qquad \qquad \alpha(t) = {\epsilon^{2n+1} \over (-T/M_{\rm pl}^{2})^{2n+1}}>0, \end{equation}

\noindent we write ($\ref{eq:hu1}$) as

\begin{equation}\label{eq:hu2} 6n{d^{2}  \over dt^{2}}(u \alpha) + 18n H{d  \over dt}(u\alpha) - {\epsilon \over \alpha^{1/(2n+1)}}\left[{1 \over u^{1/(2n+1)}}-1 \right] + {n+1 \over 3n}{T \over M_{\rm pl}^{2}}\alpha u^{2n/(2n+1)}\approx 0 .\end{equation}

\noindent Next, by introducing a fast time coordinate

\begin{equation} \lambda = {1 \over \epsilon^{n+1/2}} \int \alpha^{-1/2(2n+1)}\left({-T \over M_{\rm pl}^{2}}\right)^{n+1/2} dt ,\end{equation}

\noindent and defining $x = u-1$, we can write ($\ref{eq:hu2}$) in terms of $u$ and $\lambda$ as

\begin{equation} \label{eq:hu11} x'' + \left({6n+2 \over 2(2n+1)} (\log\alpha)' + 3 \bar{H}\right)x'  - {\epsilon \over 6n} \left({1 \over (1+x)^{1/(2n+1)}}-1 \right) + {n+1 \over 3n}{T \over M_{\rm pl}^{2}}\alpha (1+x)^{2n/(2n+1)} \end{equation} \begin{equation*}+ \left( {\alpha'' \over \alpha}- {n+1 \over 2n+1} \left({\alpha' \over \alpha}\right)^{2} + 3\bar{H} (\log\alpha)'\right)(1+x)   \approx 0 ,\end{equation*}

\noindent where primes denote differentiation with respect to $\lambda$, and $\bar{H} = a'/a$. This is a non-linear, inhomogeneous second order differential equation for $x$, and by solving for $x$ we obtain the Ricci scalar from the relation $R = -T/(M_{\rm pl}^{2}(1+x)^{1/(2n+1)})$. As before, we can deduce both the oscillatory component of $R$ and the drift away from $R_{\rm GR}$.

We now take as a specific example a pure matter era and set $n=1$, in which case we can write $\alpha = \epsilon^{3}/R_{\rm GR}^{3}$ and $\lambda = -16/27\epsilon^{2}t^{3}$. We also make the assumption that $H \approx H_{\rm GR}$, which implies that

\begin{equation} \label{eq:hu123} x'' - {10 \over 3 \lambda} x'  - {\epsilon \over 6} \left({1 \over (1+x)^{1/3}}-1 \right) + {14 \over 3 \lambda^{2}}(1+x)   \approx 0 ,\end{equation}

\noindent where we have neglected the third term on the left hand side of ($\ref{eq:hu11}$), which is negligible.

\subsubsection{Drift}

As in the AB model, an approximate expression for the drift can be found by solving the full inhomogeneous equation ($\ref{eq:hu123}$). We take as an ansatz

\begin{equation}\label{eq:hu125} x =  -{84 \over \epsilon \lambda^{2}} .\end{equation}

\noindent with this choice, $x'$ and $x''$ are of order $x' \sim  \lambda^{-3}$ and $x'' \sim \lambda^{-4}$, and hence the derivative terms in ($\ref{eq:hu123}$) can be neglected if we only consider terms of order ${\cal O}\left(\lambda^{-2}\right)$. As a result, it reduces to an algebraic expression for $x$, which is solved by ($\ref{eq:hu125}$) (this can be verified by direct substitution).

\subsubsection{Oscillations}

Having calculated the drift by solving ($\ref{eq:hu123}$), the oscillatory component of $x$ can be obtained by neglecting the last term on the right hand side of ($\ref{eq:hu123}$) (since it is of order $\sim \lambda^{-2}$), and solving the equation

\begin{equation} \label{eq:hu127}x'' - {10 \over 3 \lambda} x'  - {\epsilon \over 6} \left({1 \over (1+x)^{1/3}}-1 \right)  \approx 0 .\end{equation}

\noindent This equation describes the non-linear oscillations of $x$ around $x=0$, that is around the General Relativistic limit $R = -T / M_{\rm pl}^{2}$.  We have solved equation ($\ref{eq:hu127}$), using the initial conditions $x_{\rm i} = (R_{\rm GR}^{i} / (R_{\rm GR}^{i} + 0.1))^{3}-1$, $x'_{\rm i} = 0$, over the range $\lambda = (-3790,-69260)$ and taking $\epsilon =0.1$, and the results are exhibited in fig.($\ref{fig:fg10}$). We compare the Ricci scalar obtained in this section, given by $R = R_{\rm GR}/(1+x)^{1/3}$, to the Ricci scalar found in section \ref{sec:s10}, and find that they are in close agreement.

\subsubsection{Existence of Singularity}

As in the AB model, by using this method we find that we can only evolve $x$ backwards over a very limited dynamical range, and this is once again due to the presence of a singularity. In fig.$\ref{fig:lll26}$, we see that as we evolve $x$, after a finite time $x \to -1$, and since $R = R_{\rm GR}/(1+x)^{1/3}$, we find that as $x \to -1$, $R \to \infty$. This singularity arises due to the initial conditions that we have imposed.

As in the AB model, to evade this singularity we will instead choose initial conditions at an early time and evolve forwards through the matter era. By doing so, $x$ will remain regular throughout. To find the initial conditions that $x$ may take in order to evade this singularity, we perform the same steps as in the AB model. We first observe that $x_{\rm in} > -1$ is a lower bound, and the upper bound can be obtained from the expression

\begin{equation}\label{eq:hu130} {(x')^{2} \over 2} - (1+x)^{2n/(2n+1)} \left( {2n+1 \over 2n} - (1+x)^{1/(2n+1)}\right) = E ,\end{equation}

\noindent where we have neglected the damping term and terms of order $\sim \lambda^{-2}$ in ($\ref{eq:hu127}$). By substituting $x \to -1$, we find that $E = 0$, and hence the upper bound for $x_{\rm i}$ is found by solving ($\ref{eq:hu130}$) for $E=0$. Doing so, we arrive at $x_{i} < ((2n+1)/2n)^{2n+1}-1$, which for $n=1$ is given by $x_{\rm i} = (3/2)^{3}-1 =2.375$. In terms of $R = R_{\rm GR}/(1+x)^{1/3}$, this corresponds to the choice $2n R_{\rm GR}/(2n+1) < R_{\rm in} < \infty$ at the beginning of the matter era.

With this in mind, we now solve ($\ref{eq:hu127}$) by evolving forwards in the fast time coordinate $\lambda$, which corresponds to evolving forwards in the $t$ coordinate. We solve ($\ref{eq:hu127}$) over the range $\lambda = (-69260,-475)$, with initial conditions $x_{\rm i} = (R^{\rm i}_{\rm GR}/ (R^{\rm i}_{\rm GR} + 2.6))^{3}$ and $x'_{\rm i} = 0$. By solving ($\ref{eq:hu127}$) we obtain the oscillatory component $x_{\rm osc}$ of $x$, and hence $x \approx x_{\rm osc} + x_{\rm dr}$, where $x_{\rm dr}$ is given in ($\ref{eq:hu125}$). In fig.(\ref{fig:fg11}) we compare $x$ obtained in this section to $x = (R_{\rm GR}/R)^{3}$, where $R$ is the Ricci scalar obtained solving the full field equations numerically. As in the AB model, there are two effects; the non-linear oscillations of $x$ (and hence $R$) which decay as we evolve forwards in time, and a drift in $x$ away from its General Relativistic limit $x=0$. It is clear that the approximate solution obtained in this section closely mimics the full solution of section \ref{sec:s10}.

\section{Determination of the Hubble parameter}

In section \ref{sec:2}, we have calculated the Ricci scalar for the HSS and AB models, and have found that it will undergo non-linear oscillations and will drift away from the General Relativistic limit $R = -T/M_{\rm pl}^{2}$ as one goes back in time. From the Ricci scalar we can now calculate the Hubble parameter for the two models, using $\dot{H} + 2 H^{2} = R/6$. We look for a solution of the form $H = H_{\rm GR} + \delta H$, where $\delta H \ll H_{\rm GR}$. A solution of this form must exist in order for these models be viable; they must have a matter era for which $H \approx 2/3t$ for example, in order for normal structure formation to take place. Linearizing in $\delta H$, we find the following expression

\begin{equation}\label{eq:H10} {d  \over dt}\delta H + 4 H_{\rm GR} \delta H = {\delta R \over 6} .\end{equation}

\noindent Next, by using the time coordinate $\lambda$, where $\dot{\lambda} \propto (F'')^{-1/2}$, we obtain

\begin{equation} \label{eq:H11} {d  \over d \lambda} \left( a_{\rm GR}^{4} \delta H \right) = { \delta R a_{\rm GR}^{4}\over 6 \dot{\lambda}} \end{equation}

\noindent where $a_{\rm GR}$ is the General Relativistic scale factor. This expression can be integrated to obtain $\delta H$,

\begin{equation} \label{eq:H12} \delta H = {C \over a_{\rm GR}^{4}} + {1 \over 6 a_{\rm GR}^{4}} \int {a_{\rm GR}^{4} \delta R \over \dot{\lambda}} d \lambda \approx  {C \over a_{\rm GR}^{4}} + {1 \over 6 \dot{\lambda}} \int  \delta R  d \lambda ,\end{equation}

\noindent where $C$ is a constant of integration, and we have taken the slowly varying factor $a^{4}_{\rm GR}/\dot{\lambda}$ outside the integral, which is a good approximation for $\dot{\lambda} \gg 1$. We see that $\delta H$ contains two terms; one describing the oscillations of $H$ due to $\delta R$, and a term that goes like $\delta H \sim C a_{\rm GR}^{-4}$. The oscillatory component is suppressed by a factor of $\dot{\lambda}^{-1}$, and hence the oscillations of $R$ will not have a significant impact on the Hubble parameter. We note in obtaining $\delta H$, we have not had to specify either $\delta R$ or $\lambda$, and hence ($\ref{eq:H12}$) is valid for any model for which $R F''(R) \ll 1$, including the HSS and AB models.

In a similar manner, we can also consider the scale factor. By writing $H = \dot{a}/ a $ and expanding as $a = a_{\rm GR} + \delta a$, we have at linear order

\begin{equation}\label{eq:A1} H_{\rm GR} + \delta H = {\dot{a} \over a } \approx {\dot{a}_{\rm GR} \over a_{\rm GR}} + {\delta \dot{a} \over a_{\rm GR} } - {\dot{a}_{\rm GR} \delta a \over a_{\rm GR}^{2}} .\end{equation}

\noindent Next, by introducing the fast time coordinate $\lambda$, we write ($\ref{eq:A1}$) as

\begin{equation} \label{eq:A2} {d \over d \lambda} \left({ \delta a \over a_{\rm GR}} \right) \approx { \delta H \over \dot{\lambda}} ,\end{equation}

\noindent and by using ($\ref{eq:H12}$) and integrating, we arrive at the following approximate expression for $\delta a$,

\begin{equation} \label{eq:A3} \delta a \approx D a_{\rm GR} + a_{\rm GR} \int {\delta H \over \dot{\lambda}} d \lambda \approx D a_{\rm GR} + {a_{\rm GR} \over 6\dot{\lambda}^{2}} \int \delta R d \lambda + C a_{\rm GR} \int {d \lambda \over a_{\rm GR}^{4} \dot{\lambda}} ,\end{equation}

\noindent where $D$ is a constant of integration. $\delta a$ possesses an oscillatory term due to the rapid oscillations of $\delta R$, which is suppressed by a factor of $\dot{\lambda}^{-2} \ll 1$. We conclude that the rapid oscillations of the Ricci scalar has no significant impact on the scale factor for models in which $R F''(R) \ll 1$.

\section{\label{sec:3} Discussion and Conclusions}

In this paper we have solved the gravitational field equations for the HSS and AB models numerically and by using an alternative perturbative analysis. We have shown that the oscillations of the Ricci scalar are inherently non-linear, although they can be modeled as linear waves in a certain regime. However, for the AB model this linear regime is particularly restricted. Rather than using the earlier perturbative analysis, we have re-written the trace of the gravitational field equations as a damped, driven, non-linear oscillator. By solving this new equation, we have found that the Ricci scalar does indeed oscillate with high frequency, as predicted in ref.\cite{st}. As the amplitude and frequency of these waves grow to the past, we see non-linear behaviour becoming increasingly important. We have also found that the Ricci scalar does not exactly oscillate around its General Relativistic limit, but rather there is a highly suppressed drift.

In section \ref{sec:s10}, we solved the full gravitational field equations for both models numerically. By specifying the initial conditions at late times and evolving backwards, we observed that in general the Ricci scalar would evolve to a singularity after a finite time. This is another important consequence of the non-linear terms in the field equations, since no such behaviour was observed when using the linearized approach. By using our oscillator equation, we were able to explain why this singularity occurs, and found that it can be avoided by choosing initial conditions at some early time and evolving forwards. We were able to derive the allowed initial conditions for both the AB and HSS models. We have found that although the singularity could potentially be evaded with a suitable choice of initial conditions, the resulting Ricci scalar is likely to be unstable to perturbations away from a perfect matter era. To obtain a viable modified gravity model, it is clear that some method of regularizing this singularity is required.

Finally, we have considered the effect of the oscillations of $R$ on the Hubble parameter and scale factor. We have found that the oscillatory components of $H$ and $a$ are suppressed by factors of $\dot{\lambda}^{-1}$ and $\dot{\lambda}^{-2}$ respectively, where $\dot{\lambda} \propto \sqrt{F''}$. We have concluded that $\delta H$ and $\delta a$ will remain small, in spite of the potentially large oscillations of $R$. This conclusion is generic to models which have $R F'' \ll 1$.

Although we studied specifically the HSS and AB models in this paper, we believe that many of the results obtained are generic to $F(R)$ theories of gravity which satisfy $R F'' \ll 1$. We expect that the procedure adopted in section \ref{sec:2}, that is writing the trace of the gravitational field equations as a non-linear oscillator equation, will generalize to all models for which $R F'' \ll 1$, and if that is the case then the singularity encountered in this paper will be a common feature of $F(R)$ models.

\section{Acknowledgements}

SA is supported by a STFC studentship. We would like to thank A. Starobinsky for helpful comments on an earlier draft of the paper. During the preparation of this manuscript we became aware that A. Frolov has arrived at similar conclusions \cite{afq1}.

\end{document}